\documentclass[12pt]{article} 
\usepackage{epsfig}
\def\a{\alpha} \def\b{\beta}  \def\d{\delta}
   \def\h{\eta}
   \def\l{\lambda} \def\m{\mu}
\def\n{\nu} \def\o{\omega} \def\p{\pi} \def\q{\theta} 
\def\s{\sigma} \def\t{\tau}   \def\z{\zeta}
\def\D{\Delta} \def\F{\Phi} \def\G{\Gamma}

\def\ca{{\cal A}}      
      
\def\co{{\cal O}}

\def\bo{{\raise.15ex\hbox{\large$\Box$}}}               
\def\ltap{\raisebox{-.4ex}{\rlap{$\sim$}} \raisebox{.4ex}{$<$}}   
\def\face{{\raise.2ex\hbox{$\displaystyle \bigodot$}\mskip-2.2mu \llap
    {$\ddot
        \smile$}}}                                      
\def\dg{\sp\dagger}                                     

\def\Bar#1{\overline{#1}}                       
\def\leftrightarrowfill{$\mathsurround=0pt \mathord\leftarrow
  \mkern-6mu \cleaders\hbox{$\mkern-2mu \mathord- \mkern-2mu$}\hfill
        \mkern-6mu \mathord\rightarrow$}       
      \def\dvec#1{\vbox{\ialign{##\crcr
            \leftrightarrowfill\crcr\noalign{\kern-1pt\nointerlineskip}
        $\hfil\displaystyle{#1}\hfil$\crcr}}}           

\def\beq{\begin{equation}} \def\eeq{\end{equation}}
\def\beqa{\begin{eqnarray}} \def\eeqa{\end{eqnarray}}

\def\beqx{\begin{displaymath}} \def\eeqx{\end{displaymath}}

\def\beql{\begin{eqnarray}} \def\eeql{\end{eqnarray}}
\def\NO{\nonumber} \def\msb{\overline{\rm MS}}

\newcommand{\opi}{\rm\scriptstyle 1PI}
\newcommand{\pim}{\rm\scriptstyle PIM}

\newlength{\dinwidth} \newlength{\dinmargin}
\begin{document}

\begin{flushright}
  FSU-HEP-20010501\\
  NIKHEF/2001-01\\
  TTP01-03\\
  LBNL-43176\\
  \today\\
\end{flushright}

\vspace{8mm}
\begin{center}
  {\Large\bf\sc Sudakov Resummation and Finite Order Expansions\\ }
  \vspace{5mm} {\Large\bf\sc of Heavy Quark Hadroproduction Cross
    Sections}
\end{center}

\vspace{1cm}
\begin{center}
  {\large Nikolaos Kidonakis$^a$, Eric Laenen$^b$, Sven Moch$^{c}$, Ramona Vogt$^d$}\\
  \vspace{12mm}
  $^a${\it Department of Physics\\
    Florida State University, Tallahassee, FL 32306-4350, USA} \\
  \vspace{2.5mm}
  $^b${\it NIKHEF Theory Group\\
    P.O. Box 41882, 1009 DB Amsterdam, The Netherlands} \\
  \vspace{2.5mm}
  $^c${\it Institut f\"{u}r Theoretische Teilchenphysik \\
   Universit\"{a}t Karlsruhe, D--76128 Karlsruhe, Germany} \\
  \vspace{2.5mm}
  $^d${\it Nuclear Science Division,\\
    Lawrence Berkeley National Laboratory, Berkeley, CA 94720, USA \\
  and \\
  Physics Department,\\
    University of California at Davis, Davis, CA 95616, USA}
\end{center}
\vspace{1cm}

\begin{abstract}
  We resum Sudakov threshold enhancements in heavy quark
  hadroproduction for single-heavy quark inclusive and pair-inclusive
  kinematics.  We expand these resummed results and derive analytical
  finite-order cross sections through next-to-next-to-leading order.
  This involves the construction of next-to-leading order matching
  conditions in color space.  For the scale dependent terms we derive
  exact results using renormalization group methods. We study the
  effects of scale variations, scheme and kinematics choice on the
  partonic and hadronic cross sections, and provide estimates for top
  and bottom quark production cross sections.
\end{abstract} 
\thispagestyle{empty} \newpage \setcounter{page}{2}

\section{Introduction}\label{sec:introduction}

Long- and short-distance dynamics in inclusive hadronic hard-scattering cross sections 
are factorized in Quantum Chromodynamics (QCD)
into universal, non-perturbative parton distribution functions and
fragmentation functions, and perturbatively calculable hard scattering functions.
Remnants of long-distance dynamics
in a hard scattering function can, however, become large
in regions of phase space near partonic threshold
and dominate higher order corrections.
Such Sudakov corrections assume the form of 
distributions that are singular at partonic threshold.  
Threshold resummation organizes these double-logarithmic 
corrections to all orders, thereby extending
the predictive power of QCD to these phase space regions.

Early on \cite{Sterman:1987aj,Catani:1989ne}, the organization of such
corrections to arbitrary logarithmic accuracy was achieved for the
Drell-Yan cross section.  An equivalent level of understanding for
general QCD processes with more complex color structures at the Born
level has been achieved more recently
\cite{Contopanagos:1997nh,Kidonakis:1996aq,Kidonakis:1997gm,Kidonakis:1998bk, 
Kidonakis:1998nf,Bonciani:1998vc}.  The resummation of Sudakov
corrections in such processes to next-to-leading logarithmic (NLL)
accuracy requires understanding how these
structures mix under soft gluon radiation.
Many NLL-resummed cross sections 
have been calculated: heavy
quark hadro- \cite{Kidonakis:1996aq,Kidonakis:1997gm,Bonciani:1998vc,
Kidonakis:1997zd,Kidonakis:1998ed,Kidonakis:1998ei,Laenen:1998qw} 
and electroproduction \cite{Laenen:1998kp,Eynck:2000gz}, dijet production
\cite{Kidonakis:1998bk,Kidonakis:1998nf}, single-jet
production \cite{Kidonakis:2000gi}, Higgs production
\cite{Kramer:1996iq}, prompt photon production
\cite{Laenen:1998qw,Catani:1998tm,Catani:1999hs,Kidonakis:1999mk,
Kidonakis:1999hq} and
hadroproduction of electroweak bosons \cite{Kidonakis:1999ur}.
For a recent review  see Ref.~\cite{Kidonakis:1999ze}.
The formalism of Refs.~\cite{Contopanagos:1997nh,Kidonakis:1996aq,Kidonakis:1997gm,
 Kidonakis:1998bk,Kidonakis:1998nf} and the one\footnote{A recent 
study, Ref.~\cite{Sterman:2000pt}, compares these
formalisms for prompt photon production.} of Ref.~\cite{Bonciani:1998vc} both allow arbitrary
logarithmic accuracy. Processes involving
Born-level two-particle scattering may be described in either
single-particle inclusive or pair-inclusive kinematics.
NLL resummation was initially performed
in pair-inclusive kinematics \cite{Kidonakis:1996aq,Kidonakis:1997gm,
Kidonakis:1998bk, Kidonakis:1998nf}
and later extended to single-particle kinematics
\cite{Laenen:1998qw,Catani:1998tm}.

Resummed cross sections constitute an approximate sum of the complete
perturbative expansion if, at each order, the Sudakov corrections
dominate. The numerical evaluation of resummed cross sections 
requires a prescription to handle infrared renormalon
singularities.
Resummed cross sections may also be expanded to 
provide estimates of finite higher order corrections
which do not suffer from renormalon problems.
In this paper we shall employ the
resummed cross sections in the latter fashion:
as generating functionals of approximate perturbation theory.

Resummed results for heavy quark production at leading logarithmic
accuracy have been presented some time ago 
\cite{Laenen:1992af,Kidonakis:1995wz,Berger:1996ad,Catani:1996yz,
Kidonakis:1996jm,Berger:1998gz}.
In  our paper we expand the NLL resummed cross sections presented in
Refs. \cite{Kidonakis:1996aq,Kidonakis:1997gm,Laenen:1998qw}
and derive complete analytic expressions through 
next-to-next-to-leading order (NNLO)
for double-differential 
heavy quark cross sections in two different kinematics:
heavy quark pair-inclusive and single-heavy quark inclusive.
To achieve next-to-next-to-leading logarithmic (NNLL) accuracy we include 
color-coherence effects
and contributions due to soft radiation from
one-loop virtual graphs via matching conditions.

Our paper is organized as follows.
In section 2 we discuss both types of kinematics and their singular
functions.  Section 3 describes the construction of the resummed cross
sections. In section 4 we expand the resummed cross section to
NLO and NNLO and present analytical NNLL
double-differential cross sections at each order,
with both types of kinematics,
in the gluon-gluon ($gg$) and 
quark-antiquark ($q\Bar{q}$) channels.
We numerically study the inclusive partonic and hadronic cross sections
in sections 5 and 6 respectively.
Our conclusions are presented in section 7.  Appendix A 
contains the NLO matching terms for pair-inclusive kinematics.
In appendix B we collect all our explicit expressions for the 
NLO and NNLO differential cross sections.

A companion study \cite{Kidonakis:2000ui}, also addressing 
third and fourth order corrections, contributions from
subleading logarithms, and some differential distributions, 
already featuring some of 
the second order results for inclusive cross sections derived in this paper, 
was recently presented by one of us.

\section{Kinematics and cross sections}\label{sec:kinem-cross-sect}

The kinematics of inclusive heavy quark hadroproduction depend on which final
state momenta are reconstructed.  In threshold resummation this
kinematics choice manifests itself at next-to-leading logarithmic level
\cite{Laenen:1998qw}. We discuss
two types of near-elastic kinematics in heavy quark hadroproduction,
one-particle inclusive (1PI) and pair-invariant mass (PIM) kinematics.

\subsection{One-particle inclusive (1PI) kinematics}\label{sec:one-part-incl}

Heavy quark hadroproduction in 1PI kinematics is defined by
\begin{eqnarray}
  \label{eq:3}
h_1(P_1) +\, h_2(P_2) &\longrightarrow& {\rm{Q}}(p_1) +\,
X[{\Bar{\rm{Q}}}](p_X)\, , 
\end{eqnarray}
where $h_1$ and $h_2$ are hadrons, $X[\overline{\rm{Q}}]$ denotes any
allowed hadronic final state containing at least the heavy antiquark,
and ${\rm{Q}}(p_1)$ is the identified heavy quark with mass $m$.  The
hadronic invariants in this reaction are
\begin{equation}
  \label{eq:5}
S = (P_1+P_2)^2 \quad,\quad T_1 = (P_2-p_1)^2-m^2 \quad,\quad U_1 =
(P_1-p_1)^2-m^2\,, 
\end{equation}
and
\begin{equation}
  \label{eq:6}
S_4 = S+T_1+U_1\,  ,  
\end{equation}
where $S_4$ is a measure of the inelasticity of the hadronic reaction
(\ref{eq:3}).  Near threshold, reaction (\ref{eq:3}) is dominated by
the partonic subprocesses
\begin{eqnarray}
  \label{eq:7}
q(k_1) +\, {\Bar{q}}(k_2) &\longrightarrow& {\rm{Q}}(p_1) +\,
X'[{\Bar{\rm{Q}}}](p_2')\, ,
\\[1ex]
\label{eq:8}  
g(k_1) +\, g(k_2) &\longrightarrow& {\rm{Q}}(p_1) +\,
X'[{\Bar{\rm{Q}}}](p_2')\, . 
\end{eqnarray}
If $X'[{\Bar{\rm{Q}}}](p_2')={\Bar{\rm{Q}}}(\bar{p}_2)$, the reaction
is at partonic threshold and the heavy antiquark has momentum
$\bar{p}_2$.  Note that threshold production does not
mean that the heavy quarks are produced at rest.  The $qg$ and
$\Bar{q}g$ channels contribute at one order higher in $\alpha_s$ than the
reactions (\ref{eq:7}) and ({\ref{eq:8})\footnote{For example,
 for top quark pair production at the Tevatron
  $p\bar{p}$ collider these subprocesses contribute
  only 1\% of the total cross section 
\cite{Nason:1988xz,Beenakker:1989bq,Beenakker:1991ma}.}.  
The partonic invariants corresponding to
(\ref{eq:5}) are
\begin{eqnarray}
  \label{eq:9}
s\,=\,(k_1+k_2)^2\, , \hspace*{10mm}t_1\,=\,(k_2-p_1)^2-m^2\, ,
\hspace*{10mm}u_1\,=\,(k_1-p_1)^2-m^2\, .
\end{eqnarray}
The invariant $s_4 = (p'_2)^2-m^2$ which measures the inelasticity of
the partonic reactions (\ref{eq:7}) and (\ref{eq:8}) is related to the
other partonic invariants by
\begin{eqnarray}
  \label{eq:10}
s_4 &=& s + t_1 + u_1\, .  
\end{eqnarray}
The inclusive partonic cross section may be calculated from
\begin{eqnarray}
  \label{eq:11}
\sigma_{ij}(s,m^2) &=&
\int\limits_{s(1-\b)/2}^{s(1+\b)/2}\,d(-t_1) \int\limits_{0}^{s_4^{\rm
    max}}\,ds_4\,\,\, \frac{d^2\sigma_{i j}(s,t_1,s_4)}{dt_1\, ds_4}\, 
\end{eqnarray}
where $\beta=\sqrt{1-4m^2/s}$ and
\begin{eqnarray}
  \label{eq:12}
s_4^{\rm max} &=& s + t_1 + \frac{s m^2}{t_1}\, .    
\end{eqnarray}
The recoil momentum $p_2'$ may be split into the momentum at
threshold, $\bar{p}_2$, and the momentum of any additional radiation
above threshold, $k$, i.e. $p_2' = \bar{p}_2 +k$.  Then, when $k^2$ is
small, we can define a dimensionless weight $w_{\opi}$
\cite{Contopanagos:1997nh} that measures the distance from threshold
in 1PI kinematics which can in turn be expressed in terms of a vector
$\zeta_{\opi}^\mu$:
\begin{equation}
  \label{eq:13}
w_{\opi} = {s_4\over m^2} \simeq {2 \bar{p}_2\cdot k \over m^2} 
\equiv {2 \zeta_{\opi}\cdot k \over m}\,.
\end{equation}
Ref.~\cite{Beenakker:1991ma} contains an exact NLO treatment of this kinematics at the parton and hadron
levels.

\subsection{Pair-invariant mass (PIM) kinematics}\label{sec:pair-invariant-mass}

Heavy quark hadroproduction in pair-invariant mass kinematics is
defined by
\begin{eqnarray}
  \label{eq:14}
h_1(P_1) +\, h_2(P_2) &\longrightarrow& {\rm{Q}}{\Bar{\rm{Q}}}(p')
+\, X(p_X)\, .  
\end{eqnarray}
At the parton level, the important reactions are
\begin{eqnarray}
\label{q-qbar-annihilation-0_PIM}
q(k_1) +\, {\Bar{q}}(k_2) &\longrightarrow& {\rm{Q}}{\Bar{\rm{Q}}}(p')
+\, X'(k)\, ,
\\[1ex]
\label{gluon-gluon-fusion-0_PIM}
g(k_1) +\, g(k_2) &\longrightarrow& {\rm{Q}}{\Bar{\rm{Q}}}(p') +\,
X'(k)\, , 
\end{eqnarray}
where $p'^2=M^2$ is the pair-mass squared.  If $X'(k)=0$, the reaction
is at partonic threshold with $M^2=s$. Then
\begin{eqnarray}
\label{tupidef}  
t_1 &=& - \frac{M^2}{2} \left( 1 - \beta_M\, {\rm{cos}} \q \right)\, , \nonumber \\
u_1 &=& - \frac{M^2}{2} \left( 1 + \beta_M\, {\rm{cos}} \q \right)\, 
\end{eqnarray}
where $\beta_M=\sqrt{1-4m^2/M^2}$ and $\theta$ is the scattering angle
in the parton center-of-mass frame.  In PIM kinematics, the inclusive
partonic cross section may be calculated from
\begin{eqnarray}
\label{totalpartoncrs_PIM}
\sigma_{ij}(s,m^2) &=& 
\int\limits_{s(1-\b)/2}^{s(1+\b)/2}\,d(-t_1)
\int\limits_{M^2_{\rm min}}^{s}\,dM^2
\,\,\, \frac{d^2\sigma_{ij}
(s,M^2,t_1)}{dM^2\, dt_1}\, .  
\end{eqnarray}
with
\begin{equation}
  \label{eq:4}
  M^2_{\rm min} = \frac{-m^2}{(t_1/s)^2+(t_1/s)}\,.
\end{equation}
The weight $w_{\pim}$ measures the inelasticity in PIM
kinematics,
\begin{equation}
  \label{eq:17}
w_{\pim} = 1-z={s-M^2\over s} \simeq {2 \bar{p}'\cdot k \over s}
\equiv {2 \zeta_{\pim}\cdot k  \over \sqrt{s}}\, ,
\end{equation}
where $z \equiv M^2/s$. Equation~(\ref{eq:17}) defines the vector
$\zeta_{\pim}^\mu$ in terms of the heavy quark pair momentum at
threshold (indicated by the bar).  An exact NLO treatment of this
kinematics at the parton and hadron levels may be found in
Ref.~\cite{Mangano:1992jk}.

\subsection{Inclusive cross section}\label{sec:total-cross-section}

The inclusive hadronic cross section is obtained by convoluting the
inclusive partonic cross sections, Eqs.~(\ref{eq:11}) and
(\ref{totalpartoncrs_PIM}),
with a parton flux factor $\F_{ij}$,
\begin{eqnarray}
  \label{eq:19}
\F_{ij}(\t,\mu^2) &=& \t\,\, \int\limits_{0}^{1}
dx_1\,\, \int\limits_{0}^{1} dx_2\,\, \d(x_1x_2 - \t)\,\,
\phi_{i/h_1}(x_1,\mu^2)\, \phi_{j/h_2}(x_2,\mu^2)\, ,
\end{eqnarray}
where $\phi_{i/h}(x,\mu^2)$ is the density of partons of flavor $i$ in
hadron $h$ carrying a fraction $x$ of the initial hadron $h$ momentum, at
factorization scale $\mu$.  Then
\begin{eqnarray}
\label{totalhadroncrs}
\sigma_{h_1h_2}(S,m^2) &=& \sum\limits_{i,j = q,{\Bar{q}},g} \,\,
\int\limits_{4m^2/S}^{1}\,\frac{d\t}{\t}\,\,\F_{ij}(\t,\mu^2)\,\,
\sigma_{ij}(\t S,m^2,\mu^2)\,  \\
&=& \sum\limits_{i,j = q,{\Bar{q}},g} \,\,
\int_{-\infty}^{\log_{10}(S/4m^2-1)} d\log_{10}\eta \, \frac{\eta}{1+\eta} 
\ln(10) \, \F_{ij}(\eta,\mu^2)\,\,
\sigma_{ij}(\eta,m^2,\mu^2)\, \nonumber
\end{eqnarray}
where
\begin{eqnarray} 
\label{eq:etadef}
\eta = \frac{s}{4 m^2} - 1\, = \, \frac{\tau S}{4 m^2} -1\, .
\end{eqnarray}
The sum is over all massless parton flavors.  The second equality
in Eq.~(\ref{totalhadroncrs}) 
facilitates interpretation of the figures in section 5.

\subsection{Singular functions}\label{sec:singular-functions}

The double-differential hadronic cross section for reactions
(\ref{eq:3}) and (\ref{eq:14}), $d^2\s_{h_1 h_2}/dV\,dW$, enjoys the
factorization \cite{Collins:1989gx}
\begin{eqnarray}
\label{stdfact}
S^2\, \frac{d^2\s_{h_1 h_2}(S,V,W)}{dV\,dW} &=&
\sum_{i,j=q,\Bar{q},g} \,\int\limits_{x_1^-}^{1}\frac{dx_1}{x_1}
\,\int\limits_{x_2^-}^{1}\frac{dx_2}{x_2} \,\phi_{i/h_1}(x_1,\mu^2)\,
\phi_{j/h_2}(x_2,\mu^2)\,\, 
\NO\\
&\,&\times\,\, \o_{ij}(w_K,s,t_1,u_1,m^2,\mu^2,\alpha_s(\mu)) 
\;+{\cal O}
\left(
  \frac{\Lambda^2}{m^2}
\right)\;, 
\end{eqnarray}
where $\o_{ij}$ is the partonic cross section, or hard scattering function, 
whose dependence
on kinematics is indicated by the weight $w_K$ with 
$K=(\mathrm{1PI,PIM})$.  The
variables $V$ and $W$ represent either 1PI or PIM kinematic variables
such as the transverse momentum and rapidity of either a single heavy
quark or the heavy quark pair, respectively.  The parton momentum
fractions $x_1$ and $x_2$ have lower limits $x_1^-$ and $x_2^-$ which
depend\footnote{When $V,W = T_1,U_1$ in 1PI kinematics one has $x_1^-
  = -U_1/(S+T_1)$ and $x_2^-= -x_1 T_1/(x_1S+U_1)$ while for $V,W =
  M^2,Y$ in PIM kinematics one has $x_1^- = M \exp(Y)/\sqrt{S}$ and
  $x_2^-=M \exp(-Y)/\sqrt{S}$ where $Y$ is the pair rapidity.}  on $V$
and $W$.

We shall resum the higher order
logarithmic contributions to $\o_{ij}$ that are singular at threshold.
The arguments of these logarithms are the weights $w_K$.  Thus, the
1PI singular functions are plus distributions in $s_4$,
\begin{eqnarray}
\label{s4distdef}
\left[{\ln^{l}(s_4/m^2)\over s_4}\right]_+ = \lim_{\Delta \rightarrow
  0} \Bigg\{ {\ln^{l}(s_4/m^2)\over s_4} \theta(s_4 -\Delta) +
\frac{1}{l+1}\ln^{l+1}\Big({\Delta\over m^2}\Big)\, \delta({s_4})
\Bigg\}\,,
\end{eqnarray}
while the PIM singular functions are plus distributions in $1-z$
\begin{equation}
\label{1mzdistdef}
\left[{\ln^{l}(1-z)\over 1-z}\right]_+ = \lim_{\delta \rightarrow 0}
\Bigg\{ {\ln^{l}(1-z)\over 1-z} \theta(1-z -\delta) +
\frac{1}{l+1}\ln^{l+1}(\delta)\, \delta({1-z}) \Bigg\}\, .
\end{equation}
These singular functions yield finite large results when
convoluted with smooth but rapidly changing functions such as parton
densities.  Note that we have normalized the 1PI functions to have
mass dimension $-2$.  
Because our paper deals mostly with finite order cross sections,
we denote corrections as leading logarithmic (LL) if $l=2i+1$
at order $O(\alpha^{i+3}_s),\; i=0,1,\ldots$, as next-to-leading
logarithmic (NLL) if $l=2i$, etc. 

It is often convenient to work in
moment space, defined by the Laplace transform with respect to $w_K$
\begin{equation} {\tilde f}(N) = \int\limits_0^\infty dw_K
\,e^{-N w_K} f(w_K)\,.
\label{laplacetfm}
\end{equation} 
The upper limit of this integral is not very important, and may be set
to 1, where $\ln w_K=0$.  Under Laplace transformations, the plus
distributions in Eqs.~(\ref{s4distdef}) and (\ref{1mzdistdef}) 
become linear combinations
of $\ln^k(\tilde{N})$ with $k\leq l+1$ and $\tilde{N}=N\exp(\gamma_E)$
where $\gamma_E$ is the Euler constant\footnote{All the $N$'s in
the remainder of this work are actually $\tilde{N}$, unless specified
otherwise.}.  The precise correspondence to 
second order can be found in Ref.~\cite{Catani:1989ne}, and 
through fourth order in Ref.~\cite{Kidonakis:2000ui}.  
We shall almost always talk about logarithmic accuracy in 
momentum space. Occasionally in the following we denote in moment space 
leading logarithmic corrections 
at order $O(\alpha^{i+2}_s),\; i=1,2,\ldots$ to correspond to $l=2i$, NLL
ones to $l=2i-1$ etc. Although the Laplace transformation of the $l$'th 
plus distribution generates lower powers of $\ln N$ besides $\ln^{l+1} N$,
there should be no confusion in practice.

We work in axial gauge, $n \cdot A=0$, with the gauge vector
chosen as $n^\mu=\zeta_K^\mu$.  The implicit renormalization
scheme is that of Ref.~\cite{Collins:1978wz} and $m$ is a
pole mass.  The renormalization scale $\mu_R$ is assumed to be equal
to the factorization scale $\mu$, except where explicitly indicated
otherwise.
The scale dependence of the coupling constant 
is controlled by the QCD $\b$-function
 \begin{eqnarray}
\label{beta-fcts}
\mu\frac{d}{d\mu}\, \frac{\alpha_s(\mu)}{\pi} = 
\b(\a_s(\m)) = - 2b_2 \left(\frac{\a_s(\m)}{\p}\right)^2 - 2b_3
\left(\frac{\a_s(\m)}{\p}\right)^3 - \dots \,,
  \end{eqnarray}
where $b_2=(11 C_A - 2n_f)/12$, and $b_3=(34 C_A^2-2n_f(3 C_F + 5 C_A))/48$.

\section{Resummed cross sections}\label{sec:resumm-cross-sect}

Here we describe the threshold resummation of the heavy quark 
hadroproduction cross
section in PIM \cite{Kidonakis:1996aq,Kidonakis:1997gm} 
and  1PI \cite{Laenen:1998qw} kinematics.  Both resummations 
can be presented simultaneously
using the methods and results of 
Refs.~\cite{Contopanagos:1997nh,Kidonakis:1996aq,Kidonakis:1997gm,
Kidonakis:1998bk,Kidonakis:1998nf,Laenen:1998qw}.
  
\subsection{Refactorization}\label{sec:factorization}

The resummation of the singular functions in
Eqs.~(\ref{s4distdef}) and (\ref{1mzdistdef}) in the perturbative
expansion of $\omega_{ij}$ rests upon the refactorization of $\omega_{ij}$
into separate functions of the jet-like, 
soft, and off-shell quanta that contribute 
to its quantum corrections.  
This refactorization, valid in the
threshold region of phase space, is pictured in
Fig.~\ref{fig:threshconvfig}.
\begin{figure}[htbp]
  \begin{center}
\epsfig{file=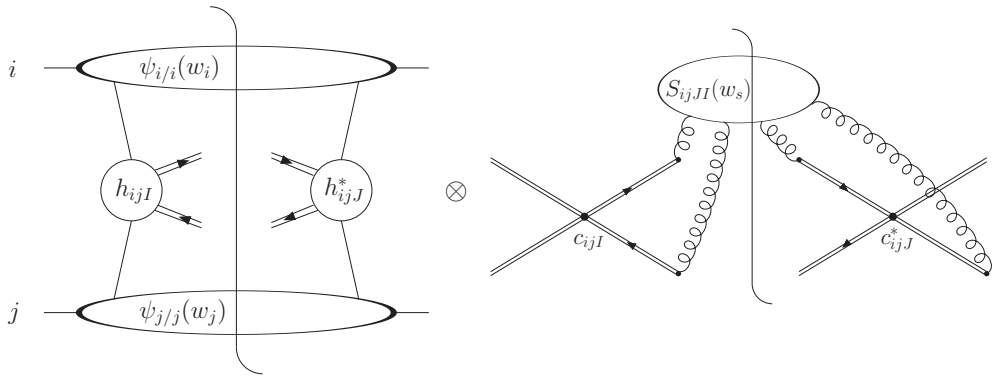,%
bbllx=8pt,bblly=550pt,bburx=575pt,bbury=775pt,angle=0,width=14cm}
    \caption{Refactorized form of heavy quark partonic
cross section near threshold. }
    \label{fig:threshconvfig}
  \end{center}
\end{figure}
Each of the functions $\psi$, $h$, and $S$
organizes large corrections
corresponding to a particular region of phase space.
The meaning of the $I,J$ indices will be given shortly
and the coefficients $c_{ij\,I}$ are given below
in Eq.~(\ref{gluon-gluon-basis}).
Factorizations of this type have been discussed earlier for deep-inelastic
scattering and Drell-Yan production \cite{Sterman:1987aj}, 
heavy quark hadro- and electroproduction 
\cite{Kidonakis:1996aq,Kidonakis:1997gm,Bonciani:1998vc,Laenen:1998kp}, 
dijet \cite{Kidonakis:1998bk} and Higgs \cite{Kramer:1996iq} production, and 
single-particle/jet inclusive cross
sections \cite{Laenen:1998qw,Kidonakis:1999ur}. They may be generalized to
include recoil effects \cite{Laenen:2000de}.

Figure~\ref{fig:threshconvfig} indicates that each factorized
function depends on its own weight function
\cite{Contopanagos:1997nh}. In essence, the choice of
working in 1PI or PIM kinematics depends on how the
total weight $w_K$ is constructed from the individual contributions
of each weight function \cite{Kidonakis:1998bk,Laenen:1998qw}:
\begin{eqnarray}
  \label{eq:kindef}
&{\rm PIM:}&\qquad\qquad w_{\pim}\, =\, w_i+w_j+w_s \\
&{\rm 1PI:}&\qquad\qquad w_{\opi}\,=\, w_i\left(\frac{-u_1}{m^2}\right)
                             +w_j\left(\frac{-t_1}{m^2}\right)+w_s \,. 
\end{eqnarray}
In the case of 1PI kinematics it is convenient to define 
\begin{eqnarray}
  \label{eq:Ndef}
&{\rm 1PI:}&\qquad\qquad N_u = N\left(\frac{-u_1}{m^2}\right),\qquad
                             N_t = N\left(\frac{-t_1}{m^2}\right) \, .
\end{eqnarray}
In both kinematics, the moments of the $ij$ partonic cross section, 
Eqs.~(\ref{eq:7}), (\ref{eq:8}), (\ref{q-qbar-annihilation-0_PIM}), and
(\ref{gluon-gluon-fusion-0_PIM}), 
can be written in refactorized form, up to ${\cal O}(1/N)$
corrections, as 
\cite{Kidonakis:1996aq,Kidonakis:1997gm,Kidonakis:1998bk,Laenen:1998qw}
\begin{eqnarray}
\label{sigresum}
\tilde{\o}_{ij}(N,s,t_1,u_1,m^2,\mu^2,\alpha_s(\mu)) \; &=& h^*_{ij\,
  J}(\zeta_K,m^2,\m^2)\; {\tilde S}_{ij\, JI}\!\left({m\over N\mu},
  \zeta_K\right)  h_{ij\, I}(\zeta_K,m^2,\m^2) \; \nonumber \\[1ex]
& &\times \, \left[ {{\tilde\psi}_{i/i}(N_u,k_1\cdot\zeta_K/\mu)\,
    {\tilde\psi}_{j/j}(N_t,k_2\cdot\zeta_K/\mu) \over
    {\tilde\phi}_{i/i}(N_u,\mu)\, {\tilde\phi}_{j/j}(N_t,\mu)} \right]
\, . \end{eqnarray}
The indices\footnote{Unless
stated otherwise, repeated indices are summed over.} $I$ and $J$ take
values in the color-tensor space
spanned by the invariant tensors
that combine the SU(3) representations of the external partons at 
threshold into a singlet. The 
vector $\zeta_K^\mu$ defines the kinematics, see section 2.
The ``incoming-jet functions''  ${\tilde \psi}_{i/i}$
describe the dynamics
of partons moving collinearly to the incoming parton $i$. The 
distributions ${\tilde \phi}_{i/i}$ are defined at fixed light-cone
momentum while the functions
${\tilde \psi}_{i/i}$ are defined at fixed $k_i\cdot \zeta_K$.
The (real) ${\tilde \psi}_{i/i}$ functions include all leading and some
next-to-leading singular functions and are diagonal in color-tensor space.
The function ${\tilde S}_{ij\,JI}$, a Hermitian matrix in color-tensor space,
summarizes the dynamics of soft gluons that are not collinear to the
incoming partons and contains the remaining next-to-leading
contributions.  Note that (next-to-leading) singularities 
associated with soft radiation from the outgoing heavy quarks\footnote{Were we 
to treat the heavy quarks as 
massless, each heavy quark would be assigned its own jet function
\cite{Kidonakis:1998bk,Kidonakis:1998nf,Laenen:1998qw,Catani:1998tm}. However, 
because the heavy quark mass prevents
collinear singularities, all singular functions arising from the
final state heavy quarks are due to soft gluons.} 
are included in ${\tilde S}_{ij\,JI}$ \cite{Kidonakis:1996aq,Kidonakis:1997gm}.
Finally, the Hermitian matrix $H_{ij IJ}\equiv h_{ij I}h^*_{ij J}$ 
incorporates the effects of far off-shell partons and contains 
no singular functions. 

The jet and soft functions in Eq.~(\ref{sigresum}) can each be
represented as operator matrix elements
\cite{Kidonakis:1996aq,Kidonakis:1997gm,Kidonakis:1998bk,
Kidonakis:1998nf,Laenen:1998qw}. 
The refactorization of the cross section in
Eq.~(\ref{sigresum}) can in fact be seen as a separation of near-threshold 
degrees of freedom into distinct effective field
theories \cite{Contopanagos:1997nh,Kidonakis:1996aq,Kidonakis:1997gm}. In each 
effective theory a
resummation of the singular functions may be performed via appropriate
evolution equations.  The resummed cross section
is composed of contributions from all these effective theories which 
must be matched together properly,
at a specified scale and to a certain order in perturbation theory. 
In this paper we match to NLO, leading to NNLL accuracy
in our finite order expansions. Our matching scale is 
the heavy quark mass $m$. The matching procedure is 
described in section 4.

In the following we describe the resummation of the jet and soft functions.  
Most of these steps have been discussed previously
in the literature \cite{Kidonakis:1996aq,Kidonakis:1997gm,Kidonakis:1998bk,
Laenen:1998qw} so that our description is brief.

\subsection{Resummed Jet and Soft Functions}\label{sec:resummed-jet-soft}

The (exponentiated) $N$ dependence of the ratio ${\tilde \psi}_{i/i}/{\tilde
  \phi}_{i/i}$ in Eq.~(\ref{sigresum}) follows from its factorization
properties
\cite{Sterman:1987aj,Contopanagos:1997nh}.  The function ${\tilde
 \psi}_{i/i}$ obeys, beside a renormalization
group equation, an evolution equation governing the energy dependence,
expressed as gauge-dependence
\cite{Sterman:1987aj,Contopanagos:1997nh,Collins:1981uk,Li:1998is}.
Solving these equations leads to the resummed form of this ratio.
The function $\psi$ has been defined in PIM kinematics for $i= q$
\cite{Sterman:1987aj} and $g$ \cite{Kidonakis:1998bk} with the 1PI
equivalents given in Refs.~\cite{Laenen:1998kp,Laenen:1998qw}.  The
resummation of Sudakov logarithms in 1PI kinematics was verified
\cite{Laenen:1998kp} to trace those in PIM kinematics
\cite{Sterman:1987aj}.  Expressed in terms
of $w_i$, the ${\tilde
  \psi}_{i/i}$ are the same in PIM and 1PI kinematics.

The resummed ratio
${\tilde \psi}_{i/i}/{\tilde \phi}_{i/i}$ is 
\begin{eqnarray}
\label{psi/phi}
\ln {\tilde{\psi}_{i/i}\left(N, 2k_i\cdot\zeta_K / \mu \right) \over
  \tilde{\phi}_{i/i}(N,\mu)} \Big|_{\mu=2k_i\cdot\zeta_K} &=&
\tilde{E}_i\left(N,2 k_i\cdot\zeta_K\right) \,,
\end{eqnarray} 
with the $\overline{\rm MS}$ exponent
\begin{eqnarray}
\label{omegaexp-MSbar}
\tilde{E}_i\left(N,2 k_i\cdot\zeta_K\right) &=& \int\limits^\infty_0
dw\,\frac{(1-e^{-N w})}{w}\; \Bigg \{\int\limits^{1}_{w^2}
\frac{d\lambda}{\lambda} A_i\left[\alpha_s(\sqrt{\lambda} \,
  2k_i\cdot\zeta_K)\right]
\\
&\ & \mbox{} + \frac{1}{2}\nu^i\left[\alpha_s\left(w\,
2k_i\cdot\zeta_K\right )\right]\,
\Bigg \}\, .\NO
\end{eqnarray}
The functions $A_i$ and $\nu^i$ differ by color factors for $i=q$ and $g$
so that
\begin{eqnarray}
A_q(\alpha_s) & = & {\alpha_s\over \pi} C_F +\left({\alpha_s\over \pi}\right)^2
  \left(\frac{1}{2} C_F\,K \right)
  +\ldots , \hspace{5mm}
A_g(\alpha_s)\; = \;   {\alpha_s\over \pi}C_A+\left({\alpha_s\over \pi}\right)^2
 \left(\frac{1}{2}C_A\, K\right)
 +\ldots ,\nonumber \\
\label{nudef}
\nu^q\left(\alpha_s\right) & = & \frac{\a_s}{\p} (2 C_F)+\ldots , 
\hspace{35mm} 
\nu^g(\alpha_s)\;=\; \frac{\a_s}{\p} (2 C_A)+\ldots ,
\end{eqnarray}
with $K= C_A\; \left ( {67/ 18}-{\pi^2/ 6 }\right ) - {5n_f/9}$
\cite{Kodaira:1982nh}. 

To incorporate the effects of scale changes in
finite-order $\a_s$ expansions, we note that 
the ratio ${\tilde{\psi}}_{i/i}/{\tilde{\phi}}_{i/i}$
transforms under renormalization scale, 
$\mu_R$, and factorization scale, $\mu$, as 
\begin{eqnarray}
\label{rge_psi/phi_mur}
\mu_R\frac{d}{d\mu_R} \ln {\tilde{\psi}_{i/i}\left(N, 2k_i\cdot\zeta_K / \mu_R
  \right) \over \tilde{\phi}_{i/i}(N,\mu)} &=& \gamma_{\psi_i}
 \; = \;2 \gamma_{i}\left(\alpha_s(\mu_R)\right)\,, \\
\label{rge_psi/phi_muf}
\mu\frac{d}{d\mu} \ln {\tilde{\psi}_{i/i}\left(N, 2k_i\cdot\zeta_K / \mu_R
  \right) \over \tilde{\phi}_{i/i}(N,\mu)} &=& 
- 2\gamma_{i/i}\left(N,\alpha_s(\mu)\right) \,,
\end{eqnarray}
where $\gamma_i$ and $2\gamma_{i/i}$ are, respectively, the anomalous dimension of the quantum field
$i$ and of the operator whose
matrix element represents \cite{Curci:1980uw,Collins:1982uw}
the $\overline{\rm MS}$-density ${\tilde \phi}_{i/i}$.
The anomalous dimension for the renormalization scale 
dependence of $\tilde\psi_{i/i}$ \cite{Kidonakis:1998bk} in
Eq.~(\ref{rge_psi/phi_mur}) does not depend on the moment 
$N$ because the ultraviolet (UV) divergences of $\tilde\psi_{i/i}$ are only 
due to wavefunction 
renormalization of parton $i$.  Hence $\gamma_{\psi_i} = 2\gamma_i$ where
$\gamma_i$ is calculated in the axial gauge.
The anomalous dimension $\gamma_{i/i}$ controlling the factorization scale
dependence does depend on $N$.
The axial gauge anomalous dimensions are (neglecting $\co(1/N)$ terms)
\begin{eqnarray}
\label{psianomdim_q}
\gamma_q\left(\alpha_s\right) &=&  \frac{3}{4}\, C_F\, \frac{\a_s}{\p}+\ldots\, ,\\
\label{MSbaranomdim_q}
\gamma_{q/q}\left(N,\alpha_s\right) &=& - \frac{\a_s}{\p}
\left(C_F \ln N -
  \frac{3}{4}\, C_F \,\right)- \left(\frac{\a_s}{\p}\right)^2
\left(\frac{1}{2}C_F K\ln N\right) +\ldots\,, \\
\label{psianomdim_g}
\gamma_g\left(\alpha_s\right) &=&  b_2\, \frac{\a_s}{\p}+\ldots \, , \\
\label{MSbaranomdim_g}
\gamma_{g/g}\left(N,\alpha_s\right) &=& - \frac{\a_s}{\p}
\Biggl( C_A \ln N - b_2 \Biggr)-\left(\frac{\a_s}{\p}\right)^2\left(\frac{1}{2}C_A K\ln N\right)
 +\ldots\, , \end{eqnarray} where
$b_2$ is given below Eq.~(\ref{beta-fcts})
and $K$ is given below Eq.~(\ref{nudef}).

The composite operator \cite{Kidonakis:1998nf} that defines ${\tilde S}_{ij\,JI}$ 
contains UV divergences
beyond the self energies of its external legs.
An additional renormalization removes these extra divergences.
However, the product $H_{ij IJ} {\tilde S}_{ij JI}$ 
has no UV divergencies beyond those
taken into account by the renormalization of the elementary fields and
couplings of the theory, because 
the renormalization-group invariant cross section and this product only
differ by the functions ${\tilde \psi}_{i/i}$ which 
are renormalized as if they were (the ``square'' of) elementary 
fields (\ref{rge_psi/phi_mur}). 
Hence the extra UV
divergencies of ${\tilde S}_{ij JI}$ 
are balanced by similar ones in $H_{ij IJ}$. Therefore, the
unrenormalized expressions ${\tilde S}_{ij JI}^{\rm bare}$ and $H_{ijIJ}^{\rm 
bare}$ renormalize multiplicatively
\cite{Kidonakis:1996aq,Kidonakis:1997gm,Kidonakis:1998bk} 
\begin{eqnarray}
\label{rge_behaviour_hard_qq}
H^{\rm{bare}}_{ij\, IJ} &=& Z_i^{-1}\,Z_{j}^{-1}\,
\Bigl(Z_{ij\,S} \Bigr)_{IK}^{-1} H_{ij\, KL}
\left(Z_{ij\,S}^{\dg}\right)_{LJ}^{-1}\, , \\[1ex]
\label{rge_behaviour_soft}
{\tilde S}^{\rm{bare}}_{ij\,JI} &=& \left(Z_{ij\,S}^{\dg}\right)_{JK} 
{\tilde S}_{ij\,
  KL} \Bigl(Z_{ij\,S} \Bigr)_{LI}\, .  \end{eqnarray} 
The factors $Z_i$ denote the renormalization constants of the external fields
$i=q,\bar{q},g$. For a given parton pair $ij$, the $(Z_{ij\,S})_{IJ}$ 
constitute a matrix, in color-tensor space, of renormalization constants 
for the overall renormalization of
the soft function. Note that
$(Z_{ij\,S})_{IJ}$ also includes the wave function
renormalization of the external heavy quark legs.

Equation (\ref{rge_behaviour_soft}) leads directly to a renormalization group 
equation for the matrix $\tilde{S}_{ij JI}$ 
\cite{Kidonakis:1996aq,Kidonakis:1997gm,Kidonakis:1998bk}
\begin{eqnarray}
\label{softevol_1}
\m_R \frac{d}{d\m_R}\, {\tilde{S}}_{ij\, JI} &=& - \G^{ij\,\dg}_{S,JK}\,
{\tilde{S}}_{ij\, KI} - {\tilde{S}}_{ij\, JK}\, \G^{ij}_{S,KI}\, . 
\end{eqnarray} 
The soft anomalous dimension matrix $\G^{ij}_{S,IJ}$ is obtained
from the renormalization constants $(Z_S)_{IJ}$, computed in 
$d=4-\epsilon$ dimensions, by \cite{Botts:1989kf}
\begin{equation}
  \label{eq:sadgam}
  \G^{ij}_{S,IJ} = -{g_s\over 2} {\partial\over \partial g_s} {\rm Res_{\epsilon\rightarrow 0}}
 (Z_{ij\,S})_{IJ}(\alpha_s,\epsilon)\,.
\end{equation}
with $g_s^2 = 4 \pi \alpha_s$.
The solution of Eq.~(\ref{softevol_1}) is in general expressed in terms of
path-ordered exponentials \cite{Kidonakis:1996aq,Kidonakis:1997gm,
Kidonakis:1998bk}, see below.  

An explicit expression for $\G^{ij}_{S,IJ}$ requires a choice of
basis tensors in color-tensor space. We choose an $s$-channel
singlet-octet basis 
\begin{eqnarray}
\label{gluon-gluon-basis}
(c_{q\Bar{q}\,I})_{mnkl} &=& \left( \d_{mn}\d_{kl}\, , (T^c_F)_{nm}\,(T^c_F)_{kl}
\right)   , \nonumber \\
(c_{gg\,I})_{abkl} &=& \left( \d_{ab}\d_{kl}\, , d_{abc}(T^c_F)_{kl}\, ,
  {\rm{i}} f_{abc}(T^c_F)_{kl} \right) \, ,  
\end{eqnarray}
where $T^c_F$ are the SU(3) generators in the fundamental representation,
where the indices $m,n,k,l$ take values.  The indices $a,b,c$ are adjoint 
indices while $d_{abc}$ and $f_{abc}$ are the totally symmetric and
antisymmetric SU(3) invariant tensors, respectively.
In this basis, the one-loop soft anomalous dimension matrix for the 
$q\bar{q}$ process
is \cite{Kidonakis:1996aq,Kidonakis:1997gm,Kidonakis:1999ze} 
\begin{equation}
\Gamma^{q\bar{q}}_S=\left[\begin{array}{cc}
\Gamma^{q\bar{q}}_{11} & \Gamma^{q\bar{q}}_{12} \\
\Gamma^{q\bar{q}}_{21} & \Gamma^{q\bar{q}}_{22}
\end{array}
\right] \, ,
\label{matrixqqQQ}
\end{equation}
with matrix elements
\begin{eqnarray}
\Gamma^{q\bar{q}}_{11}&=&-\frac{\alpha_s}{\pi}C_F \, [L_{\beta}
+\ln(2\sqrt{\nu_1\nu_2})+\pi i] \, ,
\nonumber \\
\Gamma^{q\bar{q}}_{21}&=&\frac{2\alpha_s}{\pi}
\ln\left(\frac{u_1}{t_1}\right) \, ,
\nonumber \\
\Gamma^{q\bar{q}}_{12}&=&\frac{\alpha_s}{\pi}
\frac{C_F}{C_A} \ln\left(\frac{u_1}{t_1}\right) \, ,
\nonumber \\
\Gamma^{q\bar{q}}_{22}&=&\frac{\alpha_s}{\pi}\left\{C_F
\left[4\ln\left(\frac{u_1}{t_1}\right)
-\ln(2\sqrt{\nu_1\nu_2})
-L_{\beta}-\pi i\right]\right.
\nonumber \\ &&
\left.{}+\frac{C_A}{2}\left[-3\ln\left(\frac{u_1}{t_1}\right)
-\ln\left(\frac{m^2 s}{t_1u_1}\right)+L_{\beta}+\pi i \right]\right\}\, .
\label{GammaqqQQ}
\end{eqnarray}
The function $L_\beta$ is defined as
\begin{equation}
L_{\beta}=\frac{1-2m^2/s}{\beta}\left(\ln\frac{1-\beta}{1+\beta}
+\pi i \right)\, . 
\end{equation}
The variables $\nu_i,\, i=1,2$ are
\begin{equation}
  \nu_{i}= \frac{2(k_{i}\cdot\zeta_K)^2}{s}\,.
\end{equation}
The one-loop soft anomalous dimension matrix 
for the $gg$ process \cite{Kidonakis:1997gm,Kidonakis:1999ze}
can be written as
\begin{equation}
\Gamma^{gg}_S=\left[\begin{array}{ccc}
\Gamma^{gg}_{11} & 0 & \frac{1}{2}\Gamma^{gg}_{31} \vspace{2mm} \\
0 & \Gamma^{gg}_{22} & \frac{N_c}{4} \Gamma^{gg}_{31} \vspace{2mm} \\
\Gamma^{gg}_{31} & \frac{N_c^2-4}{4N_c}\Gamma^{gg}_{31} & \Gamma^{gg}_{22}
\end{array}
\right] \, ,
\label{GammaggQQ33}
\end{equation}
where the three independent matrix elements are
\begin{eqnarray}
\Gamma^{gg}_{11}&=&\frac{\alpha_s}{\pi}\left [-C_F(L_{\beta}+1)
+C_A\left(-\frac{1}{2}\ln\left({4\nu_1\nu_2}\right)+1-\pi i\right)\right
],
\label{GammaggQQ} \\
\Gamma^{gg}_{31}&=&\frac{2 \alpha_s}{\pi}\ln\left(\frac{u_1}{t_1}\right) \, ,
\nonumber \\
\Gamma^{gg}_{22}&=&\frac{\alpha_s}{\pi}\left\{-C_F(L_{\beta}+1)
+\frac{C_A}{2}\left[-\ln\left(4\nu_1\nu_2\right)
+2+\ln\left(\frac{t_1 u_1}{m^2 s}\right)+L_{\beta}-\pi i
\right]\right\}.
\nonumber 
\end{eqnarray}
Note that we did not absorb the function $\nu^{i}$ 
of Eq.~(\ref{omegaexp-MSbar}) into $\Gamma_S$.

Combining the solution of Eq.~(\ref{softevol_1}) with the 
resummed jet functions and the hard function, 
and using matrix notation for
$H_{ij}$ and ${\tilde S}_{ij}$,
the full resummed partonic cross section is
\begin{eqnarray}
\label{ij-resummed}
&& \hspace{-5mm}
\tilde{\o}^{\rm res}_{ij}(N,s,t_1,u_1,m^2,\mu^2,\alpha_s(\mu))
\,=\,{\rm Tr}\Bigg\{
H_{ij}(\zeta_K,m^2,m^2) \\ 
\nonumber & &\hspace{0mm} \times
{\rm \bar{P}} \exp\left[\int_m^{m/N} {d\mu'\over\mu'} 
(\Gamma^{ij}_S)^{\dg}\left(\alpha_s(\m^{\prime})\right)\right]
{\tilde S}_{ij}\!\left(1,\zeta_K \right)
{\rm P} \exp\left[\int_m^{m/N} {d\mu'\over\mu'} 
\Gamma^{ij}_S\left(\alpha_s(\m^{\prime})\right)\right] \Bigg\} \\
\nonumber & & \times
\exp\left(\tilde{E}_{i}(N_u,\mu,\mu_R)\right)\, 
\exp\left(\tilde{E}_{j}(N_t,\mu,\mu_R)\right)\,\, \exp\Bigg\{ 2\,
\int\limits_{\mu_R}^{m}{d\mu'\over\mu'}\,\, \Bigl(
\gamma_i\left(\alpha_s(\m^{\prime})\right) +
\gamma_j\left(\alpha_s(\m^{\prime})\right) \Bigr) \Bigg\}\, , 
\end{eqnarray}
where the trace is in color-tensor space, and
$\rm P$ refers to path-ordering in $\mu'$ such
that $\Gamma_S^{ij}(\alpha_s(m))$ is ordered to the far right while
$\Gamma_S^{ij}(\alpha_s(m/N))$ is ordered to the far left. The operation
$\rm \bar{P}$ orders in the opposite way. 
The exponential $\exp(\tilde{E}_{i})$ is given by
\begin{eqnarray}
\label{DY-exponent}
\exp(\tilde{E}_{i}(N_u,\mu,\mu_R)) &=& \exp \Bigg \{ 
\tilde{E}_i\Big( N_u,2k_i\cdot\zeta_K \Big) \Bigg\} \\[1ex]
& &\hspace{3mm} \times\exp\Bigg \{ -2\int\limits_{\mu_R}^{2 k_i\cdot
  \zeta_K}{d\mu'\over\mu'}
\gamma_i\left(\alpha_s(\m^{\prime})\right) +
2\int\limits_{\mu}^{2 k_i\cdot
  \zeta_K}{d\mu'\over\mu'}
\gamma_{i/i}\Big(
  N_u,\alpha_s(\m^{\prime})\Big) \Bigg\}\, . \NO 
\end{eqnarray}
The exponential $\exp(\tilde{E}_{j})$ is identical except for 
obvious relabelling.
In Eq.~(\ref{ij-resummed}) we used the renormalization group 
behaviour, derived from Eqs.~(\ref{rge_behaviour_hard_qq}) and (\ref{rge_behaviour_soft}),
of the product $H_{ij\, IJ}{\tilde{S}}_{ij\, JI}$
\begin{eqnarray}
\label{rge_hardXsoft}
\mu_R\frac{d}{d\mu_R} \ln \left[ H_{ij\, IJ}(\mu_R)
  {\tilde{S}}_{ij\, JI}\!\left(N,\mu_R\right)
\right] &=& - 2 \gamma_{i}\left(\alpha_s(\mu_R)\right) - 2
\gamma_{j}\left(\alpha_s(\mu_R)\right)\, ,
\end{eqnarray} 
where we have only indicated the $\mu_R$ and $N$ dependence.
Notice that all $\mu_R$ dependence cancels in Eq.~(\ref{ij-resummed}).

So far, the results have all been given in the $\msb$-scheme.  
The $q \overline{q}$
results can also be presented in the DIS scheme, in which
the function $\tilde{E}_q$ of Eq.~(\ref{omegaexp-MSbar}) is
\begin{eqnarray}
\label{omegaexp_q-DIS}
\tilde{E}_q\left(N,2 k_i \cdot \zeta_K\right) \Biggr|_{\rm{DIS}} &=&
\tilde{E}_q\left(N,2 k_i \cdot \zeta_K\right) \Biggr|_{\rm{\msb}}
+\int\limits^\infty_0 dw\,\frac{(1-e^{-Nw})}{w}\; \Bigg
\{-\int\limits^{1}_{w} \frac{d\lambda}{\lambda}
A_q\left[\alpha_s(\sqrt{\lambda s})\right]  \NO \\
&\ & \hspace{45mm} \mbox{} 
{}+ g^{{\rm DIS}}_q\left[\alpha_s(\sqrt{ws})\right]\,
\Bigg \}\, ,
\end{eqnarray}
with
\begin{eqnarray}
\label{g2def_q}
g^{{\rm DIS}}_q\left(\alpha_s\right) = - \frac{3}{4}\, C_F
\frac{\a_s}{\p}+\ldots\, .
\end{eqnarray} 
\medskip

In some cases, e.g. for the unexpanded resummed cross section,
it is preferable to diagonalize the matrix $\G^{ij}_{S,IJ}$ 
\cite{Kidonakis:1998nf,Kidonakis:1998ei,Kidonakis:1999ze,Kidonakis:1998qe}
by a change of basis in color-tensor space, 
\begin{equation}
\G_{R,S} = R^{-1}\, \G_S\, R \, ,
\end{equation}
written in matrix notation with $\G_{R,S}$ diagonal.  The channel labels have
been
suppressed. The columns of the matrix $R$ are the eigenvectors of $\Gamma_S$.
This leads to ordinary exponentials in
the solution of Eq.~(\ref{softevol_1}), rather than path-ordered ones.
The elements on the diagonal are the (complex) eigenvalues $\l_I$.
In the new basis, indicated by the subscript $R$,
the matrices ${\tilde S}$ and $H$ are 
\begin{eqnarray}
\label{unique-match-fix}
{\tilde S}_R \,=\, R^{\dg} {\tilde S} R\, , \,\,\, \,\,\, H_R \,=\,
R^{-1} H \left( R^{-1}\right)^{\dg} \, ,
\end{eqnarray} 
again in matrix notation. Eq.~(\ref{softevol_1}) then becomes 
\begin{eqnarray}
\label{softevol_2}
\m \frac{d}{d\m}\, {\tilde{S}}_{R,JI} &=& - \left( \l_I
  + \l^*_J \right){\tilde{S}}_{R,JI}\, .  
\end{eqnarray} 
The matrix $R_{q{\Bar{q}}}$  is given in 
Refs.~\cite{Kidonakis:1998ei,Kidonakis:1999ze}
while $R_{gg}$ appears in Ref.~\cite{Kidonakis:1999ze}.
The solution of Eq.~(\ref{softevol_2}) is
\begin{eqnarray}
\label{soft_solution}
{\tilde S}_{R,JI}\!\left({m\over N
    \mu},\zeta_K\right)\!  &=& \!{\tilde
    S}_{R,JI}\!\left(1,\zeta_K \right)\;
\exp\left[\,\int\limits_{\mu}^{m/N}{d\mu'\over\mu'}\!\Bigl\{
  \l_I(\alpha_s(\mu')) + \l^*_J(\alpha_s(\mu')) \Bigr\} \right]
\,\,\,\,\,\,\, 
\end{eqnarray} 
so that (reinstating the channel labels)
\begin{eqnarray}
\label{ij-R-resummed}
&& \hspace{-5mm}
\tilde{\o}^{\rm res}_{ij}(N,s,t_1,u_1,m^2,\mu^2,\alpha_s(\mu))
\,=\,\sum\limits_{I,J}\,\,
H_{R,ij\, IJ}(\zeta_K,m^2,m^2) \\
\nonumber & & \times {\tilde S}_{R,ij\,
  JI}\!\left(1,\zeta_K\right)\; \exp\Big\{ {\tilde{E}_{ij\,
    JI}} \Big\} \,\, \exp\Bigg[ 2\,
\int\limits_{\mu}^{m}{d\mu'\over\mu'}\,\, \Bigl\{
\gamma_i\left(\alpha_s(\m^{\prime})\right) +
\gamma_j\left(\alpha_s(\m^{\prime})\right) \Bigr\} \Bigg]\, . 
\end{eqnarray}
The exponential in Eq.~(\ref{ij-R-resummed})
carrying color-tensor indices is
\begin{eqnarray}
\label{ijR-exponent}
\exp(\tilde{E}_{ij \, JI}) &=& \exp(\tilde{E}_{i}(N_u,\mu,\mu_R))\, 
\exp(\tilde{E}_{j}(N_t,\mu,\mu_R))\, \\[1ex]
&\, & \hspace{10mm}\times\,
\exp\left[\,\int\limits_{m}^{m/N}{d\mu'\over\mu'}
  \Bigl\{ \l_I(\alpha_s(\mu')) + \l^*_J(\alpha_s(\mu')) \Bigr\} \right]\, ,
\nonumber
\end{eqnarray}
with $\exp(\tilde{E}_{i})$ and  $\exp(\tilde{E}_{j})$ 
given in Eq.~(\ref{DY-exponent}).
We derived the bulk of our subsequent results in both bases.

\section{NNLL-NNLO expansions for partonic cross sections}\label{sec:nnll-nnlo-expansions}

In this section we derive analytical expressions for 
partonic double-differential heavy quark cross sections up to NNLO
by expanding the resummed cross section of the previous section.
We concentrate here on the derivation;
our explicit results are collected in the appendices.
We obtain expressions for both 1PI and PIM
kinematics and in both $\msb$ and DIS factorization schemes.  
Our aim is NNLL accuracy, as defined in
section \ref{sec:singular-functions}, including
the scale dependent sector where we distinguish between
the renormalization scale $\m_R$ and mass factorization 
scale $\mu$. This means that for coefficients of 
$\ln^i(\mu^2/m^2)$ or $\ln^i(\mu_R^2/m^2)$
we include the most singular plus distribution and the
two next-most-singular ones.

To achieve NNLL accuracy we must derive
NLO matching conditions for certain functions
in the resummed cross section, Eq.~(\ref{ij-resummed}).
To this end we expand the factors in this 
equation and identify 
order by order the perturbative coefficients
of the functions in Eq.~(\ref{ij-resummed}).
The coefficients can in
part be explicitly computed and in part inferred 
by matching to exact NLO results.

\subsection{Expansions}
\label{sec:expansions}

Let us expand each factor in Eq.~(\ref{ij-resummed}) in powers of
$\a_s$. We set $\mu_R=m$ here
because Eq.~(\ref{ij-resummed}) is independent of $\mu_R$, leading 
to a cross section expansion in $\alpha_s(m)$ which
can easily be changed back to $\alpha_s(\mu_R)$.
Neglecting $1/N$ terms, the two-loop expansion of 
$\exp(\tilde{E}_{i})$ may be written as
\begin{eqnarray}
  \label{eq:23}
\exp(\tilde{E}_{i}(N_u,\mu,m))
\simeq 1 + \frac{\a_s}{\pi}\left(\sum_{k=0}^2 C^{i,(1)}_{k}\ln^k(N_u)  \right)
+ \left(\frac{\a_s}{\pi}\right)^2\left(
\sum_{k=0}^4 C^{i,(2)}_{k}\ln^k(N_u)
\right)
+\ldots
\end{eqnarray}
The coefficients $C^{i,(n)}_k$ 
can be computed from the results in section \ref{sec:resumm-cross-sect}.
To NNLL accuracy they are
\begin{eqnarray}
  \label{eq:32}
C^{i,(1)}_2 &=& A_i^{(1)},\qquad C^{i,(1)}_1 = \frac{1}{2}\kappa_i +  A_i^{(1)}  l_\mu  ,
\qquad  C^{i,(1)}_0 = \zeta_2C^{i,(1)}_2- d_i  l_\mu \, ,\nonumber \\
C^{i,(2)}_4 &=& \frac{1}{2}\left(A_i^{(1)}\right)^2,\qquad
    C^{i,(2)}_3 = \frac{2}{3}b_2A_i^{(1)} +\frac{1}{2}\kappa_i A_i^{(1)}+\left(A_i^{(1)}\right)^2 l_\mu \, ,
\nonumber \\
C^{i,(2)}_2 &=& A_i^{(2)} +\zeta_2\left(A_i^{(1)}\right)^2+\frac{1}{8}\left(\kappa_i\right)^2
  +\frac{1}{2} b_2 \kappa_i  +  \frac{1}{2}\left(A_i^{(1)}\right)^2 l_\mu^2  
\nonumber \\
& & + \mbox{} \frac{1}{2} A_i^{(1)} \kappa_i l_\mu - d_i A_i^{(1)}  l_\mu\,,\nonumber \\
C^{i,(2)}_1 &=& \left(\zeta_2\left(A_i^{(1)}\right)^2- \frac{1}{2}d_i \kappa_i+A_i^{(2)}
\right) l_\mu + \left( - \frac{1}{2}b_2  A_i^{(1)}-d_i A_i^{(1)} \right) l_\mu^2 \,,\nonumber \\
C^{i,(2)}_0 &=&\left( \frac{1}{2}b_2 d_i +  \frac{1}{2}d_i^2 \right) l_\mu^2 \,,
\end{eqnarray}
where we have used the  notation
\begin{eqnarray}
  \label{eq:33}
  \kappa_i &=& \nu^{i,(1)} -2 \nu^{i,(1)}\ln\left(\frac{\sqrt{2\nu_i s}}{m} \right), \qquad
  l_\mu = \ln\left(\frac{\mu^2}{m^2} \right) \, , \nonumber \\
d_q &=& \frac{3}{4}C_F,\qquad d_g = b_2 \, , 
\end{eqnarray}
and where $A_i^{(n)}$  and  $\nu^{i,(n)}$ are the coefficients of 
$(\alpha_s/\pi)^n$ in the expansion of the functions $A_i$ and $\nu^i$ in Eq.~(\ref{nudef}).
The two-loop expansion of the path-ordered exponential in
Eq.~(\ref{ij-resummed}) reads
\begin{eqnarray}
\label{eq:1}
{\lefteqn{{\rm P}\exp\left[\int_m^{m/N} {d\mu'\over\mu'} 
\Gamma_S^{ij}\left(\alpha_s(\m^{\prime})\right)\right]_{IJ}
\,=\,\delta_{IJ} -  \frac{\alpha_s(m)}{\pi}\, \ln(N)
     \Gamma^{ij\, (1)}_{S,IJ} }} \nonumber \\
&+& \left(\frac{\alpha_s(m)}{\pi}\right)^2\, \left[
\frac{1}{2}\ln^2(N)
\left[\Gamma^{ij\, (1)}_S\times\Gamma^{ij\, (1)}_S\right]_{IJ}
\,-\, b_2\,\ln^2(N) 
\Gamma^{ij\, (1)}_{S,IJ} \right] + \ldots\, ,\quad\quad
\end{eqnarray}
where $\Gamma^{ij\, (1)}_S$ is the coefficient of $\a_s/\p$ 
in the one-loop soft anomalous dimensions of Eqs.~(\ref{matrixqqQQ}) and
(\ref{GammaggQQ33}).
The $\bar{\rm P}$ ordered exponential 
is identical to ${\cal O}(\a_s^2)$ up to replacing
$\Gamma_S$ by $\Gamma_S^{\dg}$. 
The matrices
$H_{ij\,IJ}(t_1,u_1,m^2,m^2)$ 
and ${\tilde S}_{ij\,IJ}\!\left(1,\zeta_K,\beta_i,\beta_j \right)$
expand to one loop as
\begin{eqnarray}
  \label{eq:24}
  H_{ij\, IJ} &=& H^{(0)}_{ij\, IJ} + \frac{\a_s}{\pi} H^{(1)}_{ij\, IJ}\,, \nonumber \\
  {\tilde S}_{ij\, IJ} &=& {\tilde S}^{(0)}_{ij\, IJ} + \frac{\a_s}{\pi} 
{\tilde S}^{(1)}_{ij\, IJ} \,.
\end{eqnarray}
Let us for the moment choose PIM kinematics
by putting $N_t=N_u = N$. We substitute Eqs.~(\ref{eq:23}),~(\ref{eq:1}), and
(\ref{eq:24}) into Eq.~(\ref{ij-resummed}), finding the
moment space expression for the NNLL-NNLO expanded cross section:
\begin{eqnarray}
  \label{eq:25}
  & &{\lefteqn{ {\tilde{\o}}_{ij}^{\rm NNLO}(N,s,t_1,u_1,m^2,\mu^2,\alpha_s(\mu))\,=\,
    {\rm Tr}\left\{ H^{(0)}_{ij} {\tilde S}^{(0)}_{ij}\right\}    
     }} \\
 & &+\, \frac{\a_s}{\pi}\, \Biggl[ \left(( C^{i,(1)}_2+C^{j,(1)}_2 )
\ln^2(N) + ( C^{i,(1)}_1+C^{j,(1)}_1 )\ln(N)\right)\,  
         {\rm Tr}\left\{ H^{(0)}_{ij} {\tilde S}^{(0)}_{ij}\right\} \nonumber \\[1ex]
& & \hspace{15mm}
  -\ln(N)\,  {\rm Tr}\left\{ H^{(0)}_{ij} \left(\Gamma^{ij\, (1)}_S\right)^{\dg}  {\tilde S}^{(0)}_{ij}+
                             H^{(0)}_{ij} {\tilde S}^{(0)}_{ij}\Gamma^{ij\, (1)}_S\right\} 
 \nonumber \\[1ex]
& & \hspace{15mm}+\, {\rm Tr}\left\{ H^{(0)}_{ij}  {\tilde S}^{(1)}+
                             H^{(1)} {\tilde S}^{(0)}_{ij}\right\} 
+( C^{i,(1)}_0+C^{j,(1)}_0 ) {\rm Tr}\left\{ H^{(0)}_{ij} {\tilde S}^{(0)}_{ij}\right\}    
\Biggr] \nonumber \\[1ex]
& &+\, \left(\frac{\a_s}{\pi}\right)^2\, 
\Biggl[ \Bigg(\left(C^{i,(2)}_4+C^{j,(2)}_4+ C^{i,(1)}_2C^{j,(1)}_2\right)
\ln^4(N)\nonumber \\[1ex] 
& & \hspace{5mm} 
+ \left( C^{i,(2)}_3+C^{j,(2)}_3
+ C^{i,(1)}_1 C^{j,(1)}_2  + C^{i,(1)}_2 C^{j,(1)}_1 
\right)\Bigg)\ln^3(N)\, 
 {\rm Tr}\left\{ H^{(0)}_{ij} {\tilde S}^{(0)}_{ij}\right\} \nonumber \\[1ex]
& &\hspace{5mm}- 
\left( C^{i,(1)}_2+C^{j,(1)}_2 \right)\ln^3(N)\, {\rm Tr}\left\{ H^{(0)}_{ij}
\left(\Gamma^{ij\, (1)}_S\right)^{\dg} {\tilde S}^{(0)}_{ij}+
                   H^{(0)}_{ij} {\tilde S}^{(0)}_{ij}\Gamma^{ij\, (1)}_S\right\}  \nonumber \\[1ex]
& &\hspace{5mm}+
\,\ln^2(N)\, \left( C^{i,(2)}_2+C^{j,(2)}_2 +\sum_{k=0}^2 C^{i,(1)}_k C^{j,(1)}_{2-k}\right) 
{\rm Tr}\left\{ H^{(0)}_{ij} {\tilde S}^{(0)}_{ij}\right\}  \nonumber \\[1ex]
& & \hspace{5mm}
-(b_2 + C^{i,(1)}_1+C^{j,(1)}_1)\ln^2(N)\, {\rm Tr}\left\{ H^{(0)}_{ij} 
   \left(\Gamma^{ij\, (1)}_S\right)^{\dg} {\tilde S}^{(0)}_{ij}+
             H^{(0)}_{ij} {\tilde S}^{(0)}_{ij}\Gamma^{ij\, (1)}_S\right\}  \nonumber \\[1ex]
& &\hspace{5mm}+\,(C^{i,(1)}_2+C^{j,(1)}_2 )\ln^2(N)\, {\rm Tr}\left\{ H^{(0)}_{ij} 
 {\tilde S}^{(1)}_{ij}+
                             H^{(1)}_{ij} {\tilde S}^{(0)}_{ij}\right\} \nonumber \\[1ex]
& &\hspace{5mm}+\ln^2(N)\,{\rm Tr}\Biggl\{
  \frac{1}{2} H^{(0)}_{ij} \left(\Gamma^{ij\, (1)}_S\right)^{\dg} 
                           \left(\Gamma^{ij\, (1)}_S\right)^{\dg}
 {\tilde S}^{(0)}_{ij}+
 H^{(0)}_{ij} \left(\Gamma^{ij\, (1)}_S\right)^{\dg}{\tilde S}^{(0)}_{ij}\Gamma^{ij\, (1)}_S
 \nonumber \\[1ex]
& & \hspace{5mm}
 + \frac{1}{2} H^{(0)}_{ij} {\tilde S}^{(0)}_{ij}\Gamma^{ij\, (1)}_S \Gamma^{ij\, (1)}_S
\Biggr\}
 \Biggr]\,. \nonumber
\end{eqnarray}
With 1PI kinematics we use Eq.~(\ref{eq:Ndef}) and expand
$\ln(N_u) = \ln(N)+\ln(-u_1/m^2)$ etc.
To transform from moment to momentum space we use
results given in the appendices of Refs.~\cite{Catani:1989ne,Laenen:1998kp,
Kidonakis:2000ui}.
Note that we do not keep subleading terms in our expansions
when we invert to momentum space \cite{Kidonakis:2000ui}.

\subsection{Matching}\label{sec:matching-conditions}

We can identify the functions we must determine from
Eq.~(\ref{eq:25}).

The coefficients $C_k^{i,(n)}$ are given in
Eq.~(\ref{eq:32}).
The matrices $\Gamma^{ij\, (1)}_{IJ}$ are given
in Eqs.~(\ref{GammaqqQQ}) and (\ref{GammaggQQ}).
The remaining unknowns are
\begin{eqnarray}
  \label{eq:2}
 S^{(0)}_{ij\,IJ}\,\,,\;\; H^{(0)}_{ij\,IJ}\,,\;\;
{\rm Tr}\left\{ H^{(0)}_{ij}  {\tilde S}^{(1)}_{ij}+
   H^{(1)}_{ij} {\tilde S}^{(0)}_{ij}\right\}\,.
\end{eqnarray}
The leading-order soft function ${\tilde S}_{IJ}^{(0)}$ is simply
\begin{eqnarray}
\label{soft_lo}
\tilde{S}^{(0)}_{ij\,IJ} &=& {\rm{tr}} \left( c^{{\dg}}_I\, c_J \right) \,,  
\end{eqnarray}
where $\{c_I\}$ is our basis in color-tensor space.
Recall that we have chosen the $s$-channel singlet-octet basis, 
Eq.~(\ref{gluon-gluon-basis}).
Note that $\tilde{S}^{(0)}={S}^{(0)}$.
The lowest-order hard function $H^{(0)}_{ij\,IJ}=h_{ij\,I}^{(0)}
h^{* (0)}_{ij\,J}$ is calculated
by projecting the lowest order  amplitude $\ca_{ij}$
onto this color basis:
 \begin{eqnarray}
\label{hard_lo}
h_{ij\,I}^{(0)} \,=\, \left(\tilde{S}^{(0)}_{ij} \right)^{-1}_{IK} \, {\rm{tr}} \left(
  c^{{\dg}}_K \, \ca_{ij} \right) \, , \,\,\,\,\,\,\,\,\, h^{* (0)}_{ij\,J} \,=\,
{\rm{tr}} \left( \ca_{ij}^{{\dg}} \, c_K \right) \, \left(\tilde{S}^{(0)}_{ij}
\right)^{-1}_{KJ}\, , \end{eqnarray} 
where the trace acts in ordinary color space. 
Note that at this stage the function
$h^{(0)}_{ij\,I}$ may still depend on all other quantum numbers of the
process under consideration, such as
Lorentz-spinor and spin degrees of freedom.
We are however not interested here
in spin-dependent observables and we trace over all remaining degrees
of freedom.  
The matrix $H^{(0)}_{ij\,IJ}$ is real and symmetric but
may have zero eigenvalues if the Born process does not span the full
color-tensor space, as is the case for the $q\bar{q}$ channel.
For the $q{\Bar{q}}$ channel \cite{Kidonakis:1998ei} the results are
\begin{eqnarray}
\label{soft-fct_q,hard-fct_q}
\tilde{S}^{(0)}_{q{\Bar{q}}\, IJ} &=& \left(
\begin{array}{cc} N_c^2  & 0  \\[1ex]
0  & \displaystyle \frac{N_c^2-1}{4}
\end{array}
\right) \, , \hspace*{20mm} H^{(0)}_{q{\Bar{q}}\, IJ} \,=\, \left(
\begin{array}{cc} 0  & 0  \\
 0  & H_{q{\Bar{q}}\, 22}^{(0)}
\end{array}
\right) \, , \end{eqnarray} with $H_{q{\Bar{q}}\, 22}^{(0)}$ given by
\begin{eqnarray} H^{(0)}_{q{\Bar{q}}\, 22}(m,m)\; &=&
\frac{2 \pi\, \alpha_s^2(m)}{N_c^2}\, \left( \frac{t_1^2 +
    u_1^2}{s^2} + \frac{2 m^2}{s} \right) \, .
\end{eqnarray} 
The lowest-order results are kinematics-independent.
For the $gg$ channel we find
\begin{eqnarray}
\label{soft-fct_g}
\tilde{S}^{(0)}_{gg\, IJ} &=& \left(
\begin{array}{ccc} 
N_c (N_c^2-1)  & 0 & 0  \\[1ex]
0  &  \displaystyle (N_c^2-1) \frac{N_c^2-4}{2 N_c} &  0 \\[1ex] 
0  & 0 & \displaystyle  \frac{1}{2} N_c (N_c^2-1)
\end{array}
\right) \, , \\[2ex]
\label{hard-fct_g}
H^{(0)}_{gg\, IJ} &=& \left(
\begin{array}{ccc} 
H^{(0)}_{gg\, 11}  & N_c H^{(0)}_{gg\, 11} & H^{(0)}_{gg\, 13}  \\[1ex]
N_c H^{(0)}_{gg\, 11}  & N_c^2 H^{(0)}_{gg\, 11} & 
N_c H^{(0)}_{gg\, 13} \\[1ex]
H^{(0)}_{gg\, 13} & N_c H^{(0)}_{gg\, 13} &  H^{(0)}_{gg\, 33}
\end{array}
\right) \, , 
\end{eqnarray} where
\begin{eqnarray} H^{(0)}_{gg\,
  11}(m,m)\; &=&
\frac{\pi \alpha_s^2(m)}{2 N_c^2 (N_c^2-1)^2} B_{gg}\,, \\[1ex]
H^{(0)}_{gg\, 13}(m,m)\; &=&
\frac{\pi \alpha_s^2(m)}{2 N_c (N_c^2-1)^2} B_{gg} \frac{t_1^2-u_1^2}{s^2}\,, \\[1ex]
H^{(0)}_{gg\, 33}(m,m)\; &=& \frac{\pi \alpha_s^2(m)}{2
  (N_c^2-1)^2} B_{gg} \left(2 \frac{t_1^2+u_1^2}{s^2} - 1 \right)\, ,
\end{eqnarray} 
and
\begin{eqnarray} B_{gg} &=& \frac{u_1}{t_1} + \frac{t_1}{u_1} +
\frac{4 s m^2}{t_1 u_1}\left(1 - \frac{s m^2}{t_1 u_1}\right)\, .
\end{eqnarray} 
It now remains to determine\footnote{Note that we do not need the individual matrices
$S^{(1)}_{ij\,IJ}$ and $H^{(1)}_{ij\,IJ}$.}
\begin{eqnarray}
{\rm Tr}\left\{ H^{(0)}_{ij}  {\tilde S}^{(1)}_{ij}+
   H^{(1)}_{ij} {\tilde S}^{(0)}_{ij}\right\}\,.
\end{eqnarray}
We do this by matching. The necessary matching conditions
can be inferred by comparing the expansion in Eq.~(\ref{eq:25})
to exact results for the partonic cross section.

At lowest order, the matching condition is
\begin{eqnarray}
\label{eq:30}
 \tilde{\o}^{\rm LO}_{ij}(N,s,t_1,u_1,m^2,\mu^2,\alpha_s(\mu)) 
= H^{(0)}_{ij\,IJ} \,\, \tilde{S}^{(0)}_{ij\,JI} \, .
\end{eqnarray}
It is straightforward to check that this condition is fulfilled by
our results for $\tilde{S}^{(0)}_{ij\, IJ}$ and $H^{(0)}_{ij\,IJ}$.
Neglecting $1/N$ terms, and choosing the case $K=\mathrm{PIM}$ for the moment,
we have at NLO the matching condition 
\begin{eqnarray}
\label{eq:29}
  & &{\lefteqn{ \frac{\alpha_s}{\pi} \, \tilde{T}_{ij\, K}^{(1)}\equiv
 {\rm Tr}\left\{ H^{(0)}_{ij}  {\tilde S}^{(1)}_{ij}+
                             H^{(1)}_{ij} {\tilde S}^{(0)}_{ij}\right\}
\,=\, \tilde{\o}^{\rm NLO}_{ij}(N,s,t_1,u_1,m^2,\mu^2,\alpha_s(\mu))   
     }} \\
 & &-\, \frac{\a_s}{\pi}\, \Biggl[ \left( ( C^{i,(1)}_2+C^{j,(1)}_2 )
\ln^2(N) + ( C^{i,(1)}_1+C^{j,(1)}_1 )\ln(N)  \right)\,  
         {\rm Tr}\left\{ H^{(0)}_{ij} {\tilde S}^{(0)}_{ij}\right\} \nonumber \\[1ex]
& &\hspace{5mm}
  - \ln(N)\,{\rm Tr}\left\{ H^{(0)}_{ij} \left(\Gamma^{ij\, (1)}_S\right)^{\dg} {\tilde S}^{(0)}_{ij}+
                H^{(0)}_{ij} {\tilde S}^{(0)}_{ij}\Gamma^{ij\, (1)}_S\right\}
+( C^{i,(1)}_0+C^{j,(1)}_0) {\rm Tr}\left\{ H^{(0)}_{ij} {\tilde S}^{(0)}_{ij}\right\}    
 \Biggr] \,.
 \nonumber
\end{eqnarray}
Let us discuss this equation.
To begin, the first term on the right hand side  may be determined from
the exact results in Refs.~\cite{Beenakker:1989bq,Beenakker:1991ma,Mangano:1992jk},
converted to moment space.
To NNLL accuracy this term consists 
of all terms in the differential one-loop
partonic cross sections that either diverge
logarithmically or are constant. 

The constant terms consist, first of all, of the virtual graph contributions. 
In addition there is the soft-gluon
contribution from the radiative graphs. The virtual contribution
is kinematics independent but the soft gluon one is not because what is
defined as soft differs for 1PI and PIM kinematics. The respective
criteria are $s_4<\Delta$ and $1-z < \delta$, with $\Delta$
and $\delta$ defined in Eqs.~(\ref{s4distdef}) and (\ref{1mzdistdef}).
The virtual graphs contain various divergences. The ultraviolet
ones are cancelled by renormalization of the QCD coupling
constant and the heavy quark mass, for both
kinematics in the scheme of Ref.~\cite{Collins:1978wz}.
Infrared divergences are cancelled
by adding the virtual and soft contributions, 
so that only collinear divergences remain. These must be subtracted
via mass factorization. The corresponding subtraction
terms in general involve  convolutions of  Altarelli-Parisi
splitting functions \cite{Altarelli:1977zs} with the 
lowest order cross section. However, we must mass factorize only with
the soft plus virtual parts of these splitting functions.
These depend on kinematics, see e.g. Eqs.~(6.8) and (6.13) in
\cite{Beenakker:1989bq}.
What remains at NLO after mass factorization can now be categorized
as terms multiplying $\ln^i(\Delta/m^2)$ or $\ln^i(\delta)$,
$i=1,2$, and other terms. The logarithms $\ln(\Delta/m^2)$ or $\ln(\delta)$
should be completed to plus-distributions via Eqs.~(\ref{s4distdef}) and (\ref{1mzdistdef}).
The final result of these procedures, in moment space, is  what constitutes the first term
on the right hand side of the equals sign in Eq.~(\ref{eq:29}).
The other terms on the right hand side merely subtract 
all terms containing the singular functions in Eq.~(\ref{s4distdef}) or
(\ref{1mzdistdef}).

For 1PI kinematics there
is a slight subtlety, related to the use
of $N_t$ and $N_u$ in Eq.~(\ref{eq:Ndef}). Expanding
$\ln(N_t) = \ln(N)+\ln(-t_1/m^2)$ (and  $\ln(N_u)$ likewise) leads to 
extra terms containing $\ln(-t_1/m^2)$ 
and $\ln(-u_1/m^2)$ but not
$\ln(N)$ and one must
be careful in accounting for such terms.

We can now obtain
specific results for the matching terms in momentum space.
At NLO, $T_{ij\, K}^{(1)}$ on the left hand side of Eq.~(\ref{eq:29}) for
1PI kinematics,
\begin{eqnarray}
\label{qqbar-one-loop-const}
T^{(1)}_{ij\, \opi} (s,t_1,u_1)\equiv
\hat{T}^{(1)}_{ij\, \opi} (s,t_1,u_1)\,\delta(s_4) \, ,
\end{eqnarray}
is the expression for $s^2 d^2\sigma^{(1),{\rm S+V}}_{ij}(s,t_1,u_1)/dt_1\, du_1$ 
given by Eq.~(4.7) of Ref.~\cite{Beenakker:1991ma} for $ij=q\Bar{q}$,
and Eq.~(6.19) of Ref.~\cite{Beenakker:1989bq} for $ij=gg$, modified
by removing a factor $\alpha_s/\pi$ as well as all terms containing
$\ln^2(\D/m^2)$ or $\ln(\D/m^2)$ and terms containing
$\ln(\mu/m)$ or $\ln(\mu_R/m)$\footnote{Note 
that in these references the factorization scale is denoted by $Q$
and that the definition of $t_1$ and $u_1$ in Ref.~\cite{Beenakker:1991ma} 
for $ij=q\Bar{q}$ is interchanged as compared to this work.}.
The definition of the hatted symbol in Eq.~(\ref{qqbar-one-loop-const}) 
makes the delta function explicit and is useful when we present the 
two-loop results.

In PIM kinematics the NLO matching terms may be determined 
as follows. We start from the above 1PI results and modify
those parts that are kinematics dependent. From our
discussion following Eq.~(\ref{eq:29}), it is clear that 
these are only the soft contributions
and the mass factorization
subtraction terms. We shall label these terms by the superscript ${\rm S\!+\!MF}$.
The soft contributions to the PIM NLO
cross sections can be computed using the results of Ref.~\cite{Mangano:1992jk}.
As mentioned in the above, 
we mass-factorize these results with
only the soft plus virtual part of the splitting functions in PIM kinematics.
The same procedure is done for the 1PI case using
the results of Refs.~\cite{Beenakker:1989bq,Beenakker:1991ma}.
For channel $ij$ we then compute the $\msb$ quantity
\begin{eqnarray}
\label{eq:26}
   \Delta_{ij}^{(1)}(M^2,{\rm{cos}}\q) = s \frac{d^2\sigma^{(1),{\rm S+MF}}_{ij}(s,M^2,{\rm{cos}}\q)}{dM^2\, d{\rm{cos}}\q}
\,-\,\frac{\beta}{2} s^2 \frac{d^2\sigma^{(1),{\rm S+MF}}_{ij}(s,t_1,u_1)}{dt_1\, du_1}\Biggr|_{\rm PIM}\,
\end{eqnarray}
where the subscript PIM on the second term on the right hand side indicates
that for $t_1,u_1$ 
one must use the expressions in Eq.~(\ref{tupidef}).
Note that we may replace $\beta_M$ by $\beta$ to the accuracy at which
we are working in this paper, and we shall do so for all following results.
Explicit expressions for the terms on the right hand side of Eq.~(\ref{eq:26})
can be found in appendix A.
Then the PIM equivalent of Eq.~(\ref{qqbar-one-loop-const}) is,
accounting for an overall jacobian factor
originating from the definition of the cross section,
\begin{eqnarray}
  \label{eq:15}
T^{(1)}_{ij\, \pim}(M^2,{\rm{cos}}\q) \equiv
\hat{T}^{(1)}_{ij\, \pim}(M^2,{\rm{cos}}\q)\delta(1-z)
= \frac{\beta}{2}
T^{(1)}_{ij\, \opi}(s,t_1,u_1) \Biggr|_{\rm PIM}
+ \Delta_{ij}^{(1)}(M^2,{\rm{cos}}\q)\,.
\end{eqnarray}
Note that terms containing renormalization and factorization scale dependence
can be easily included by expanding the corresponding scale dependent 
exponentials in Eq.~(\ref{ij-resummed}) and adjusting 
the coefficients in Eq.~(\ref{eq:25}) correspondingly.
Alternatively, the scale dependence can 
be constructed  using renormalization group methods, which
we do in the next section. 

Changing to the DIS scheme at NNLL 
for the $q\Bar{q}$ channel
involves using Eq.~(\ref{omegaexp_q-DIS})
rather than Eq.~(\ref{omegaexp-MSbar}) in the expansion, Eq.~(\ref{eq:23}),
and extra constant terms at NLO as given in appendix B. 

Because of their length, we have collected all our results
in appendix B. 
In the next section we perform a numerical study of the results obtained in 
this section for the inclusive partonic cross sections.

\section{Scaling functions} \label{sec:scalingfunctions}

It is convenient to express the inclusive partonic cross sections in terms of 
dimensionless scaling functions $f^{(k,l)}_{ij}$ that depend only on  
$\h$, defined in Eq.~(\ref{eq:etadef}), as
\begin{eqnarray}
\label{scalingfunctions}
\sigma_{ij}(s,m^2,\m^2) &=& \frac{\a_s^2(\m)}{m^2}
\sum\limits_{k=0}^{\infty} \,\, \left( 4 \p \a_s(\m) \right)^k
\sum\limits_{l=0}^k \,\, f^{(k,l)}_{ij}(\h) \,\,
\ln^l\left(\frac{\m^2}{m^2}\right) \, .
\end{eqnarray} 
We derive LL, NLL, and NNLL approximations to $f^{(k,l)}_{ij}(\h)$
for both the $q{\Bar{q}}$ and $gg$ channels
using the results of the previous section,
via Eqs.~(\ref{eq:11}) and (\ref{totalpartoncrs_PIM}).
Where possible we will compare with exact results.
We also examine the results for the $q{\Bar{q}}$ channel 
in the DIS scheme. Of course, a complete 
comparison between the DIS and ${\Bar{\rm{MS}}}$ schemes also requires
the use of DIS parton densities, as discussed in the next section.

We begin by showing the scaling functions 
$f^{(k,0)}_{ij}(\h), k = 0,1,2$, in both 1PI and PIM kinematics.
In Fig.~\ref{plot-1q} we present 
the $q\Bar{q}$ Born and $\overline{\rm MS}$ NLO results
for the scaling functions $f^{(0,0)}_{q{\Bar{q}}}$
and $f^{(1,0)}_{q{\Bar{q}}}$, respectively.
A comparison shows that, in the case of 
$f^{(1,0)}_{q{\Bar{q}}}$, the LL approximations already reproduce the exact curve 
quite well in both kinematics \cite{Laenen:1992af}, but 
the NLL approximations agree much better 
with the exact results to larger $\h$. 
Adding the NNLL corrections, numerically 
dominated by the negative contributions from gluon exchange between final state
heavy quarks (Coulomb terms),
leads to excellent agreement with the exact curve over a large range of $\h$. 
Such progressive improvement was also observed, for 1PI kinematics,
in Ref.~\cite{Kidonakis:1995uu} where the absolute threshold
limit was taken.
We also see that, while the LL approximations in 1PI and PIM kinematics differ 
and under- or overestimate the true result, the NLL approximations 
in both kinematics agree better with each other, and even more so at NNLL,
at least for $\eta\;\ltap \; 0.1$.
To summarize the one loop results in this channel,
differences due to kinematics choice decrease as the
logarithmic accuracy increases.

The same conclusions hold for the NLO scaling function 
in the $gg$ channel,
$f^{(1,0)}_{gg}$, shown in Fig.~\ref{plot-1g}.
The Coulomb terms are positive here.
At large $\eta$, the agreement 
between the 1PI and PIM results is not as good as
in the $q{\Bar{q}}$ channel: the 1PI results at NNLL
remain positive while the PIM ones are negative.

\begin{figure}[htp]
\begin{center}
\epsfig{file=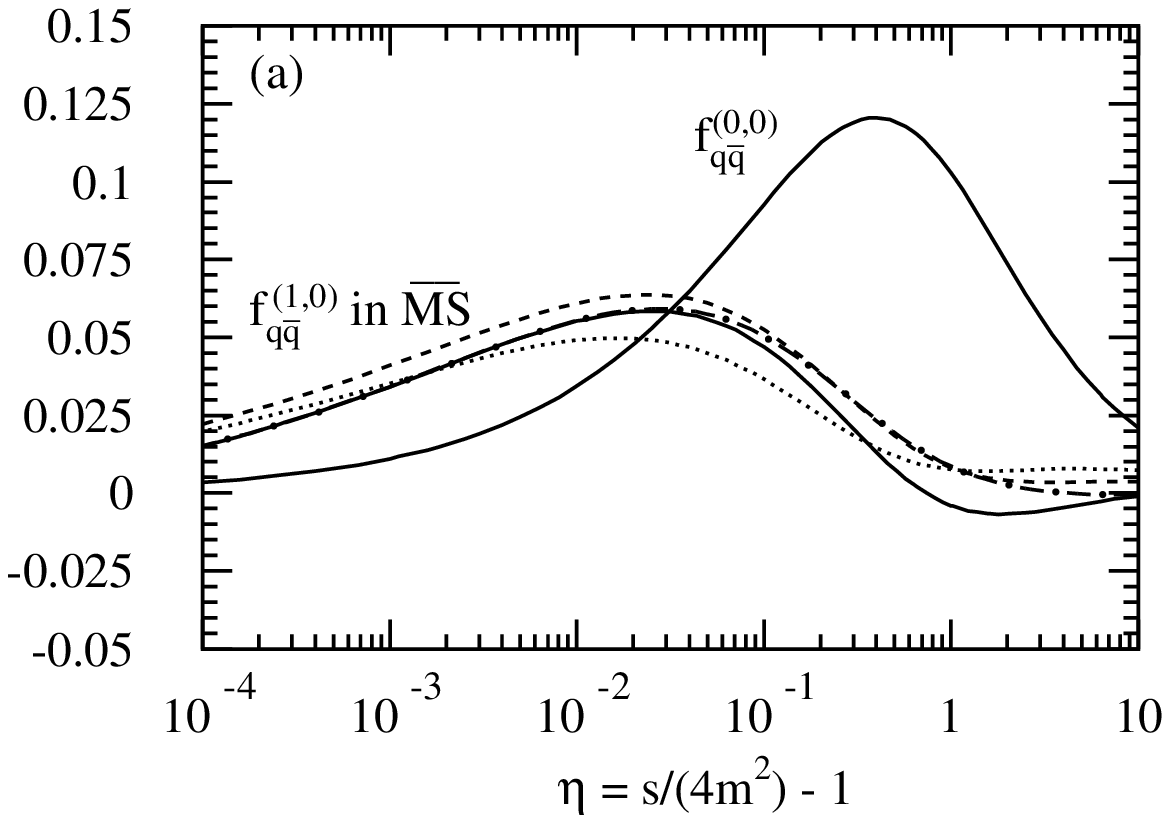,%
bbllx=50pt,bblly=110pt,bburx=285pt,bbury=450pt,angle=270,width=8.25cm}
\epsfig{file=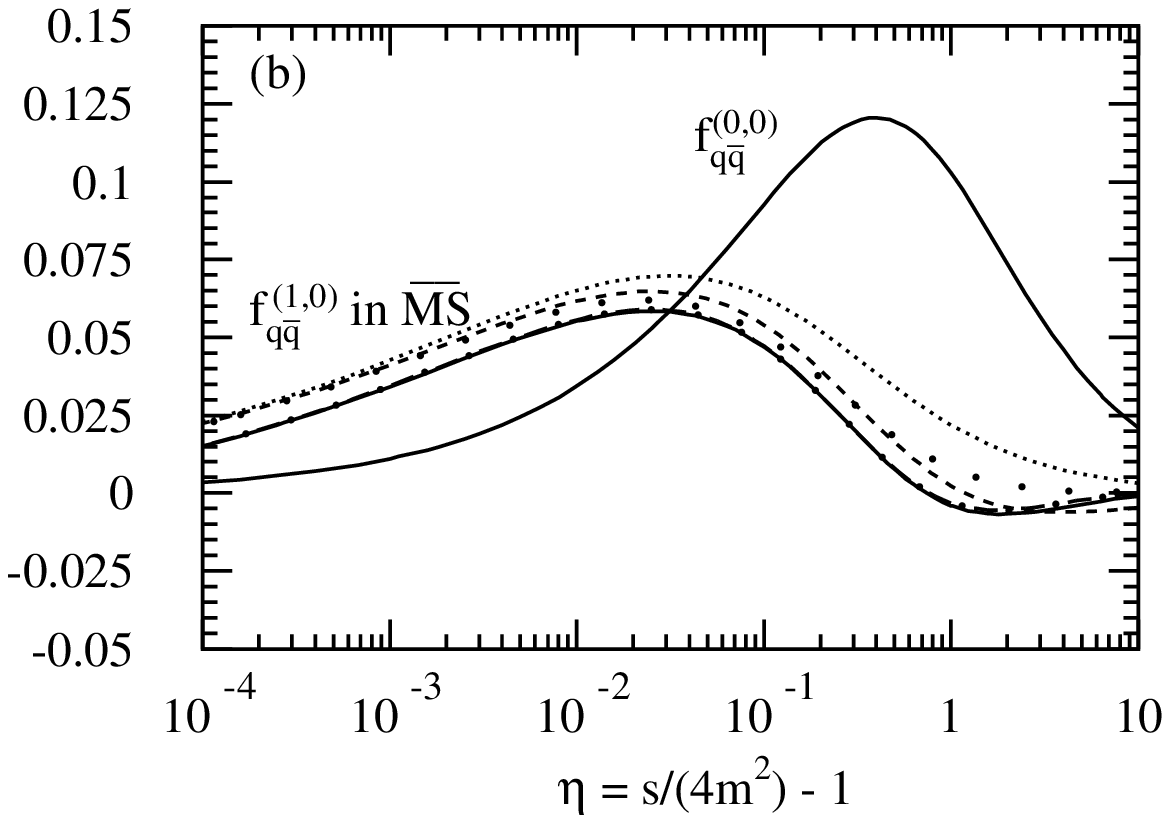,%
bbllx=50pt,bblly=110pt,bburx=285pt,bbury=450pt,angle=270,width=8.25cm}
\caption[dum]{\label{plot-1q} {\small{
(a) The $\h$-dependence of the scaling functions 
$f^{(k,0)}_{q{\Bar{q}}}(\h),\;k=0,1$ in the ${\Bar{\rm{MS}}}$-scheme 
and 1PI kinematics. 
We show the exact results for $f^{(k,0)}_{q{\Bar{q}}},\;k=0,1$ 
(solid lines), the LL approximation to $f^{(1,0)}_{q{\Bar{q}}}$ (dotted line), 
the NLL approximation to $f^{(1,0)}_{q{\Bar{q}}}$ (dashed line) and the 
NNLL approximation to $f^{(1,0)}_{q{\Bar{q}}}$ (dashed-dotted line). 
(b) The same as (a) in PIM kinematics. The spaced-dotted curve
corresponds to the approximation involving the leading two
powers of $\ln(\beta)$.
}}}
\end{center}
\end{figure}
\begin{figure}[htp]
\begin{center}
\epsfig{file=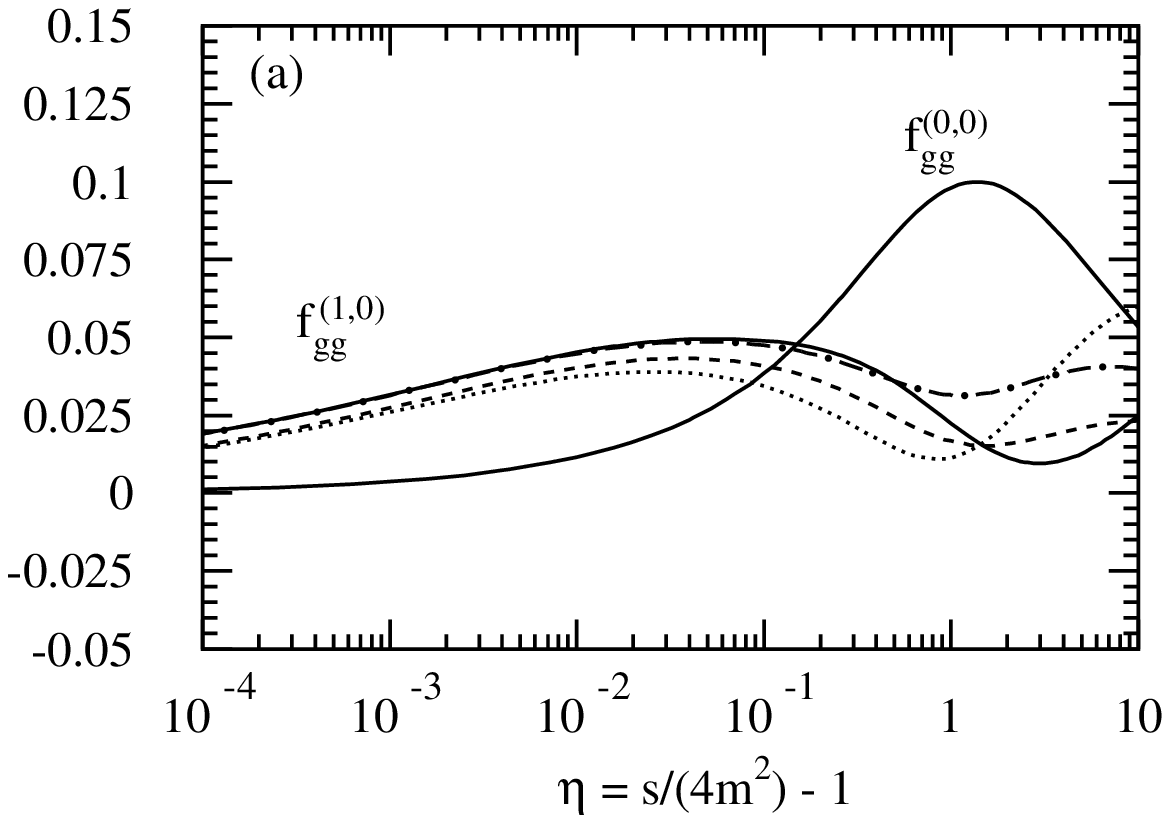,%
bbllx=50pt,bblly=110pt,bburx=285pt,bbury=450pt,angle=270,width=8.25cm}
\epsfig{file=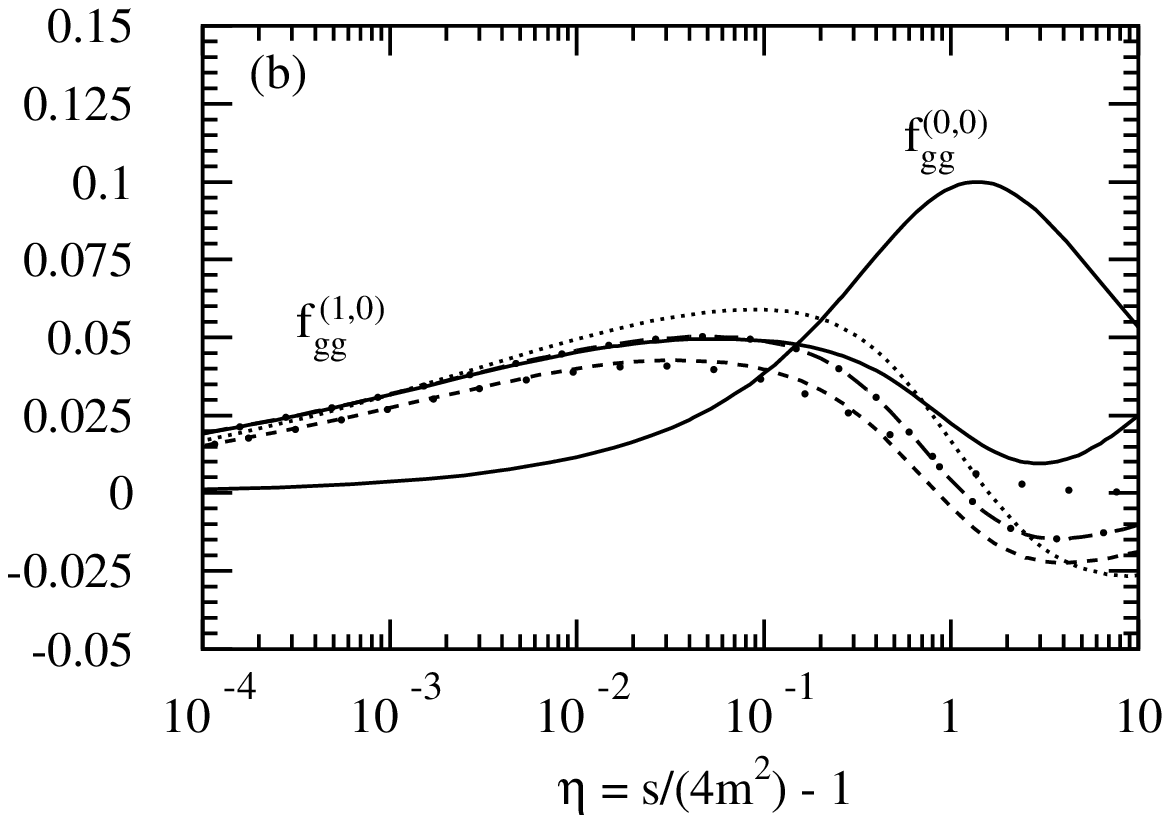,%
bbllx=50pt,bblly=110pt,bburx=285pt,bbury=450pt,angle=270,width=8.25cm}
\caption[dum]{\label{plot-1g} {\small{
(a) The $\h$-dependence of the scaling functions 
$f^{(k,0)}_{gg}(\h),\;k=0,1$ in 1PI kinematics. 
We show the exact results for $f^{(k,0)}_{gg},\;k=0,1$ 
(solid lines), the LL approximation to $f^{(1,0)}_{gg}$ (dotted line), 
the NLL approximation to $f^{(1,0)}_{gg}$ (dashed line) and the 
the NNLL approximation to $f^{(1,0)}_{gg}$ (dashed-dotted line). 
(b) The same as (a) in PIM kinematics. The spaced-dotted curve
corresponds to the approximation involving the leading two
powers of $\ln(\beta)$.
}}}
\end{center}
\end{figure}

In Figs.~\ref{plot-2q} and \ref{plot-2g}, we show the NNLO scaling functions 
$f^{(2,0)}_{q{\Bar{q}}}$ and $f^{(2,0)}_{gg}$ for both kinematics. 
Again, the LL approximations differ but there is good agreement 
between 1PI and PIM kinematics at NLL accuracy.
We also display the NNLL approximations for $f^{(2,0)}_{q{\Bar{q}}}$ and 
$f^{(2,0)}_{gg}$, presently the best estimates for these functions,
at least for $\eta \; \ltap 1$,
since exact calculations are not yet available. 
Near threshold, the NNLL approximations show some deviation from 
the NLL approximation due to interplay between the one-loop LL 
contributions and the one-loop Coulomb terms.  Note that while the $q \overline
q$ channel shows relatively good agreement between the two kinematics at large
$\eta$, the $gg$ results are quite different since the 1PI results are
positive while the PIM results are negative.

We have also calculated the behavior 
of the scaling functions $f^{(k,0)}_{ij}(\h), k = 1,2,$ 
in the threshold limit $s\rightarrow 4m^2$,
keeping only terms that grow as $\ln(\beta)$
to next-to-leading accuracy.
The resulting expressions are given in 
appendix \ref{sec:results-sright-4m2},
and shown in Figs. \ref{plot-1q}(b), \ref{plot-1g}(b), 
\ref{plot-2q}(b) and \ref{plot-2g}(b). They 
agree fairly well with the NLL PIM results.

In general, comparison of the exact and approximate 
results in the various channels and kinematics
in Figs. \ref{plot-1q} and \ref{plot-1g} 
indicates the range of $\eta$ values over which we expect the 
threshold approximation to be valid.
The large-$\eta$ behaviour of the scaling functions,
furthest away from threshold,
may be approximated by the methods of Ref.~\cite{Ellis:1990hw,
Collins:1991ty,Catani:1990xk,Catani:1991eg}.
However, the behavior of the scaling functions
in the intermediate region, $1\;\ltap \; \eta \; \ltap \; 10$, can 
only be determined by exact computation.

\begin{figure}[htp]
\begin{center}
\epsfig{file=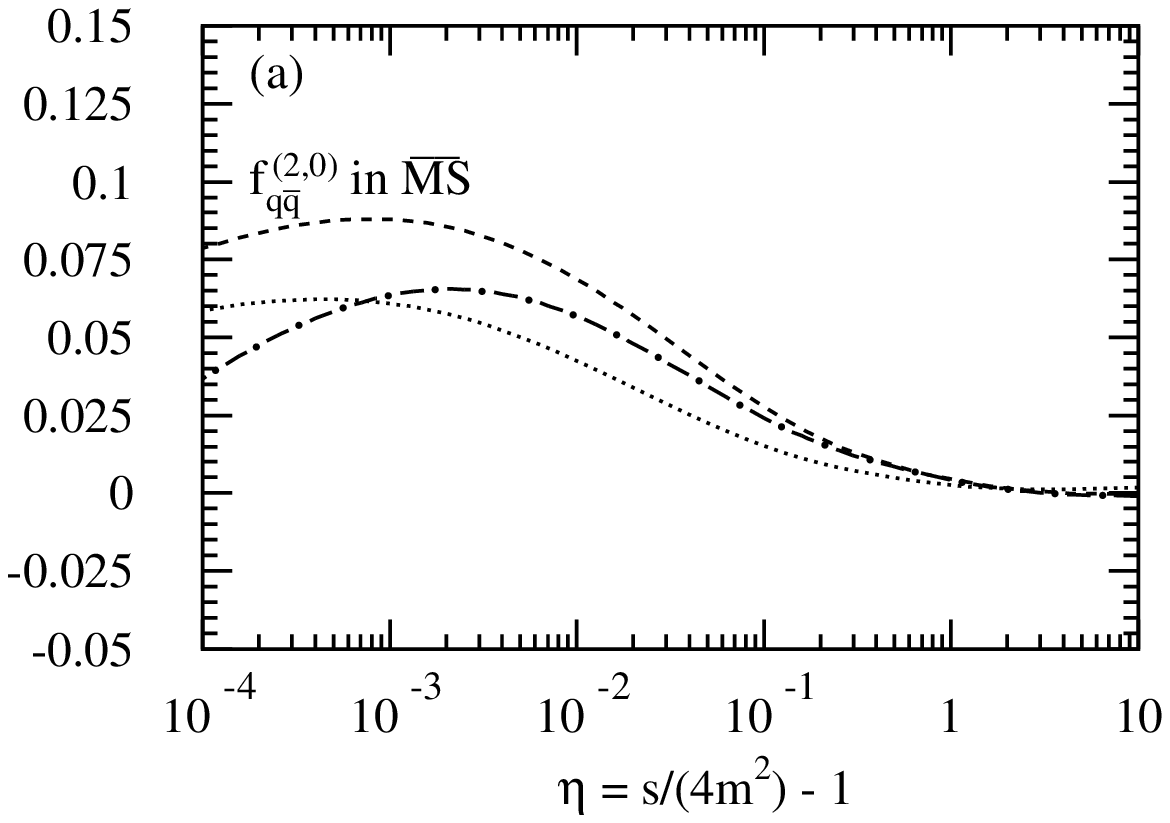,%
bbllx=50pt,bblly=110pt,bburx=285pt,bbury=450pt,angle=270,width=8.25cm}
\epsfig{file=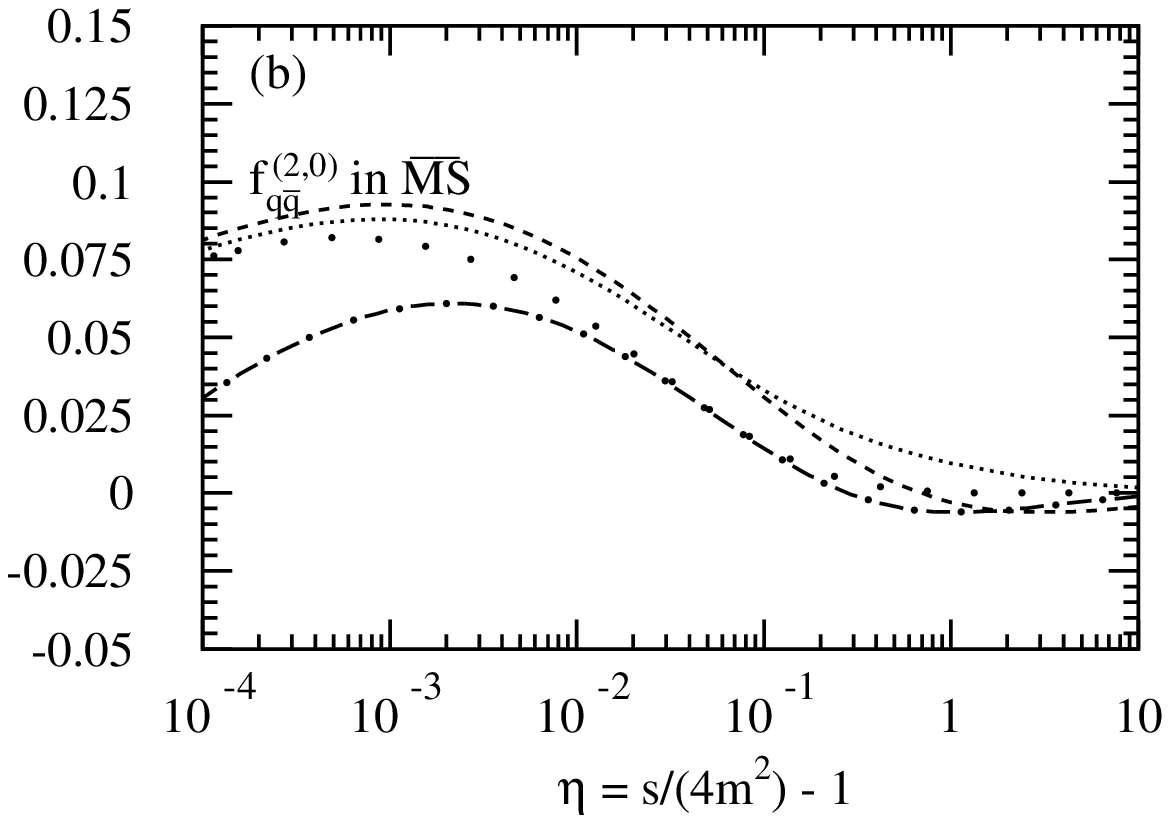,%
bbllx=50pt,bblly=110pt,bburx=285pt,bbury=450pt,angle=270,width=8.25cm}
\caption[dum]{\label{plot-2q} {\small{
(a) The $\h$-dependence of the scaling function 
$f^{(2,0)}_{q{\Bar{q}}}(\h)$ in the ${\Bar{\rm{MS}}}$-scheme 
and 1PI kinematics. 
We show the LL approximation (dotted line), 
the NLL approximation (dashed line) and 
the NNLL approximation (dashed-dotted line). 
(b) The same as (a) in PIM kinematics. The spaced-dotted curve
corresponds to the approximation involving the leading two
powers of $\ln(\beta)$.
}}}
\end{center}
\end{figure}
\begin{figure}[htp]
\begin{center}
\epsfig{file=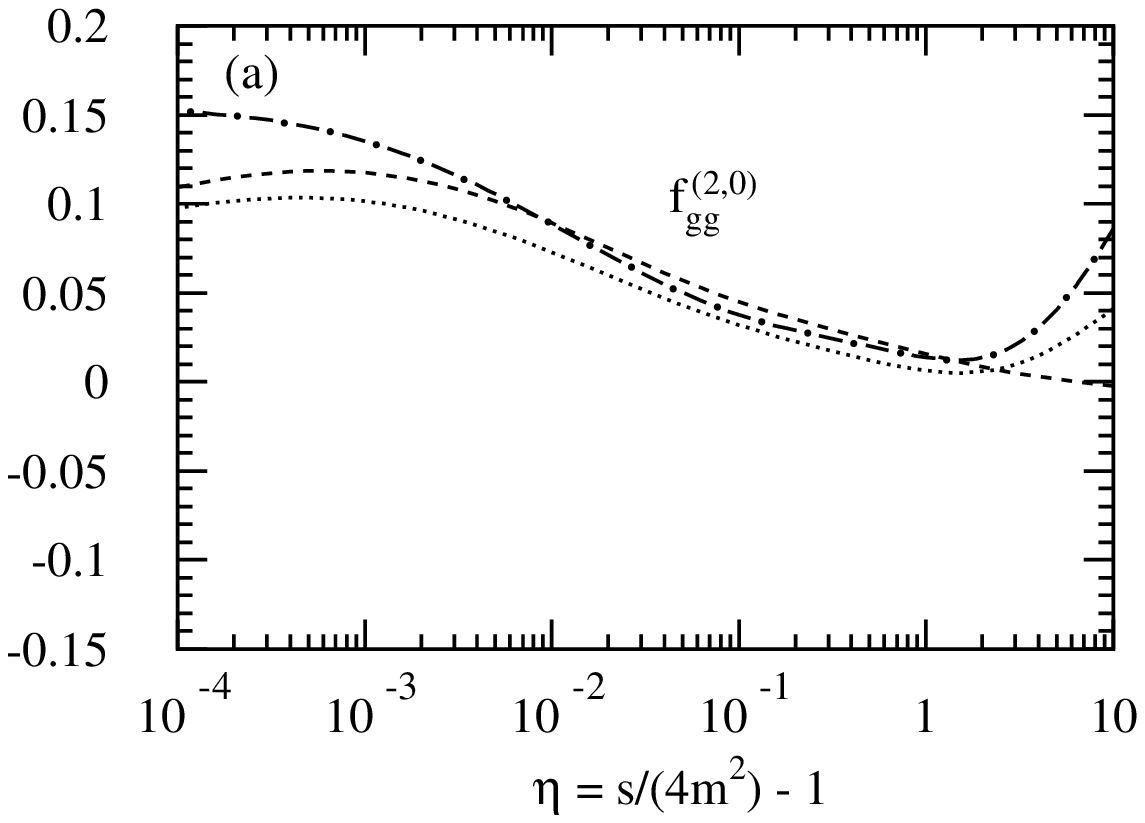,%
bbllx=50pt,bblly=110pt,bburx=285pt,bbury=450pt,angle=270,width=8.25cm}
\epsfig{file=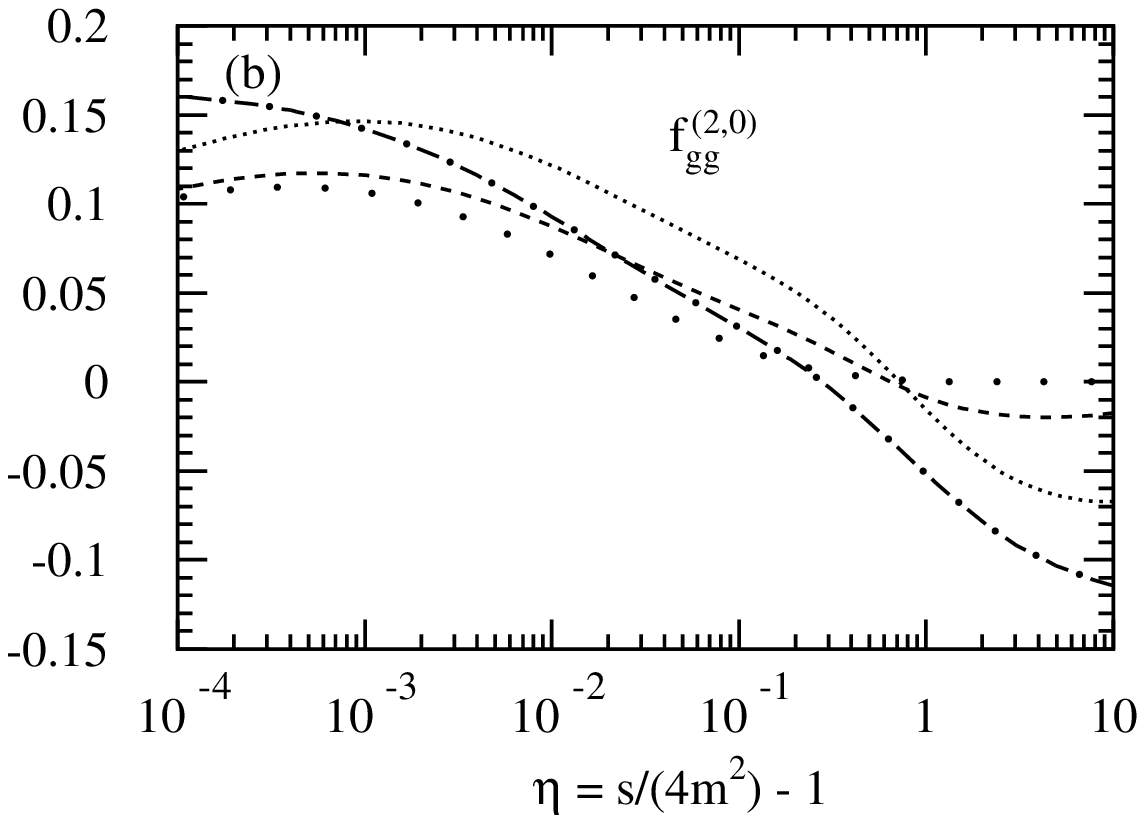,%
bbllx=50pt,bblly=110pt,bburx=285pt,bbury=450pt,angle=270,width=8.25cm}
\caption[dum]{\label{plot-2g} {\small{
(a) The $\h$-dependence of the scaling function 
$f^{(2,0)}_{gg}(\h)$ in 1PI kinematics. 
We show the LL approximation (dotted line), 
the NLL approximation (dashed line) and 
the NNLL approximation (dashed-dotted line). 
(b) The same as (a) in PIM kinematics. The spaced-dotted curve
corresponds to the approximation involving the leading two
powers of $\ln(\beta)$.
}}}
\end{center}
\end{figure}

We now check if the $q{\Bar{q}}$ ${\Bar{\rm{MS}}}$ results
are mirrored in the DIS scheme. 
Since the scaling functions 
in the two schemes differ already at LL, the higher-order scaling functions
will be quite different numerically.
Of course, this difference should be compensated in principle by
the parton densities for the 
scheme-independent hadronic cross section.

Figures \ref{plot-d1q} and \ref{plot-d2q}
show the functions $f^{(1,0)}_{q \overline q}$ and $f^{(2,0)}_{q \overline q}$
in the DIS scheme.
We see again, as in 
Figs.~\ref{plot-1q} and \ref{plot-2q}, that the exact $f^{(1,0)}_{q \overline 
q}$ results are best approximated at small $\eta$ by the NNLL calculations in both
kinematics while the LL and NLL calculations either over- or underestimate the
exact calculation.  The $\overline {\rm MS}$
results generally trace the exact curves somewhat better than those in the DIS scheme. 
In particular, for the DIS
scaling function $f^{(1,0)}_{q{\Bar{q}}}$,
only the NNLL approximations provide good agreement with the exact curve 
for both 1PI and PIM kinematics over a large range in $\h$.
This can be traced to the large delta-function terms in the scheme-changing functions 
in Eqs.~(\ref{DIS-qqbar-one-loop}) and (\ref{DIS-qqbar-one-loop_PIM}).

\begin{figure}[htp]
\begin{center}
\epsfig{file=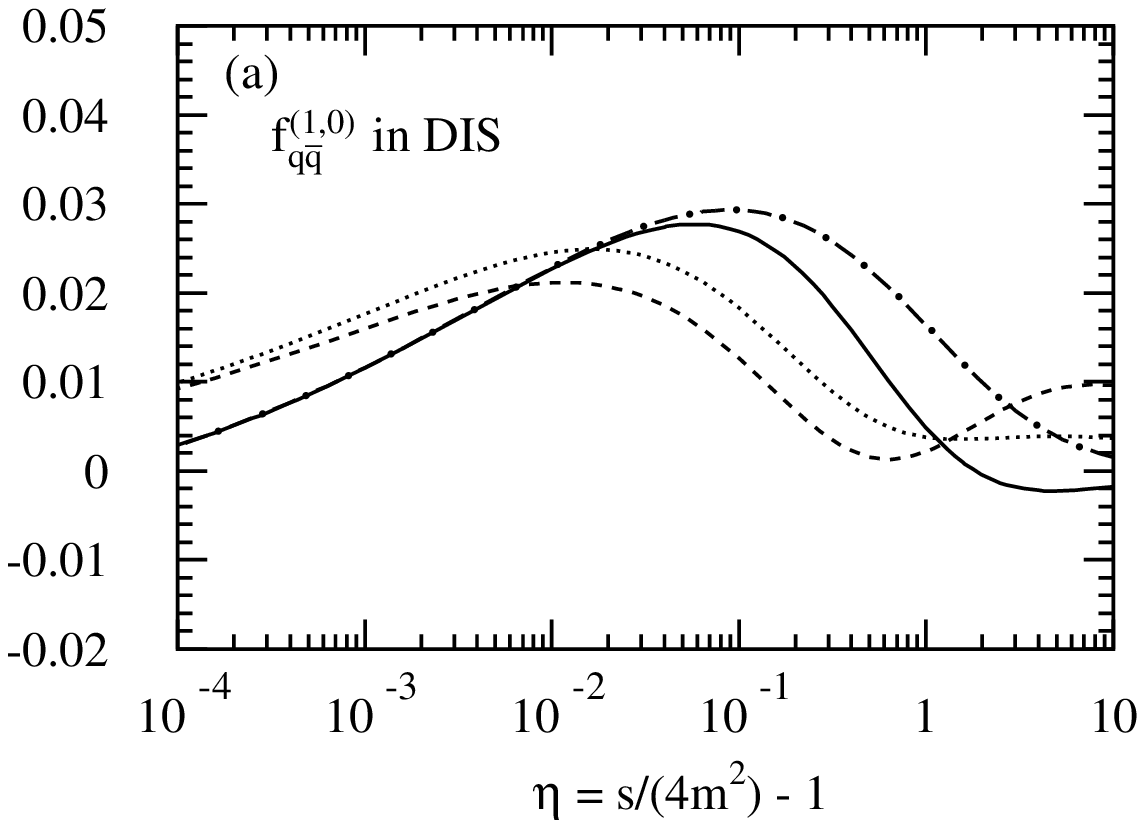,%
bbllx=50pt,bblly=110pt,bburx=285pt,bbury=450pt,angle=270,width=8.25cm}
\epsfig{file=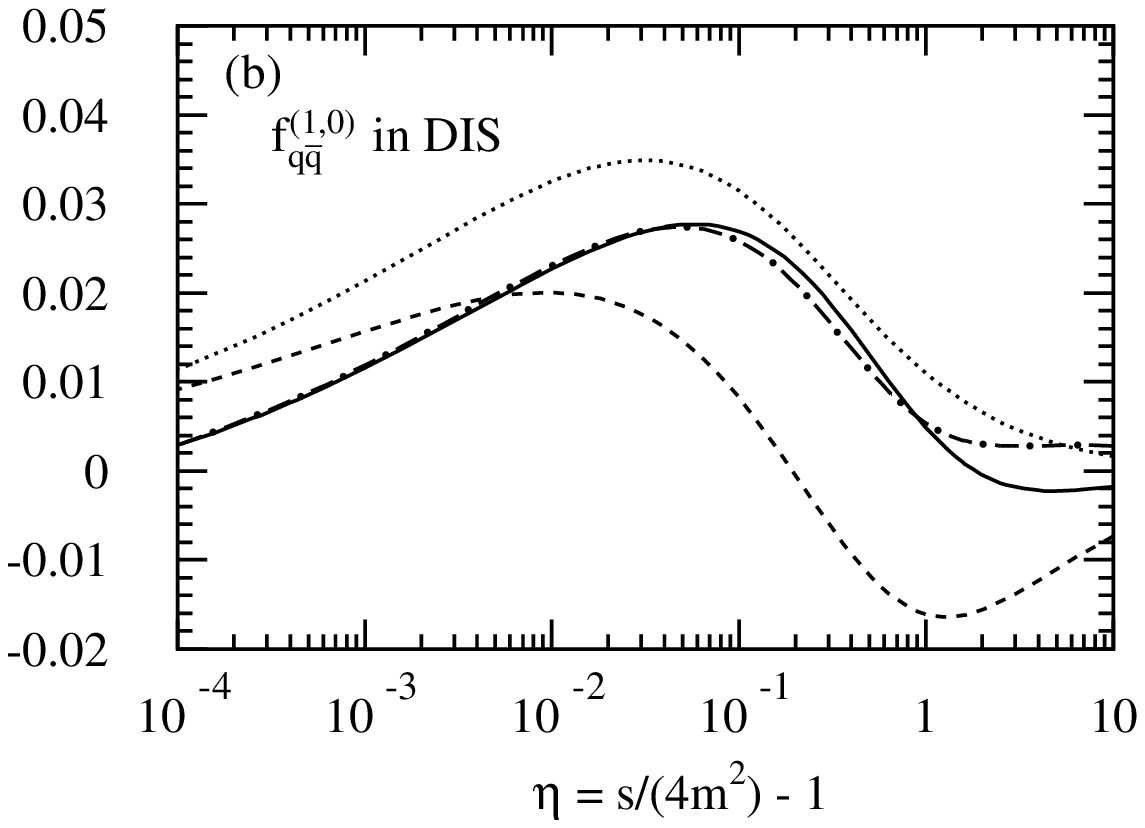,%
bbllx=50pt,bblly=110pt,bburx=285pt,bbury=450pt,angle=270,width=8.25cm}
\caption[dum]{\label{plot-d1q} {\small{
(a) The $\h$-dependence of the scaling function 
$f^{(1,0)}_{q{\Bar{q}}}(\h)$ in the DIS-scheme 
and 1PI kinematics. 
We show the exact result (solid line), the LL approximation (dotted line), 
the NLL approximation (dashed line) and the 
NNLL approximation (dashed-dotted line). 
(b) The same as (a) in PIM kinematics.
}}}
\end{center}
\end{figure}
\begin{figure}[htp]
\begin{center}
\epsfig{file=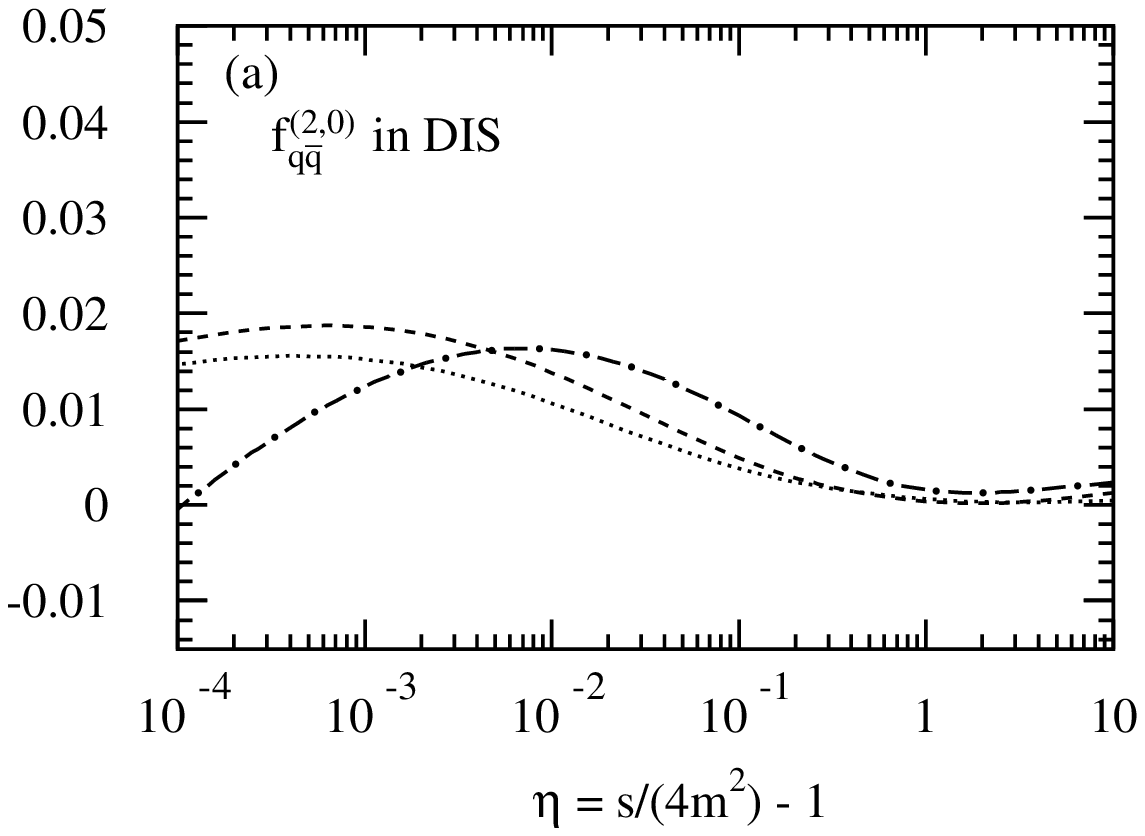,%
bbllx=50pt,bblly=110pt,bburx=285pt,bbury=450pt,angle=270,width=8.25cm}
\epsfig{file=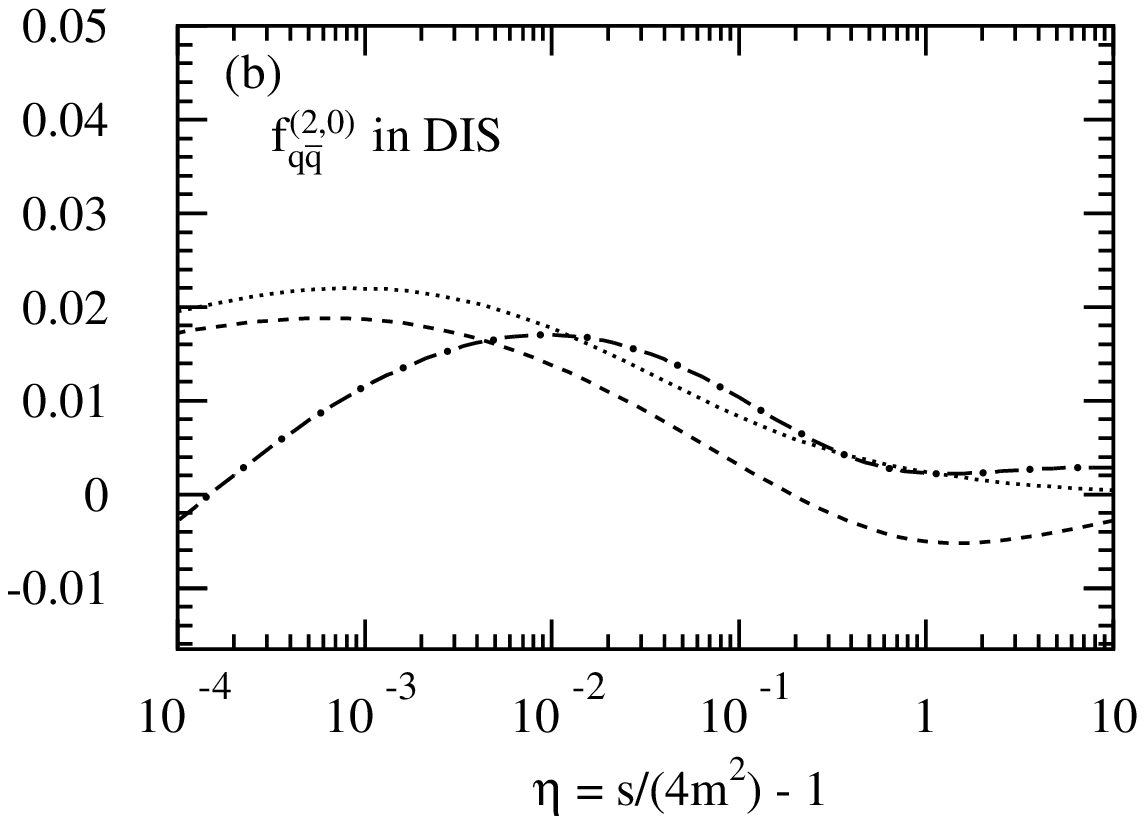,%
bbllx=50pt,bblly=110pt,bburx=285pt,bbury=450pt,angle=270,width=8.25cm}
\caption[dum]{\label{plot-d2q} {\small{
(a) The $\h$-dependence of the scaling function 
$f^{(2,0)}_{q{\Bar{q}}}(\h)$ in the DIS-scheme 
and 1PI kinematics. 
We show the LL approximation (dotted line), 
the NLL approximation (dashed line) and 
the NNLL approximation (dashed-dotted line). 
(b) The same as (a) in PIM kinematics.
}}}
\end{center}
\end{figure}

We next discuss the functions controlling the scale dependence.
They can be determined exactly at NLO and NNLO
using renormalization group methods.
The exact $f^{(1,0)}_{ij}$ are required
to construct $f^{(2,1)}_{ij}$ and $f^{(2,2)}_{ij}$.
We neglect flavor mixing terms
which are of order $1/N$, where $N$ is the moment variable. We
checked that at Tevatron energies the error
due to this approximation is less than $1 \%$ at NNLO. In this way we
obtain
\begin{eqnarray} f^{(1,1)}_{q{\Bar{q}}} &=& \frac{1}{4 \p^2} \left[ 2 b_2\,
  f^{(0,0)}_{q{\Bar{q}}} - f^{(0,0)}_{q{\Bar{q}}} \otimes P_{qq}^{(0)}
\right]\, ,
\label{ex-f11qq}\\[1ex]
f^{(2,1)}_{q{\Bar{q}}} &\simeq& \frac{1}{(4 \p^2)^2} \left[ 2 b_3\,
  f^{(0,0)}_{q{\Bar{q}}} - f^{(0,0)}_{q{\Bar{q}}} \otimes P_{qq}^{(1)}
\right]\, + \frac{1}{4 \p^2} \left[ 3 b_2\, f^{(1,0)}_{q{\Bar{q}}} -
  f^{(1,0)}_{q{\Bar{q}}} \otimes P_{qq}^{(0)} \right]\, ,
\label{ex-f21qq}\\[1ex]
f^{(2,2)}_{q{\Bar{q}}} &\simeq& \frac{1}{(4 \p^2)^2} \left[ 3 b_2^2\,
  f^{(0,0)}_{q{\Bar{q}}} - \frac{5}{2} b_2\, f^{(0,0)}_{q{\Bar{q}}}
  \otimes P_{qq}^{(0)} + \frac{1}{2} f^{(0,0)}_{q{\Bar{q}}} \otimes
  P_{qq}^{(0,0)} \right]\, ,
\label{ex-f22qq} \\[1ex]
f^{(1,1)}_{gg} &=& \frac{1}{4 \p^2} \left[ 2 b_2\, f^{(0,0)}_{gg} -
  f^{(0,0)}_{gg} \otimes P_{gg}^{(0)} \right]\, ,
\label{ex-f11gg}\\[1ex]
f^{(2,1)}_{gg} &\simeq& \frac{1}{(4 \p^2)^2} \left[ 2 b_3\,
  f^{(0,0)}_{gg} - f^{(0,0)}_{gg} \otimes P_{gg}^{(1)} \right]\, +
\frac{1}{4 \p^2} \left[ 3 b_2\, f^{(1,0)}_{gg} - f^{(1,0)}_{gg}
  \otimes P_{gg}^{(0)} \right]\, ,
\label{ex-f21gg}\\[1ex]
f^{(2,2)}_{gg} &\simeq& \frac{1}{(4 \p^2)^2} \left[ 3 b_2^2\,
  f^{(0,0)}_{gg} - \frac{5}{2} b_2\, f^{(0,0)}_{gg} \otimes
  P_{gg}^{(0)} + \frac{1}{2} f^{(0,0)}_{gg} \otimes P_{gg}^{(0,0)}
\right]\, ,
\label{ex-f22gg}
\end{eqnarray} 
where $P_{qq}^{(0)}$, $P_{gg}^{(0)}$ and $P_{qq}^{(1)}$, $P_{gg}^{(1)}$
are the one- and two-loop splitting functions 
\cite{Furmanski:1980cm,Curci:1980uw,Moch:1999eb}, 
and $\simeq$ indicates the neglect of flavor-mixing terms.   
The convolutions involving a scaling function $f^{(i,0)}_{ij}$ are defined as 
\begin{eqnarray} 
\left(
  f^{(i,0)}_{pk} \otimes P_{p^{\prime}k}^{(j)} \right) (\h(x))
&\equiv& \int\limits_{4m^2/s}^{1}\, dz\,\, f^{(i,0)}_{pk}
\left(\h(xz)\right)\, P_{p^{\prime}k}^{(j)}(z)\, 
\end{eqnarray} 
with $\h(x)
= x s/(4m^2) - 1$.  The standard convolution of two splitting functions,
$P_{qq}^{(0,0)}$ and
$P_{gg}^{(0,0)}$, in Eqs.~(\ref{ex-f22qq}) and (\ref{ex-f22gg}) are
\begin{eqnarray} P_{ii}^{(0,0)}(x) &\equiv& \int\limits_{0}^{1}\, dx_1\,
\int\limits_{0}^{1}\, dx_2\, \d(x -x_1 x_2)
P_{ii}^{(0)}(x_1) P_{ii}^{(0)}(x_2)\, , 
\end{eqnarray} 
with $i=q,g$.
Equations~(\ref{ex-f11qq}) and (\ref{ex-f11gg}) naturally agree 
with the results
for $f^{(1,1)}_{q{\Bar{q}}}$ and $f^{(1,1)}_{gg}$ in Ref.~\cite{Nason:1988xz}.
We have also checked that the above results agree to NNLL with 
the expressions in appendix B when 
integrated as in Eqs.~(\ref{eq:11}) and (\ref{totalpartoncrs_PIM}).
Note that in appendix B we have also given results for terms in the 
differential NNLO functions controlling the scale dependence
beyond NNLL accuracy, 
thus deriving all the soft plus virtual terms in these functions. 
The exact hard terms are calculated
only for the integrated cross section as above.

We begin by showing the scale-changing scaling functions 
in the $q\Bar{q}$ channel and $\overline{\rm MS}$ scheme, 
comparing 1PI and PIM kinematics. 
In Fig.~\ref{plot-3q} we show $f^{(1,1)}_{q{\Bar{q}}}$ 
and note that the PIM LL approximation reproduces the exact curve somewhat
better than the 1PI LL aproximation.
The NLL approximations agree better, even for larger $\h$.

\begin{figure}[htp]
\begin{center}
\epsfig{file=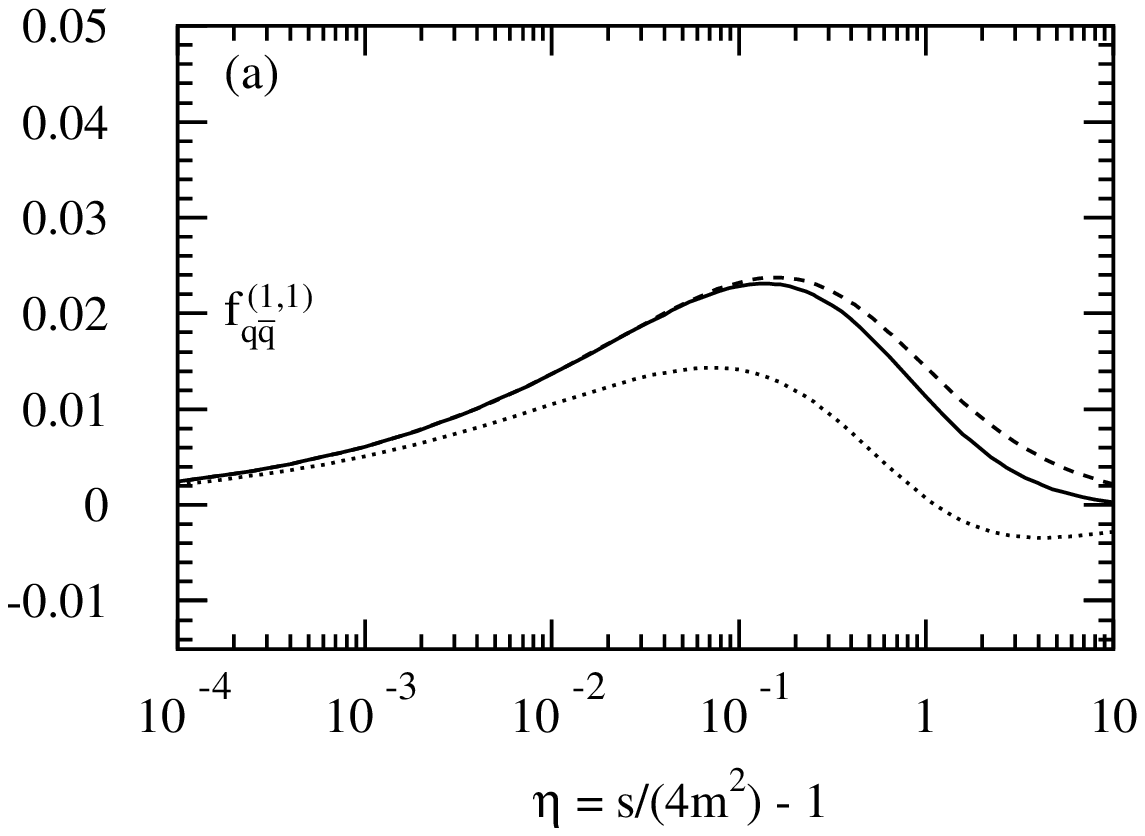,%
bbllx=50pt,bblly=110pt,bburx=285pt,bbury=450pt,angle=270,width=8.25cm}
\epsfig{file=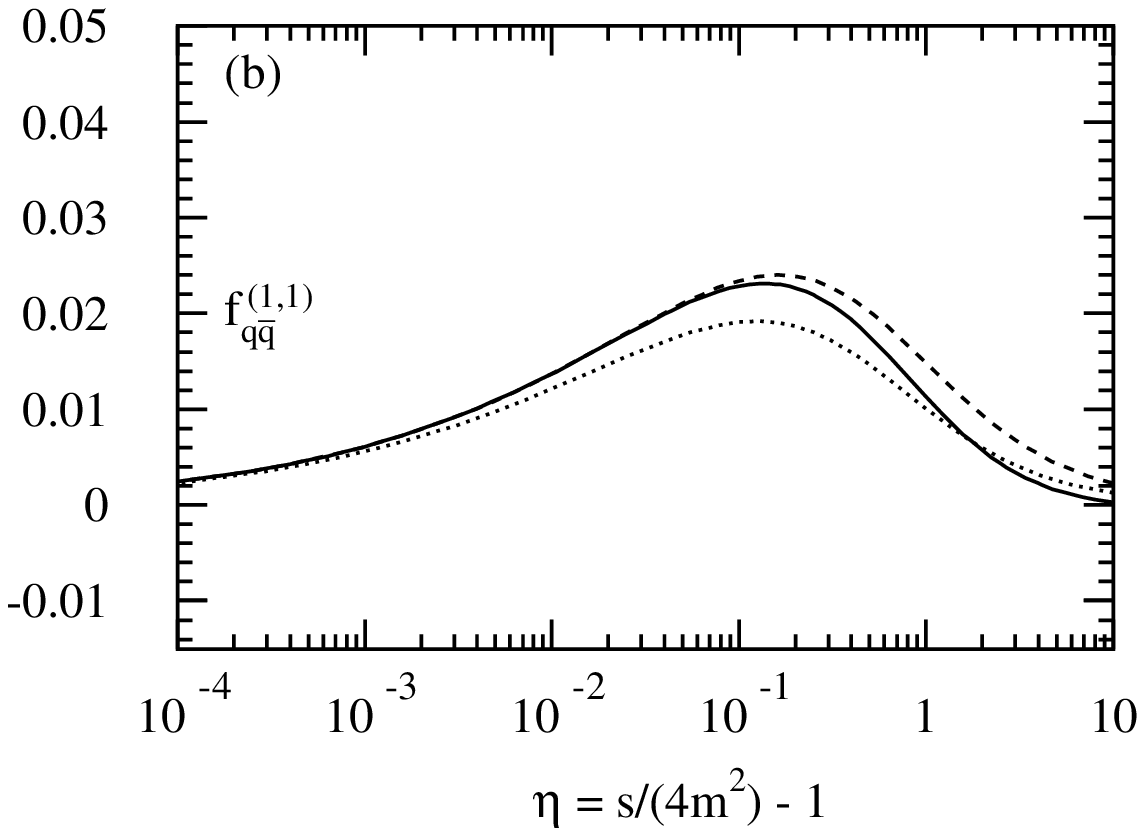,%
bbllx=50pt,bblly=110pt,bburx=285pt,bbury=450pt,angle=270,width=8.25cm}
\caption[dum]{\label{plot-3q} {\small{
(a) The $\h$-dependence of the scaling function
$f^{(1,1)}_{q{\Bar{q}}}(\h)$ in 1PI kinematics. 
We show the exact result (solid line), 
the LL approximation (dotted line) and 
the NLL approximation (dashed line). 
(b) The same as (a) in PIM kinematics.
}}}
\end{center}
\end{figure}
\begin{figure}[htp]
\begin{center}
\epsfig{file=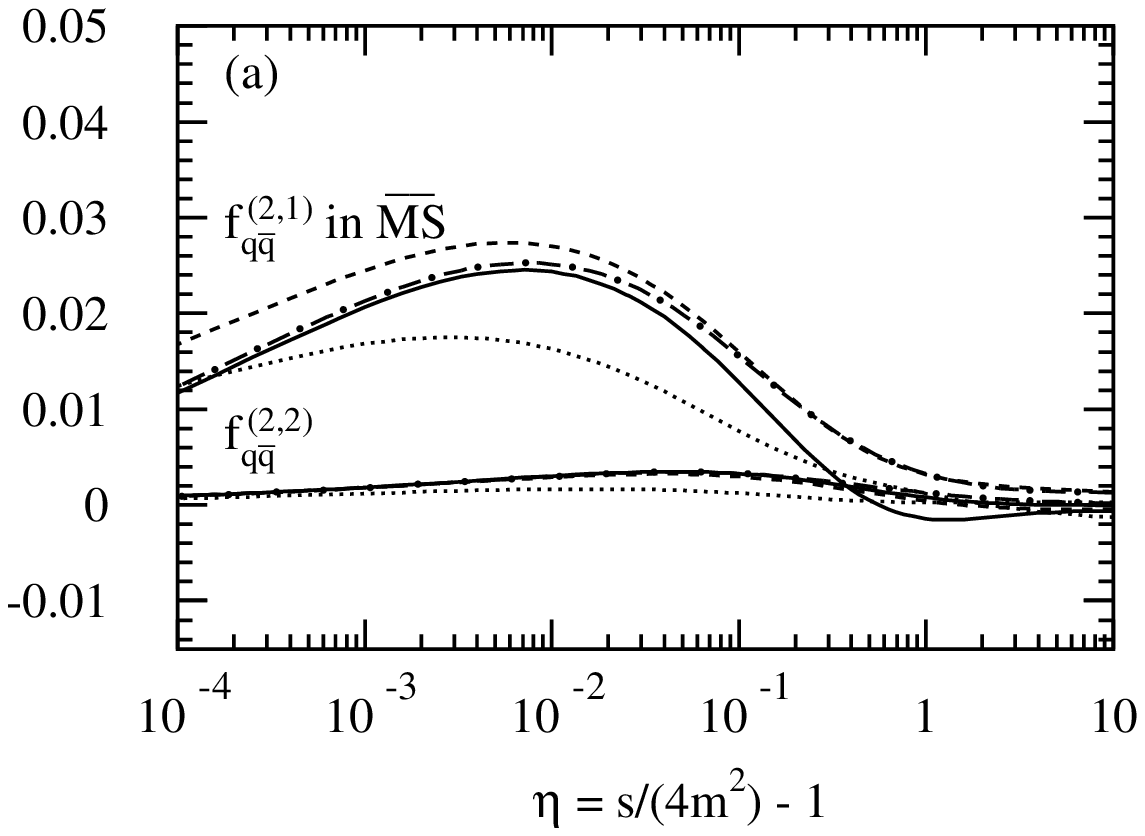,%
bbllx=50pt,bblly=110pt,bburx=285pt,bbury=450pt,angle=270,width=8.25cm}
\epsfig{file=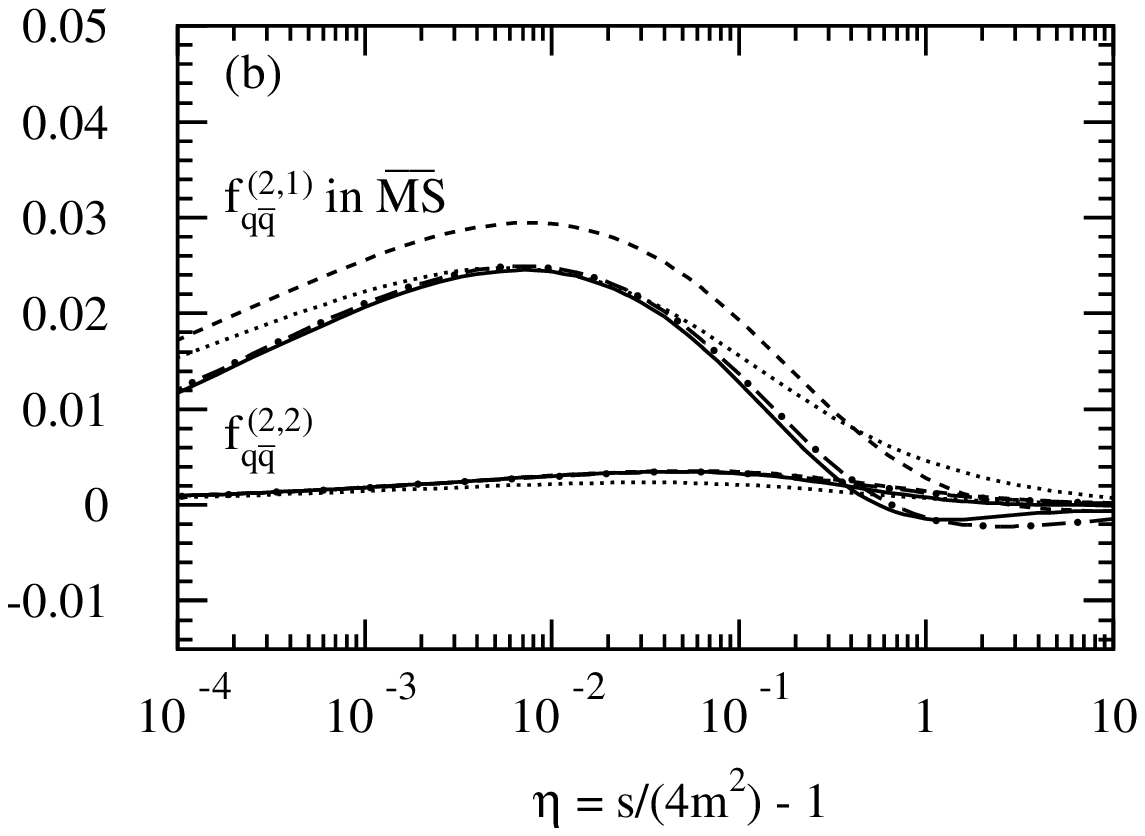,%
bbllx=50pt,bblly=110pt,bburx=285pt,bbury=450pt,angle=270,width=8.25cm}
\caption[dum]{\label{plot-4q} {\small{
(a) The $\h$-dependence of the scaling functions
$f^{(2,1)}_{q{\Bar{q}}}(\h)$ (${\Bar{\rm{MS}}}$-scheme)
and $f^{(2,1)}_{q{\Bar{q}}}(\h)$ in 1PI kinematics. 
We show the exact results (solid lines), 
the LL approximations (dotted lines), the NLL approximations (dashed lines) 
and the NNLL approximations (dashed-dotted lines). 
(b) The same as (a) in PIM kinematics.
}}}
\end{center}
\end{figure}

In Fig.~\ref{plot-4q} we show
the NNLO scaling functions $f^{(2,1)}_{q{\Bar{q}}}$ 
and $f^{(2,2)}_{q{\Bar{q}}}$.  We compare the exact curves 
calculated from Eqs.~(\ref{ex-f21qq}) and (\ref{ex-f22qq}) with 
our LL, NLL, and NNLL approximations.
Again we see that the NNLL approximations provide a 
remarkably good description 
of the exact results, both in shape and magnitude. 
The NNLL curves for 1PI and PIM kinematics are in very 
good agreement with each 
other, i.e. ambiguities
from the kinematics choice are very mild.
Similar conclusions hold for the $gg$ scaling functions 
$f^{(1,1)}_{gg}$, $f^{(2,1)}_{gg}$, and  $f^{(2,2)}_{gg}$,
shown in Figs.~\ref{plot-3g} and \ref{plot-4g}. Note however that also here
the NNLL 1PI and PIM results for $f^{(2,1)}_{gg}$ differ at large $\eta$.
The PIM NNLL scaling function differs significantly from the exact result.

\begin{figure}[htp]
\begin{center}
\epsfig{file=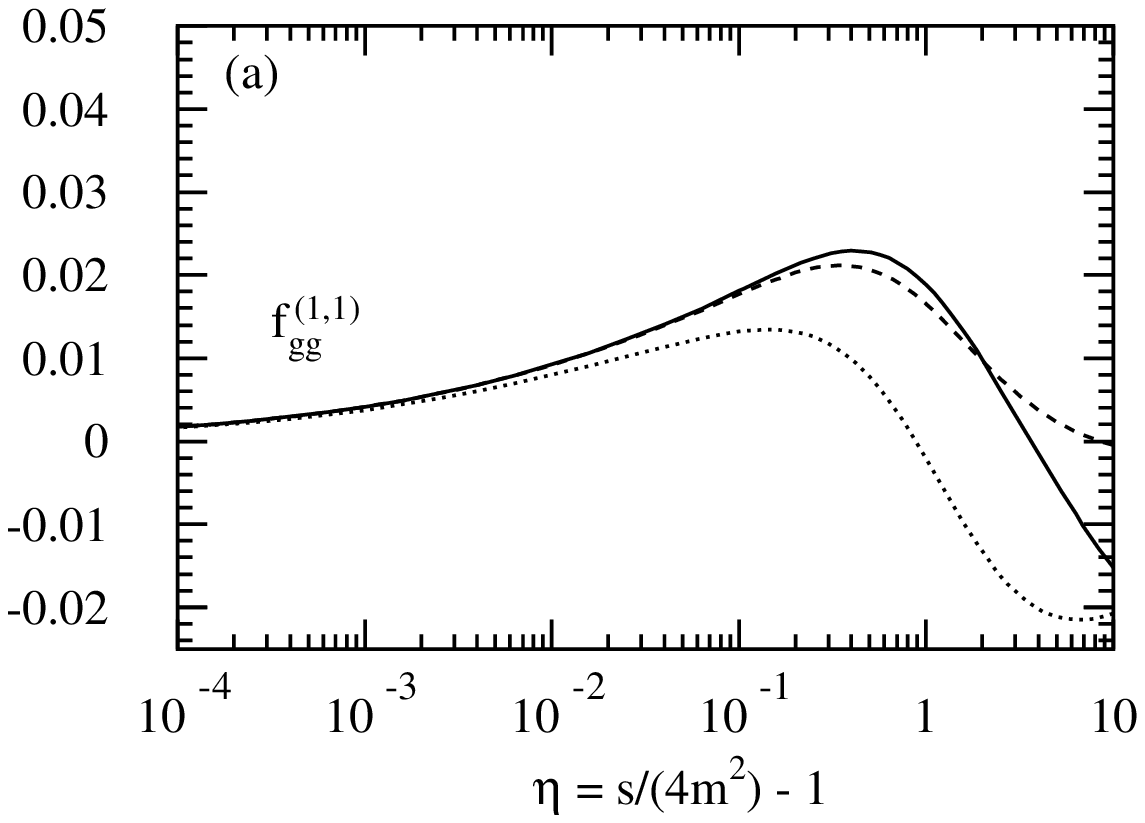,%
bbllx=50pt,bblly=110pt,bburx=285pt,bbury=450pt,angle=270,width=8.25cm}
\epsfig{file=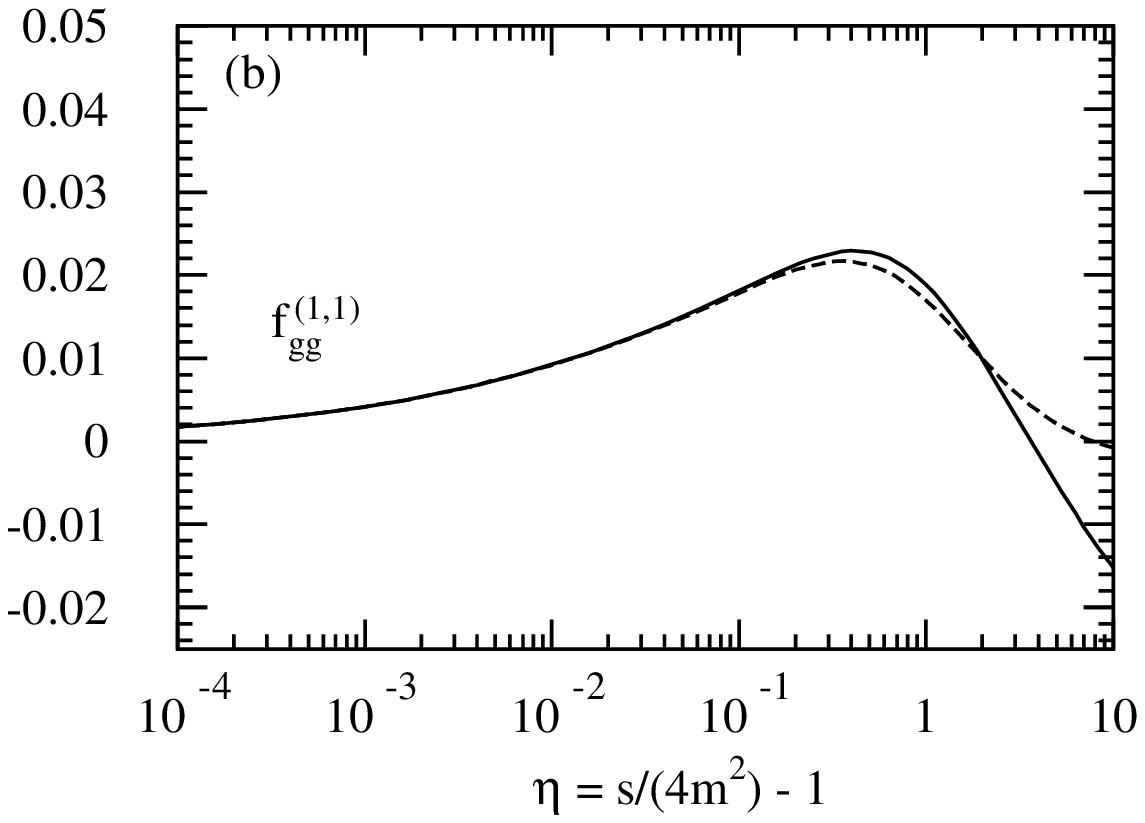,%
bbllx=50pt,bblly=110pt,bburx=285pt,bbury=450pt,angle=270,width=8.25cm}
\caption[dum]{\label{plot-3g} {\small{
(a) The $\h$-dependence of the scaling function
$f^{(1,1)}_{gg}(\h)$ in 1PI kinematics. 
We show the exact result (solid line), 
the LL approximation (dotted line) and 
the NLL approximation (dashed line). 
(b) The same as (a) in PIM kinematics; 
LL and NLL approximation coincide here.
}}}
\end{center}
\end{figure}
\begin{figure}[htp]
\begin{center}
\epsfig{file=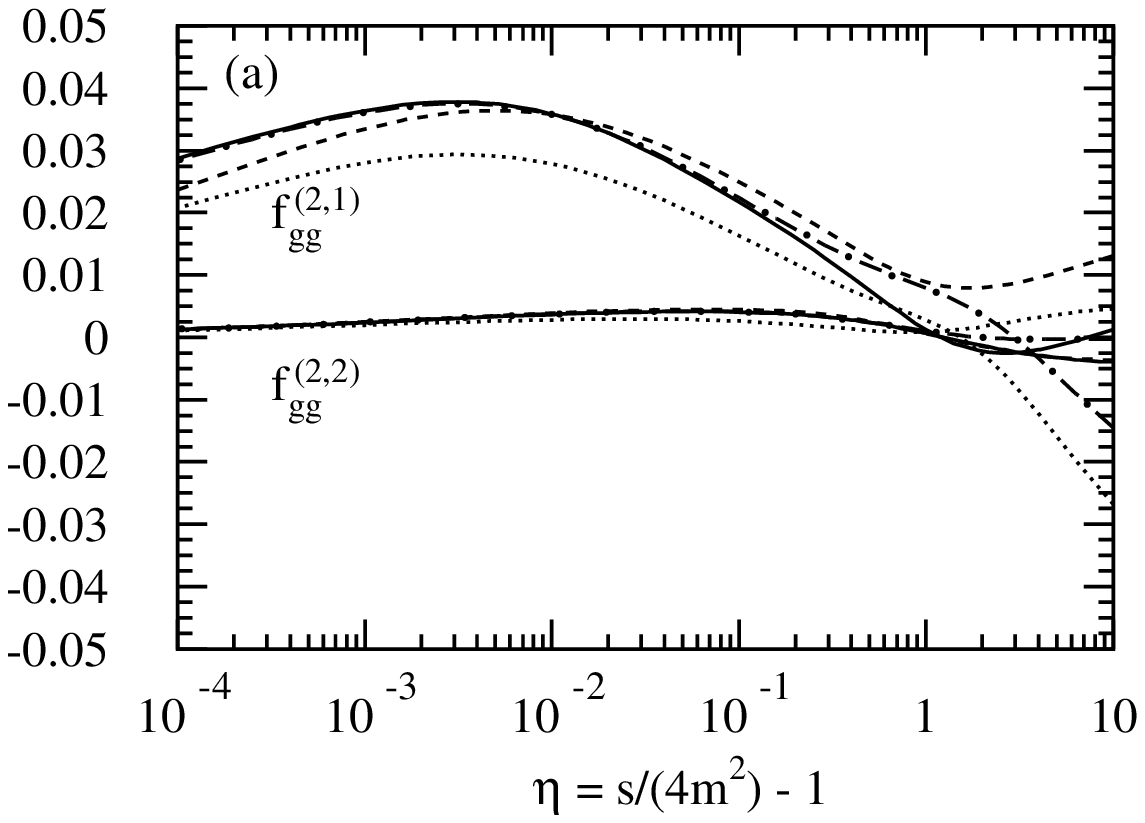,%
bbllx=50pt,bblly=110pt,bburx=285pt,bbury=450pt,angle=270,width=8.25cm}
\epsfig{file=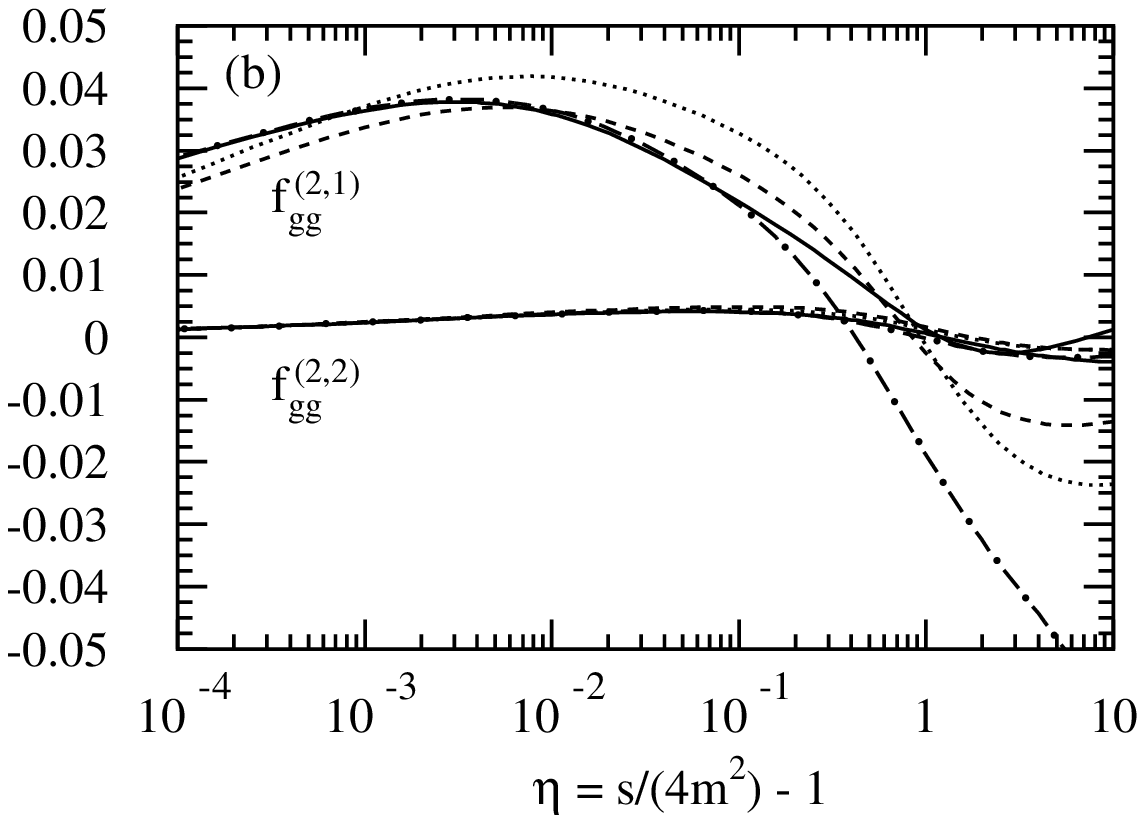,%
bbllx=50pt,bblly=110pt,bburx=285pt,bbury=450pt,angle=270,width=8.25cm}
\caption[dum]{\label{plot-4g} {\small{
(a) The $\h$-dependence of the scaling function
$f^{(2,l)}_{gg}(\h),\;l=1,2$ in 1PI kinematics. 
We show the exact results (solid lines), 
the LL approximations (dotted lines), the NLL approximations (dashed lines) 
and the NNLL approximations (dashed-dotted lines). 
(b) The same as (a) in PIM kinematics.
}}}
\end{center}
\end{figure}

For completeness, we also display the NNLO scaling functions 
$f^{(2,1)}_{q{\Bar{q}}}$ and $f^{(2,2)}_{q{\Bar{q}}}$ in the DIS scheme. 
The NLL approximations roughly trace the exact curve.
The (percentage-wise) large difference between the NLL approximation and the exact curve 
close to threshold may be attributed to the large constants in 
the one-loop scheme-changing functions in Eqs.~(\ref{DIS-qqbar-one-loop}) 
and (\ref{DIS-qqbar-one-loop_PIM}) that interfere 
with the one-loop LL terms. These are accounted for at NNLL accuracy, 
as Fig.~\ref{plot-d3q} demonstrates. However, the NNLL results for $f^{(2,1)}_{q{\Bar{q}}}$
differ more significantly between the two kinematic choices than their
$\msb$ counterparts.
\begin{figure}[htp]
\begin{center}
\epsfig{file=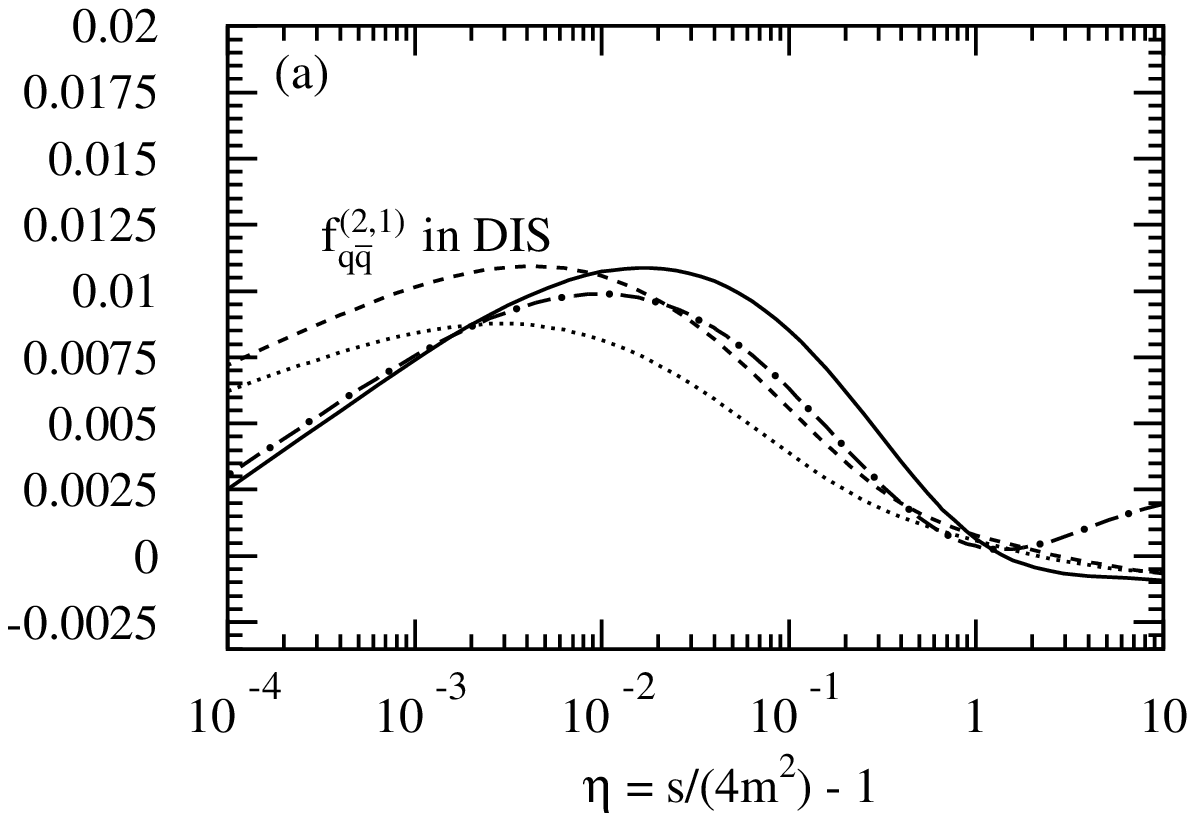,%
bbllx=50pt,bblly=110pt,bburx=285pt,bbury=450pt,angle=270,width=8.25cm}
\epsfig{file=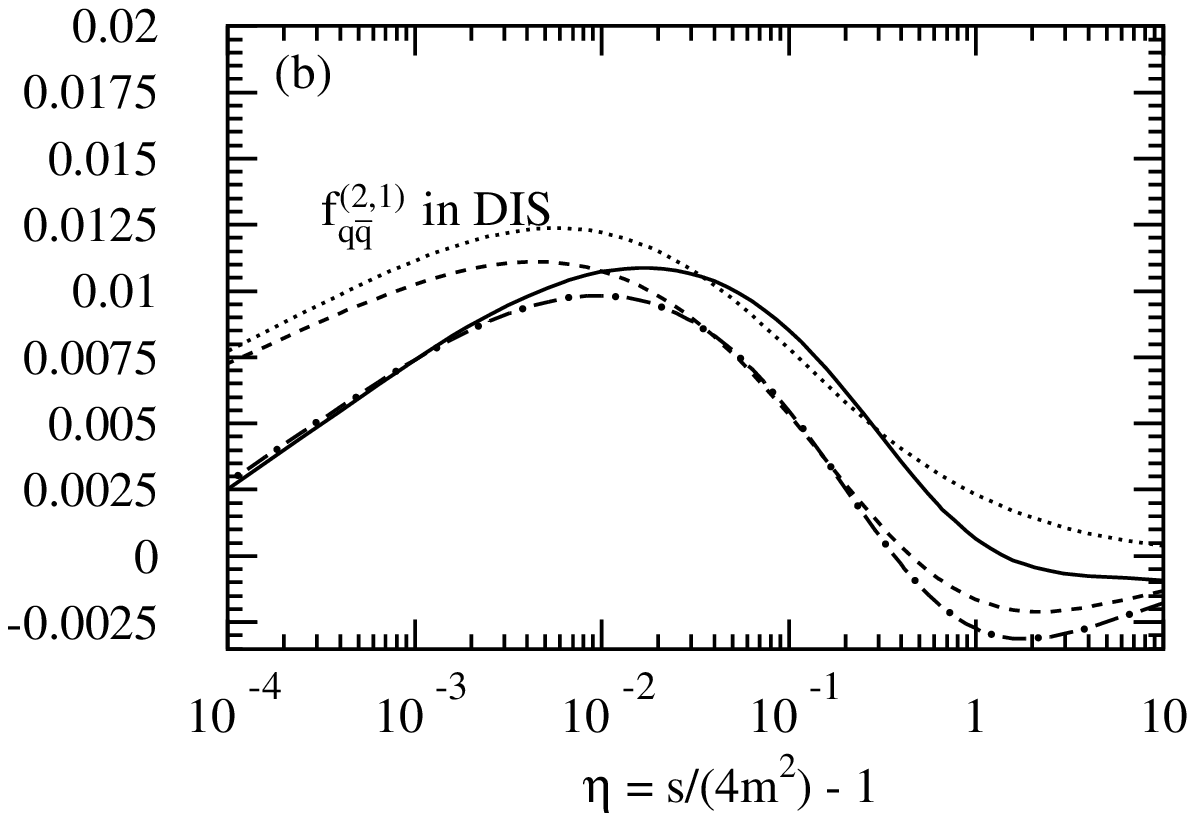,%
bbllx=50pt,bblly=110pt,bburx=285pt,bbury=450pt,angle=270,width=8.25cm}
\caption[dum]{\label{plot-d3q} {\small{
(a) The $\h$-dependence of the scaling function
$f^{(2,1)}_{q{\Bar{q}}}(\h)$ in the DIS-scheme 
and 1PI kinematics. 
We show the exact result (solid line), 
the LL approximation (dotted line), the NLL approximation (dashed line) 
and the NNLL approximation (dashed-dotted line). 
(b) The same as (a) in PIM kinematics.
}}}
\end{center}
\end{figure}

\section{Hadronic cross sections}

In the previous section we examined the quality
of our threshold approximations at the parton level.
Here we assess these approximations at the hadron
level for inclusive 
top quark production at the Fermilab 
Tevatron and bottom quark production at HERA-B.
The inclusive hadronic cross section is the convolution,
Eq.~(\ref{totalhadroncrs}), of parton distribution functions with the partonic cross
section, expressed in terms of the scaling functions, Eq.~(\ref{scalingfunctions}).
To facilitate the understanding of the results in this section in terms
of those of the previous section, we plot 
the flux factors $\Phi_{ij}(\eta,\mu^2)$,
Eq.~(\ref{eq:19}), for the above cases in Fig.~\ref{fluxxtevqq}. They show which $\eta$ values 
receive the most weight in the convolution integral.
In our numerical studies 
we use the two-loop expression of $\alpha_s$
and the CTEQ5M ($\overline {\rm MS}$ scheme) or CTEQ5D (DIS scheme) 
parametrizations of the parton distributions \cite{Lai:1999wy}
not only for the NLO results, but also for the NNLO (NNLO parton distributions
are not yet available) and LO results.
Thus in this section we keep the nonperturbative part of 
our results fixed when studying the
effect of increasing the perturbative order of our partonic cross sections.
For top quark production at the 
Tevatron and bottom quark production at HERA-B
the calculations probe the moderate to large $x$ region
where the parton distributions are well known, see Fig.~\ref{fluxxtevqq}.
We use five and four active flavors respectively for these cases and 
fix $\mu_R=\mu$.
\begin{figure}[htp]
\begin{center}
\epsfig{file=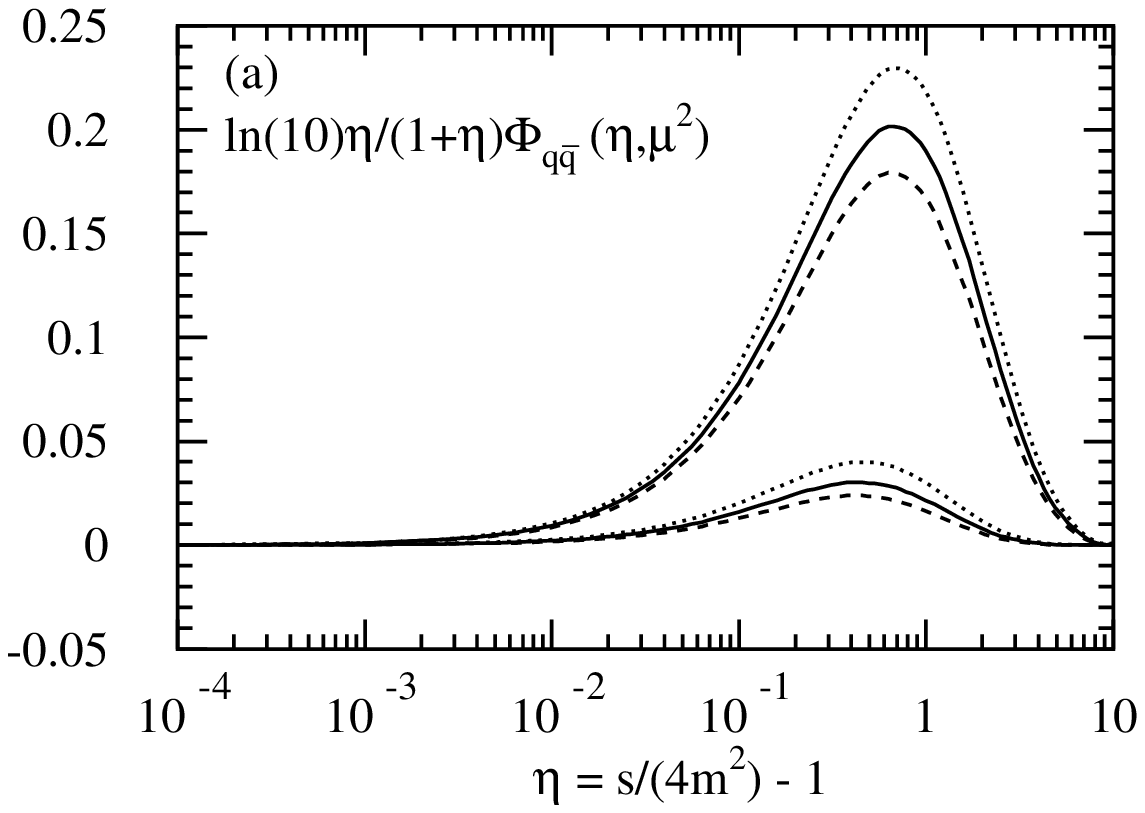,%
bbllx=50pt,bblly=110pt,bburx=285pt,bbury=450pt,angle=270,width=8.25cm}
\epsfig{file=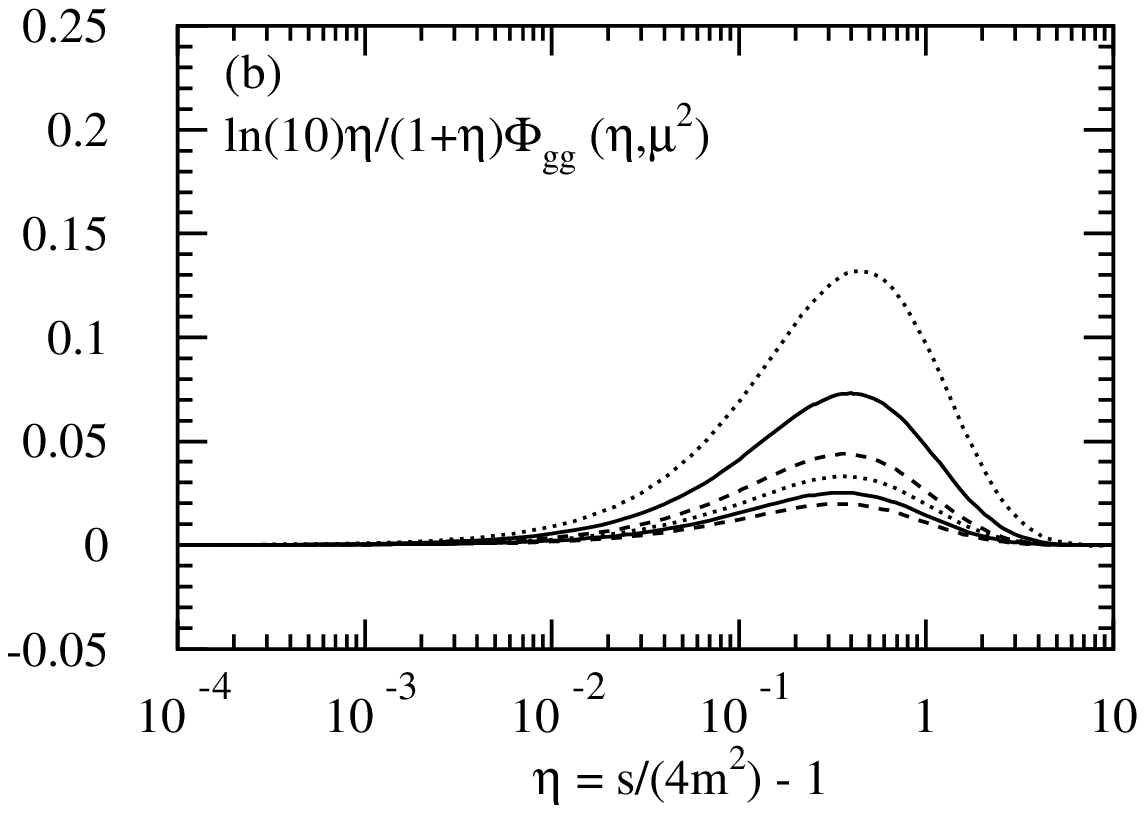,%
bbllx=50pt,bblly=110pt,bburx=285pt,bbury=450pt,angle=270,width=8.25cm}
\caption[dum]{\label{fluxxtevqq} {\small{
(a) The $q\Bar{q}$ parton flux factor $\ln(10)\eta/(1+\eta)\Phi_{q{\Bar{q}}}$
(see Eq.~(\ref{totalhadroncrs}))
for the CTEQ5M parametrization at the 
Tevatron (upper three curves, $\sqrt{S}=1.8$ TeV and $m=175$ GeV) and for
HERA-B (lower three curves $\sqrt{S}=41.6$ GeV and $m=4.75$ GeV).
We show results for 
$\mu=m$ (solid curves), 
$\mu=m/2$ (dotted curves), and 
$\mu=2m$ (dashed curves).
(b) Same as (a) for the $gg$ parton flux factor $\ln(10)\eta/(1+\eta)\Phi_{gg}$.
Now the upper set of curves correspond to HERA-B and the lower set to the Tevatron.
}}}
\end{center}
\end{figure}
Except where specified otherwise we have multiplied 
the non-exact scaling functions at NLO and NNLO with a damping factor 
$1/\sqrt{1+\eta}$, as in Ref.~\cite{Meng:1990rp}
in order to lessen the influence of the 
large $\eta$ region of the scaling functions
where threshold logarithms become less dominant
and we lose some theoretical control\footnote{In 1PI kinematics 
such a factor is effectively equivalent to including
$\theta(s_4-m^2)$ in Eq.~(\ref{eq:11}), see Ref.~\cite{Laenen:1998kp}.}.
Figure~\ref{plot-1gfudge} demonstrates that this 
factor indeed damps the large $\eta$ region of 
the NNLO scaling functions $f^{(2,0)}_{ij}$, while leaving
the small and medium $\eta$ regions unaffected. 
The effect of the damping factor will also be made explicit in the tables.
\begin{figure}[htp]
\begin{center}
\epsfig{file=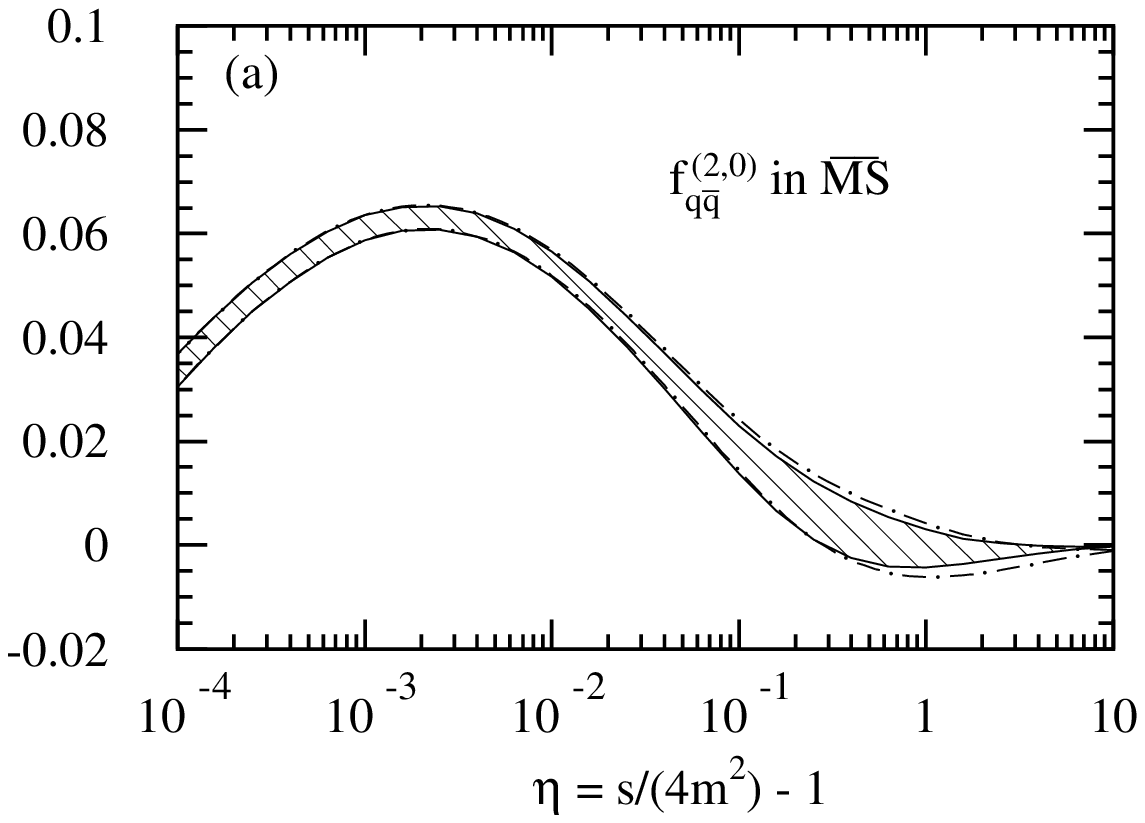,%
bbllx=50pt,bblly=110pt,bburx=285pt,bbury=450pt,angle=270,width=8.25cm}
\epsfig{file=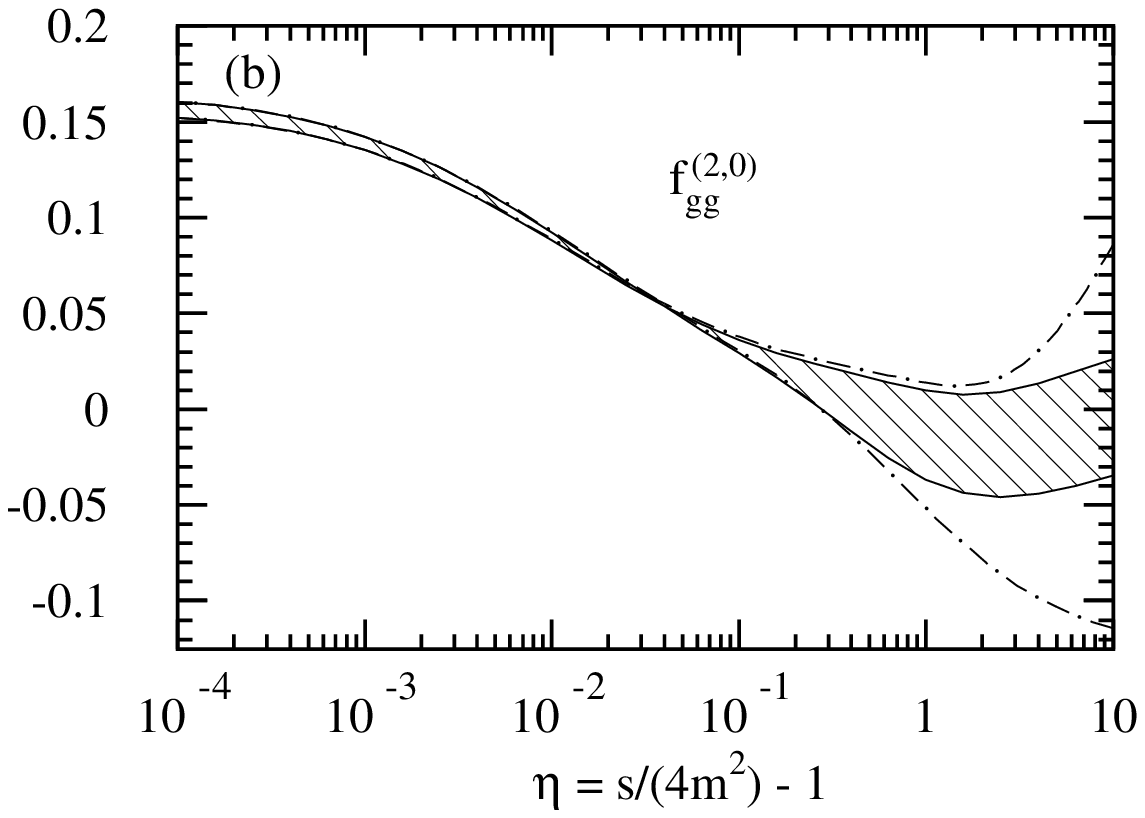,%
bbllx=50pt,bblly=110pt,bburx=285pt,bbury=450pt,angle=270,width=8.25cm}
\caption[dum]{\label{plot-1gfudge} {\small{
(a) The $\h$-dependence of the scaling functions
$f^{(2,0)}_{q{\Bar{q}}}(\h)$ in the ${\Bar{\rm{MS}}}$ scheme
with and without the damping factor $1/\sqrt{1+\eta}$.
We show the NNLL approximation to $f^{(2,0)}_{q{\Bar{q}}}$
in 1PI/PIM kinematics without the factor $1/\sqrt{1+\eta}$
(upper/lower dashed-dotted lines) and with the factor $1/\sqrt{1+\eta}$
(upper/lower boundary of shaded region).
The shaded region indicates the residual uncertainty for
$f^{(2,0)}_{q{\Bar{q}}}$ due to a particular kinematics choice.
(b) The same as (a) for $f^{(2,0)}_{gg}(\h)$.
}}}
\end{center}
\end{figure}

\subsection{Results for $t\bar{t}$ production at the Tevatron}

We first discuss top quark production at the Tevatron in proton-antiproton collisions. 
(For a review of the data see Refs.~\cite{Campagnari:1997ai, BHAT:1998CD}.)
We give results for both Run I ($\sqrt{S}=1.8$ TeV) and Run II ($\sqrt{S}=2.0$ TeV).
At the Tevatron, top quarks are mainly produced in pairs.
With a top quark mass of $175$ GeV, 
the dominant production channel is $q\bar q$ annihilation,
constituting about 90\% of the total cross section 
at the Born level for $\sqrt{S}=1.8$ TeV, with the 
$gg$ channel making up the remainder.
The NLO corrections in the $q\bar{q}$ channel are moderate,
of the order of 20\%, whereas those in the $gg$ channel are more than 80\%
so that the $gg$ channel 
gains significance at NLO\footnote{Recall that we also use
NLO parton distribution function for LO cross sections.}. 
These large corrections originate predominantly
from the threshold region. Figure~\ref{fluxxtevqq} shows 
that the range $0.1 \;\ltap \; \eta \; \ltap\; 2$ contributes the most.
At still higher orders, this trend
continues: the relative corrections to the $gg$ channel are larger
than those for the $q {\bar q}$ channel, as we shall see.
This is due to the larger color factors in the analytical expressions
for the corrections for the $gg$ channel.
As we mentioned earlier, the $qg$ and $\bar{q}g$ channels give 
negligible contributions to the total cross section and are not considered 
here.

We begin by comparing $\overline{\rm MS}$ results 
in 1PI and PIM kinematics,
including only the $q\bar{q}$ channel in Eq.~(\ref{totalhadroncrs}),
at $\sqrt{S}=1.8$ TeV.
In Fig.~\ref{hadro-m1} we show the Born cross section and 
the exact and approximate NLO corrections, the latter 
at both NLL and NNLL accuracy, for $\mu=m$
in the range $150 < m < 200$ GeV.
We see that our NLO 1PI approximations are
a little larger than the exact result while the NLO-NNLL
approximation in PIM kinematics
is indistinguishable from the exact answer.
The corrections are about 20--30\% in the
mass range shown.
In Fig.~\ref{hadro-m2} we give the 
equivalent results for the $gg$ channel.
Here the 1PI approximations agree with the exact result better 
than the PIM ones. 

Figure~\ref{hadro-m3}(a) displays the approximate $q {\bar q}$ 
NNLO corrections at NLL and NNLL accuracy for $\mu=m$
as a function of the top quark mass
in a direct comparison of the 1PI and PIM results.
The shaded area 
indicates the kinematics ambiguity at NNLL, about
0.6 pb at $m=175$ GeV. The figure shows that the
NNLL ambiguity is larger than the NLL one.
(Thus the small size of the NLL kinematics ambiguity seems
somewhat accidental.) Figure~\ref{hadro-m3}(b) shows 
similar results for the $gg$ channel.
\begin{figure}[htp]
\begin{center}
\epsfig{file=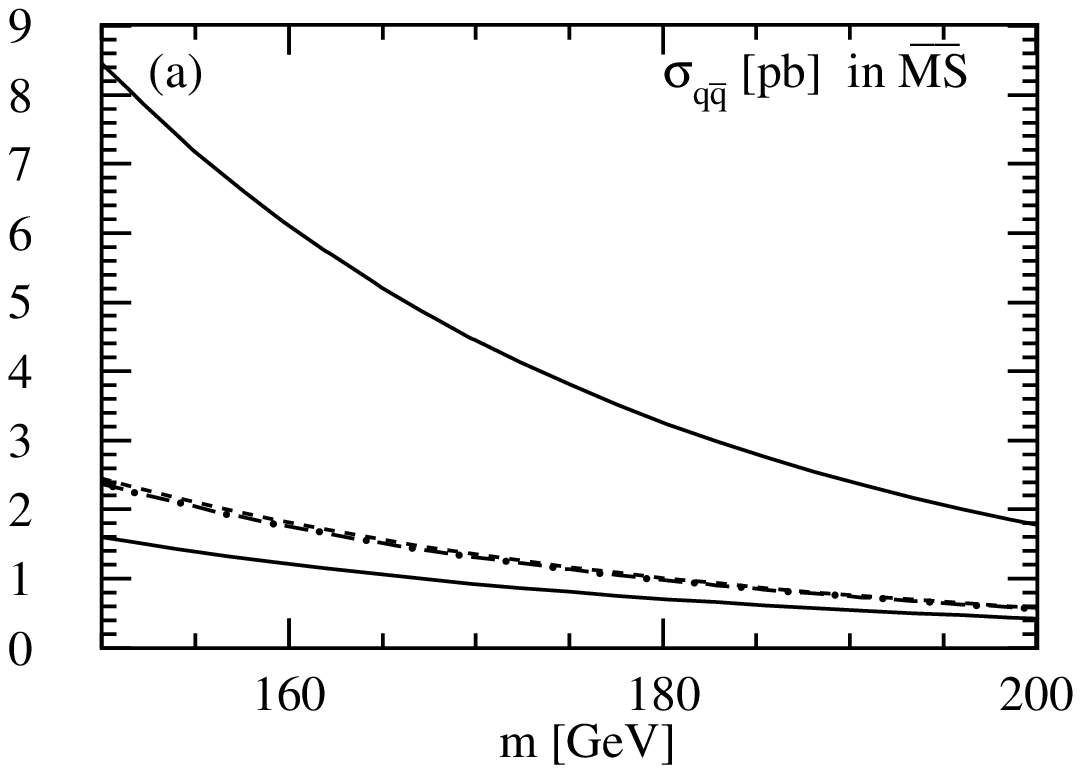,%
bbllx=50pt,bblly=110pt,bburx=285pt,bbury=450pt,angle=270,width=8.25cm}
\epsfig{file=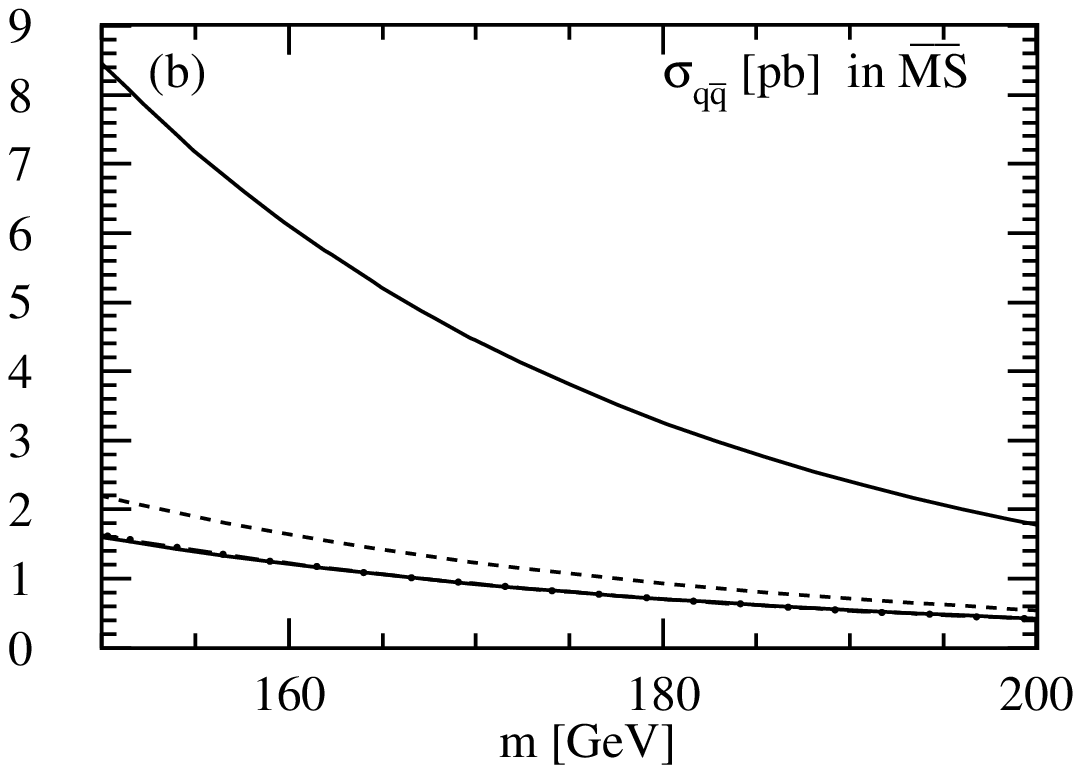,%
bbllx=50pt,bblly=110pt,bburx=285pt,bbury=450pt,angle=270,width=8.25cm}
\caption[dum]{\label{hadro-m1} {\small{
(a) The $q \bar{q}$ channel contribution to the top quark cross section 
at the Tevatron with $\sqrt{S}=1.8$ TeV and $\mu = m$ 
in the $\overline{\rm MS}$ scheme. 
We show the Born term (upper solid line),
the exact NLO corrections (lower solid line) and 
the 1PI approximate NLL (dashed line) and NNLL (dashed-dotted line)
one-loop corrections.
(b) The same as (a) in PIM kinematics.
}}}
\end{center}
\end{figure}
\begin{figure}[htp]
\begin{center}
\epsfig{file=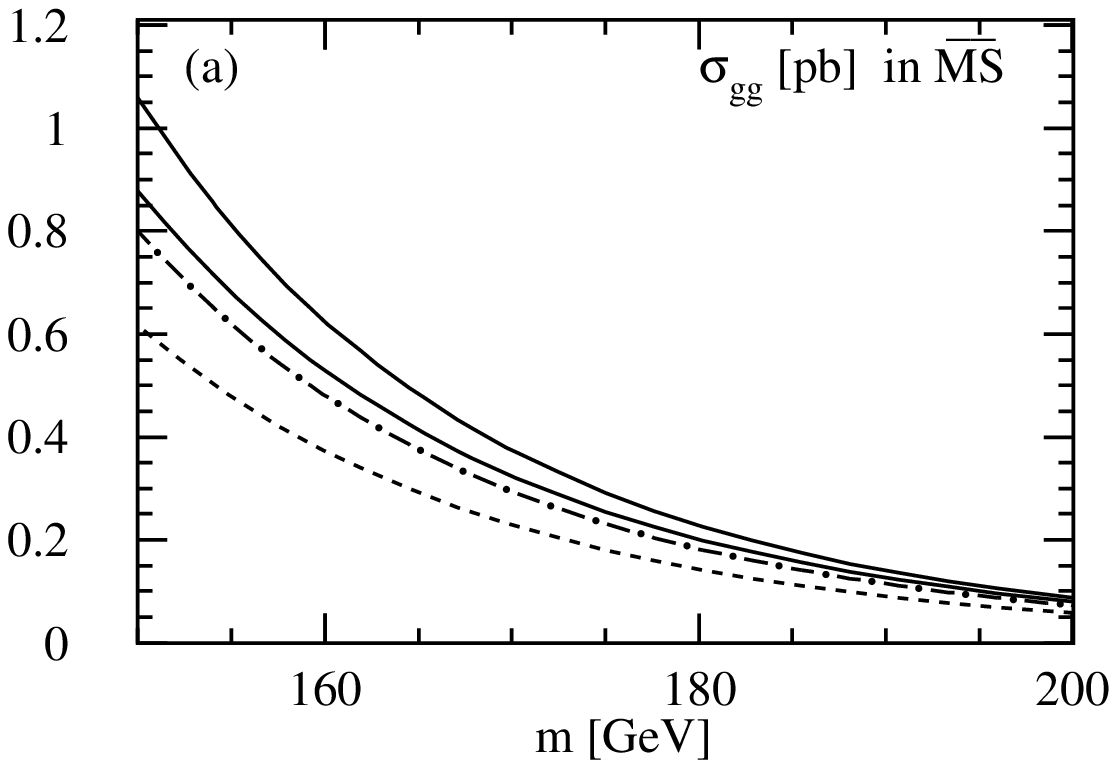,%
bbllx=50pt,bblly=110pt,bburx=285pt,bbury=450pt,angle=270,width=8.25cm}
\epsfig{file=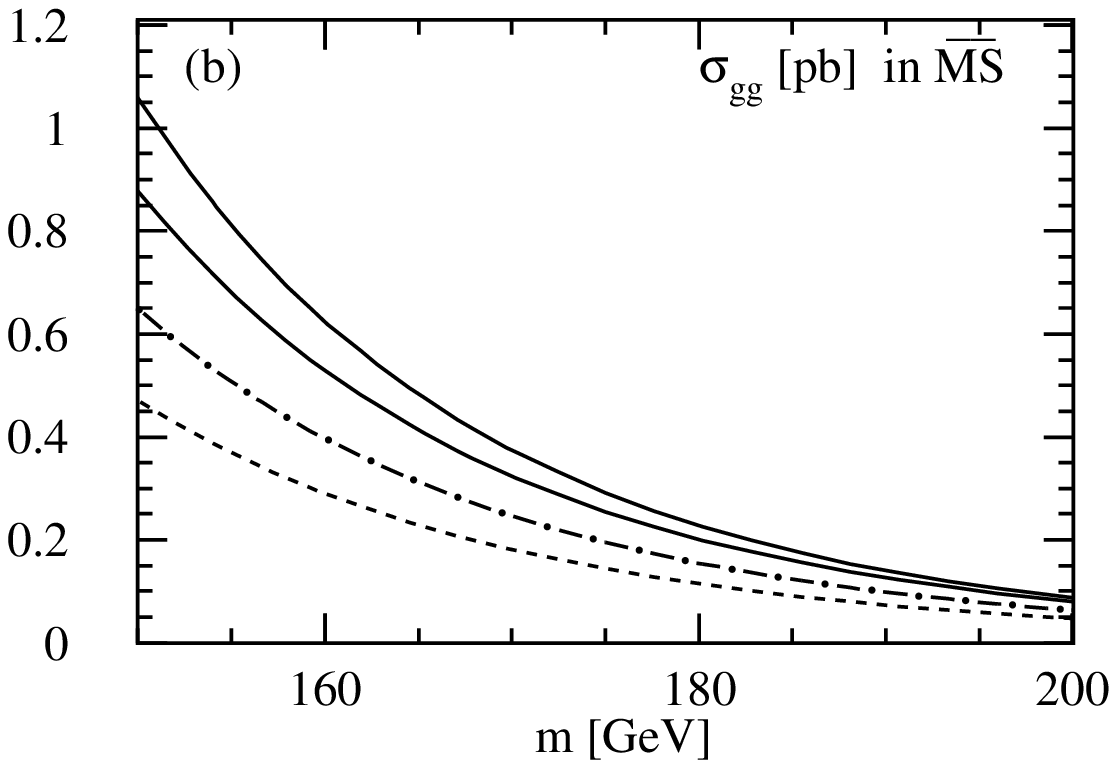,%
bbllx=50pt,bblly=110pt,bburx=285pt,bbury=450pt,angle=270,width=8.25cm}
\caption[dum]{\label{hadro-m2} {\small{
(a) The $gg$ channel contribution to the top quark cross section 
at the Tevatron with $\sqrt{S}=1.8$ TeV and $\mu = m$ 
in the $\overline{\rm MS}$ scheme. 
We show the Born term (upper solid line),
the exact NLO corrections (lower solid line) and 
the 1PI approximate NLL (dashed line) and NNLL (dashed-dotted line)
one-loop corrections.
(b) The same as (a) in PIM kinematics.
}}}
\end{center}
\end{figure}
\begin{figure}[htp]
\begin{center}
\epsfig{file=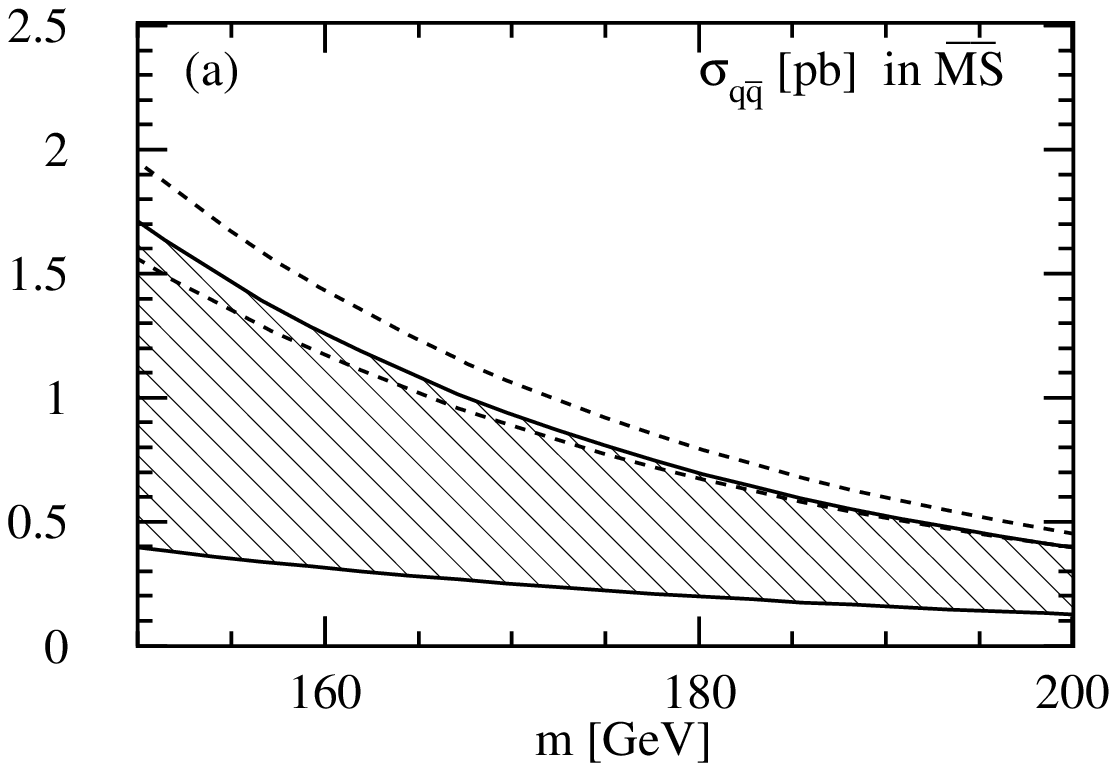,%
bbllx=50pt,bblly=110pt,bburx=285pt,bbury=450pt,angle=270,width=8.25cm}
\epsfig{file=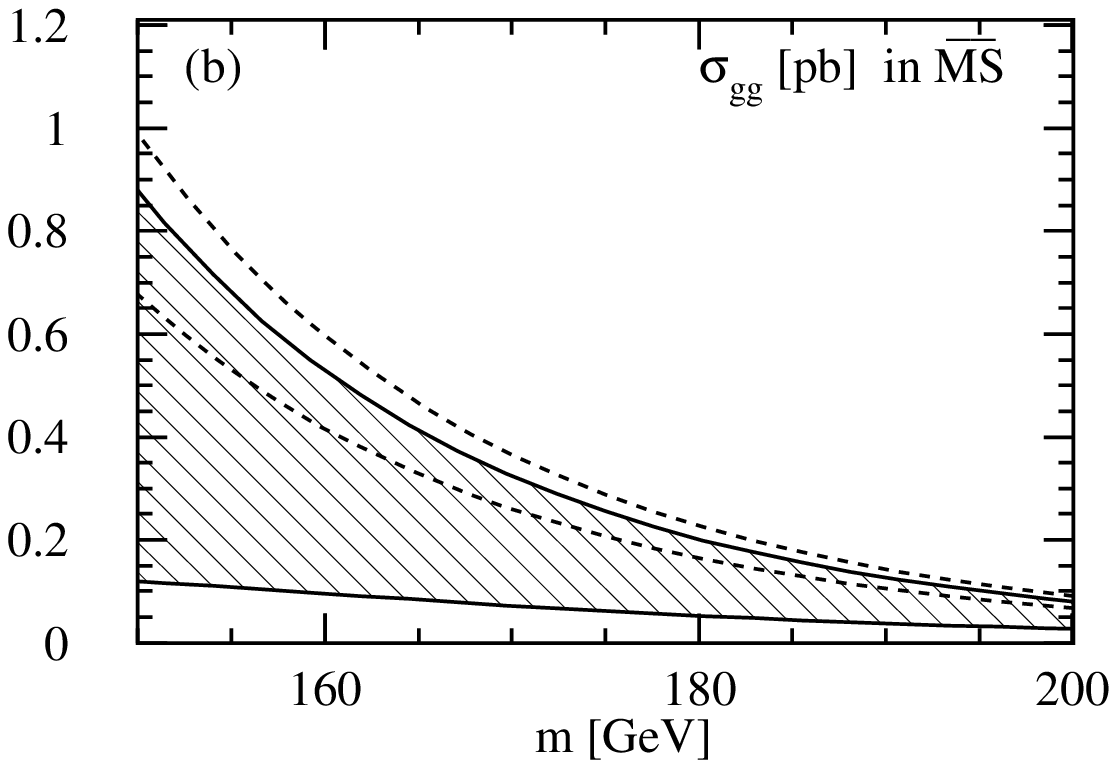,%
bbllx=50pt,bblly=110pt,bburx=285pt,bbury=450pt,angle=270,width=8.25cm}
\caption[dum]{\label{hadro-m3} {\small{
(a) The $q \bar{q}$ channel two-loop corrections to the top quark cross section 
at the Tevatron with $\sqrt{S}=1.8$ TeV and $\mu = m$ 
in the $\overline{\rm MS}$ scheme.
We show the 1PI approximate NLL (upper dashed line) and NNLL 
(upper boundary of shaded region) 
two-loop corrections 
and the PIM approximate NLL (lower dashed line) and NNLL 
(lower boundary of shaded region)
two-loop corrections.
(b) The same as (a) for the $gg$ channel two-loop corrections.
}}}
\end{center}
\end{figure}

For completeness, we show 
the corresponding results for the $q \bar{q}$ channel in the DIS scheme
in Figs.~\ref{hadro-m4} and \ref{hadro-m5}. 
Again PIM kinematics approximates the exact results somewhat better
than 1PI kinematics at NLO-NNLL.
We see that the NNLL-NNLO kinematics ambiguity in Fig.~\ref{hadro-m5},
again indicated by the shaded region, is greatly reduced compared 
to the $\overline{\rm MS}$ case, Fig.~\ref{hadro-m3}(a). 
It should however be kept in mind
that, in the DIS scheme, the parton densities absorb
large threshold logarithms
which are not properly accounted for at NNLO 
if one uses NLO  parton distributions as in Fig.~\ref{hadro-m5}. 
Therefore it seems likely that the DIS scheme kinematics ambiguity
is somewhat underestimated in Fig.~\ref{hadro-m5}.
\begin{figure}[htp]
\begin{center}
\epsfig{file=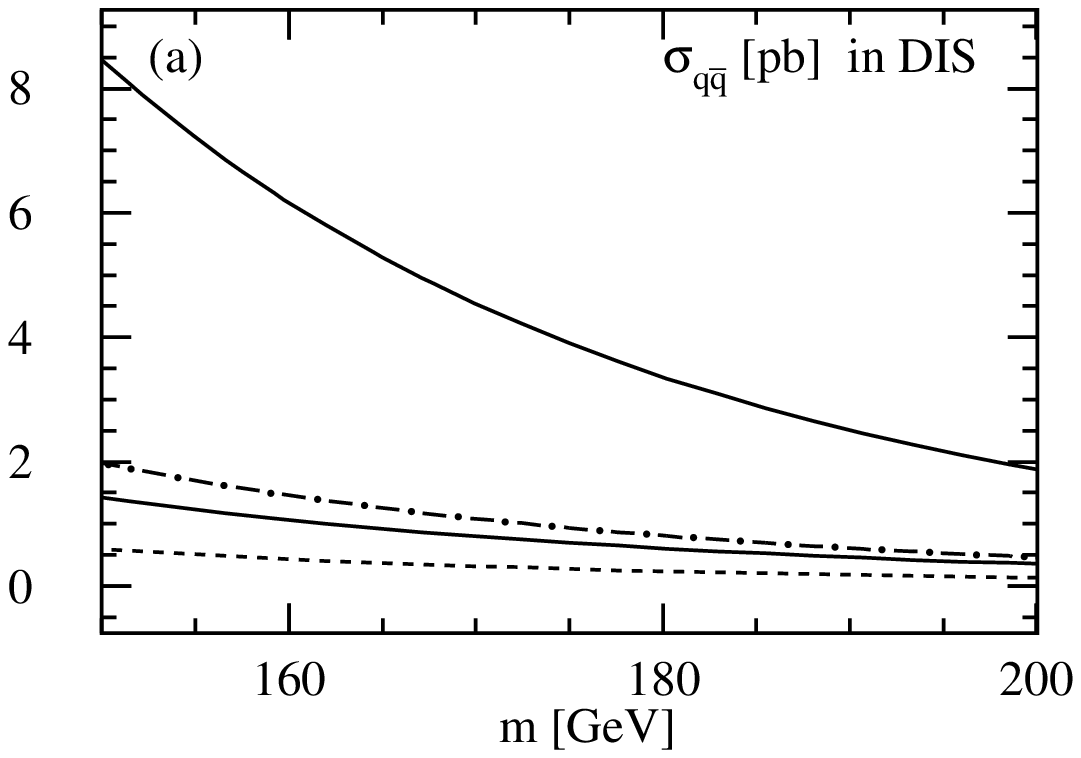,%
bbllx=50pt,bblly=110pt,bburx=285pt,bbury=450pt,angle=270,width=8.25cm}
\epsfig{file=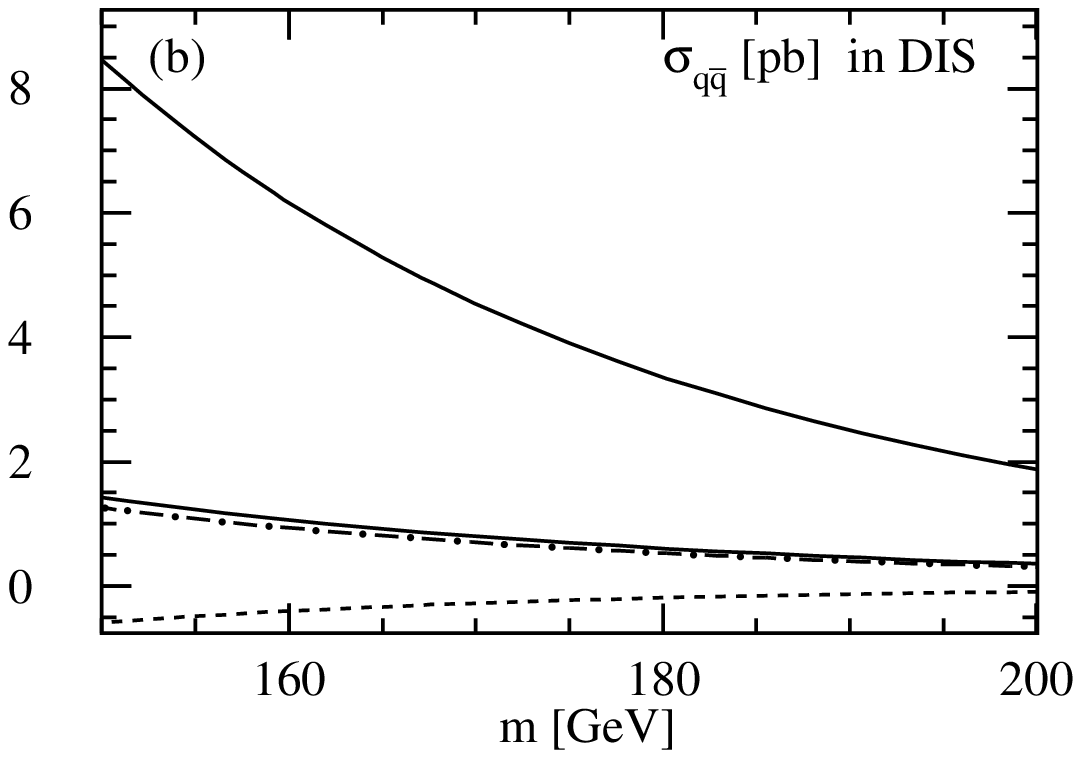,%
bbllx=50pt,bblly=110pt,bburx=285pt,bbury=450pt,angle=270,width=8.25cm}
\caption[dum]{\label{hadro-m4} {\small{
(a) The $q \bar{q}$ channel contribution to the top quark cross section 
at the Tevatron with $\sqrt{S}=1.8$ TeV and $\mu = m$ 
in the DIS scheme. 
We show the Born term (upper solid line),
the exact NLO corrections (lower solid line) and 
the 1PI approximate NLL (dashed line) and NNLL (dashed-dotted line)
one-loop corrections.
(b) The same as (a) in PIM kinematics.
}}}
\end{center}
\end{figure}
\begin{figure}[htp]
\begin{center}
\epsfig{file=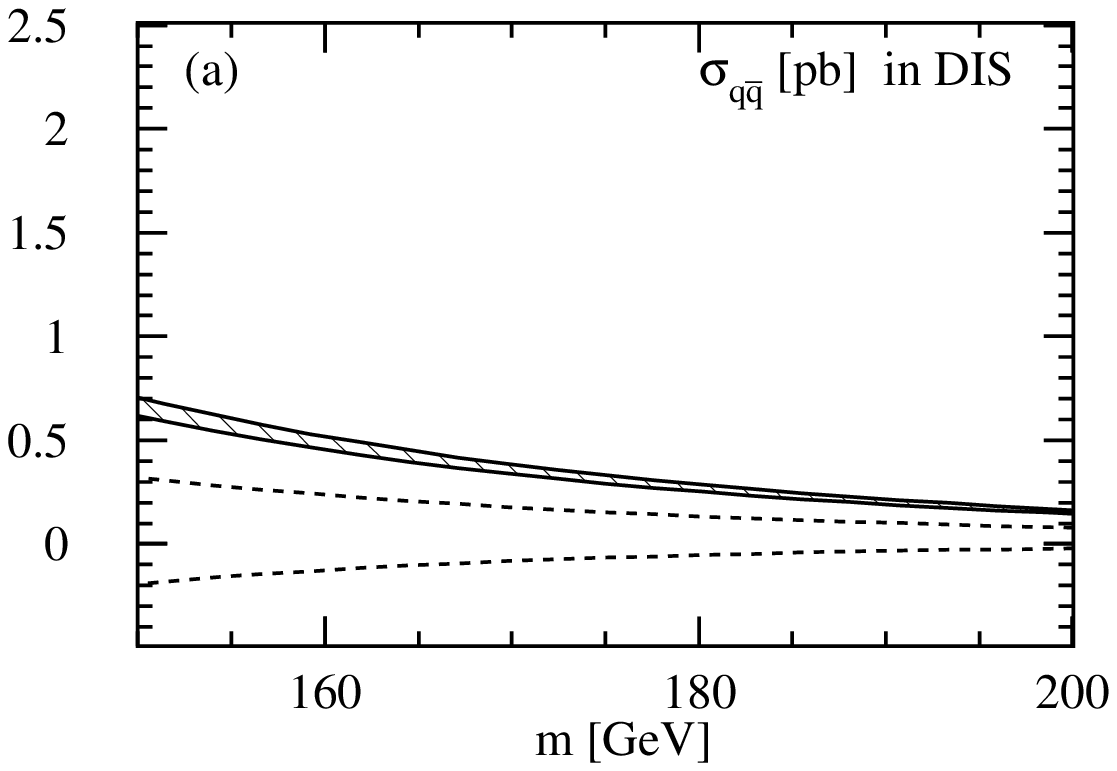,%
bbllx=50pt,bblly=110pt,bburx=285pt,bbury=450pt,angle=270,width=8.25cm}
\caption[dum]{\label{hadro-m5} {\small{
The $q \bar{q}$ channel contribution to the top quark cross section 
at the Tevatron with $\sqrt{S}=1.8$ TeV and $\mu = m$ 
in the DIS scheme. 
We show the 1PI approximate NLL (upper dashed line) and NNLL (lower boundary of
shaded region) 
two-loop corrections 
and the PIM approximate NLL (lower dashed line) and NNLL (upper boundary of
shaded region) two-loop corrections.
}}}
\end{center}
\end{figure}

In Fig.~\ref{hadro-m6} the sum of the $q\bar{q}$ and $gg$ channels in
the $\overline{\rm MS}$ scheme is shown as a function of the top quark mass.
We display the exact NLO cross section and the approximate NNLO 
cross section, which is 
the sum of the exact NLO cross section and the NNLL-NNLO corrections.
We show results for $\mu=m/2,m$, and $2m$. 
We see that the NNLO cross section is uniformly larger than the 
exact NLO one, although less so in PIM kinematics, and that the scale 
dependence of the NNLO cross section is 
considerably reduced relative to NLO. Comparing Figs.~\ref{hadro-m6}(a) 
and (b) we observe that the kinematics ambiguity is larger than the
scale uncertainty.
\begin{figure}[htp]
\begin{center}
\epsfig{file=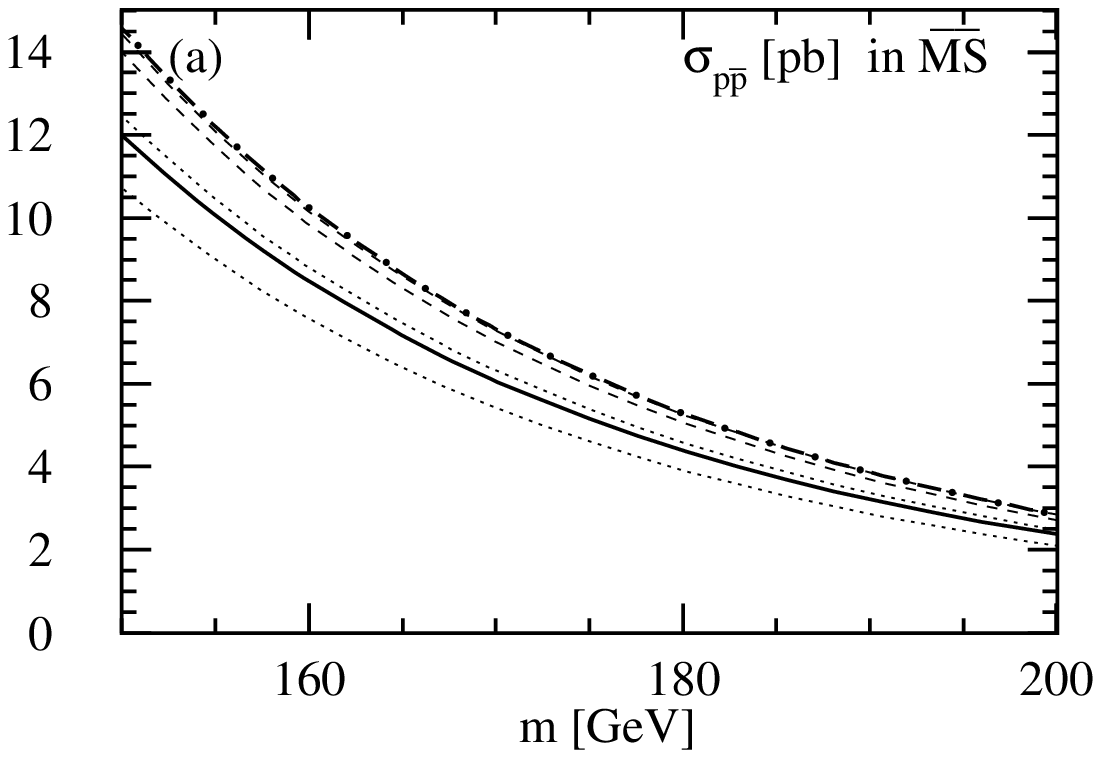,%
bbllx=50pt,bblly=110pt,bburx=285pt,bbury=450pt,angle=270,width=8.25cm}
\epsfig{file=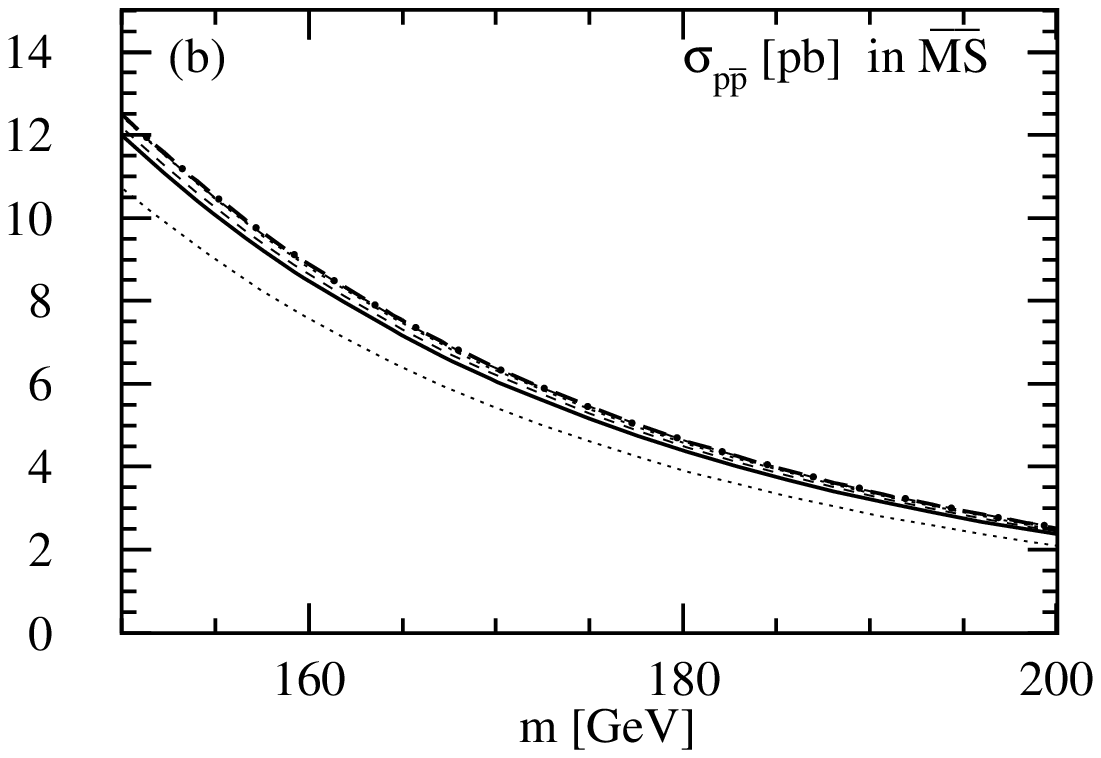,%
bbllx=50pt,bblly=110pt,bburx=285pt,bbury=450pt,angle=270,width=8.25cm}
\caption[dum]{\label{hadro-m6} {\small{
(a) The top quark cross section at the Tevatron with $\sqrt{S}=1.8$ TeV for the sum of 
the $q \bar{q}$ and $gg$ channels in the $\overline{\rm MS}$ scheme.
We show the exact NLO cross section 
for $\mu=m$ (solid line), $m/2$ (upper dotted line), and $2m$ 
(lower dotted line), and the 1PI approximate NNLL-NNLO cross section
for $\mu=m$ (dashed-dotted line), $m/2$ (upper dashed line), and $2m$ 
(lower dashed line).
(b) The same as (a) in PIM kinematics.
}}}
\end{center}
\end{figure}

We now turn to a more detailed study of the scale dependence of the inclusive
$\overline{\rm MS}$ top cross section.
We show the sum of the $q\bar{q}$ and $gg$ channels 
in Fig.~\ref{hadro-sc1} at $\sqrt{S}=1.8$ TeV and $m=$175 GeV at
several orders and to different accuracies. The Born and NLO results shown
are exact. One of the NNLO curves is constructed
by adding the contributions from the
NNLL approximate two-loop scaling functions $f^{(2,l)}_{ij},\,l=0,1,2$, 
to the exact NLO results, 
the other by adding instead the \textit{exact} $f^{(2,1)}_{ij}$ and $f^{(2,2)}_{ij}$ 
functions to the approximate $f^{(2,0)}_{ij}$, thus making
the \textit{changes} to the partonic cross section when changing $\mu$ exact.
The differences between these two NNLO curves are due 
to subleading terms and represent, for each kinematics, 
the corresponding ambiguity in the scale dependence.
At very small $\mu$ the contributions of the terms involving 
$f^{(2,1)}_{ij}$ and $f^{(2,2)}_{ij}$ are much larger that
the contribution from $f^{(2,0)}_{ij}$. The sizable difference
between the two NNLO curves in Fig.~\ref{hadro-sc1}(a) is in fact
mainly due to the difference between the exact and NNLL
1PI results for $f^{(2,1)}_{q\Bar{q}}$ at medium and large $\eta$
in Fig.~\ref{plot-4q}(a). Even if we include all soft plus
virtual terms in the approximate
1PI $f^{(2,1)}_{q\Bar{q}}$, as derived in appendix B,
there is still a sizeable difference 
from the exact result. Therefore this difference stems mainly from the hard,
i.e. ${\cal O}(1/N)$, terms in $f^{(2,1)}_{q\Bar{q}}$. 

The NNLO differences
in PIM kinematics are much smaller than in 1PI kinematics,
in correspondence with the good agreement of the exact
and NNLL PIM results for $f^{(2,1)}_{q\Bar{q}}$ 
over all relevant $\eta$, as shown in Fig.~\ref{hadro-sc1}(b).
\begin{figure}[htp]
\begin{center}
\epsfig{file=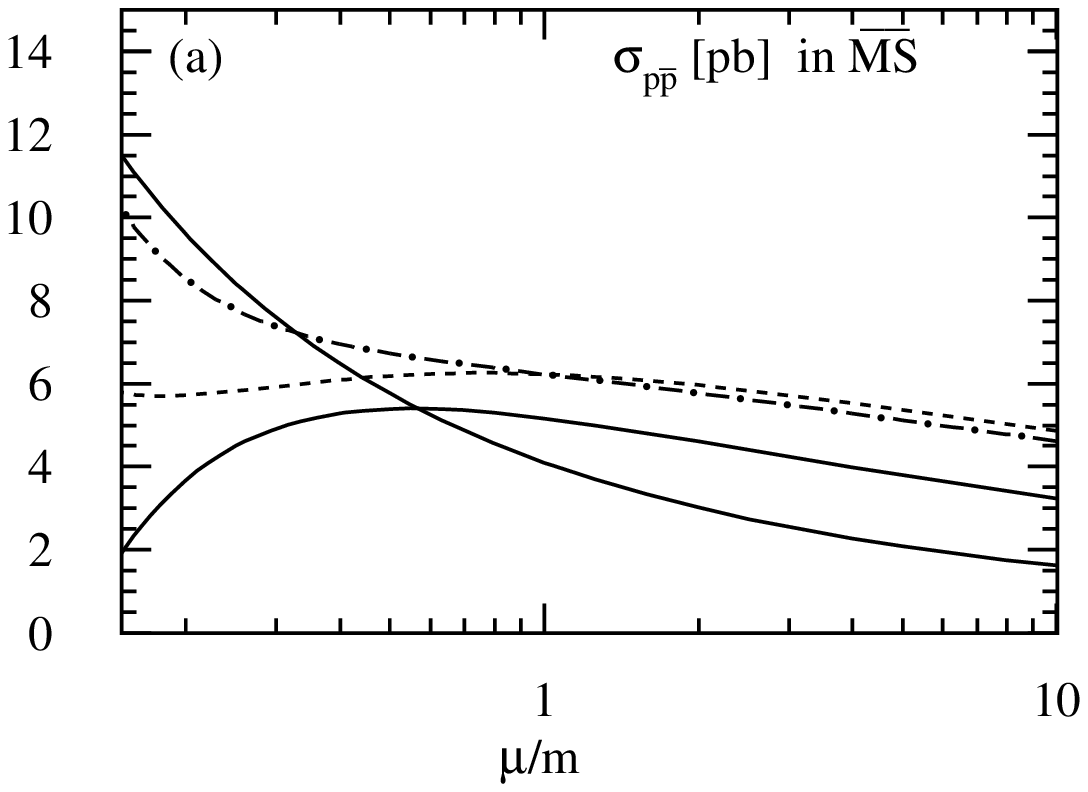,%
bbllx=50pt,bblly=110pt,bburx=285pt,bbury=450pt,angle=270,width=8.25cm}
\epsfig{file=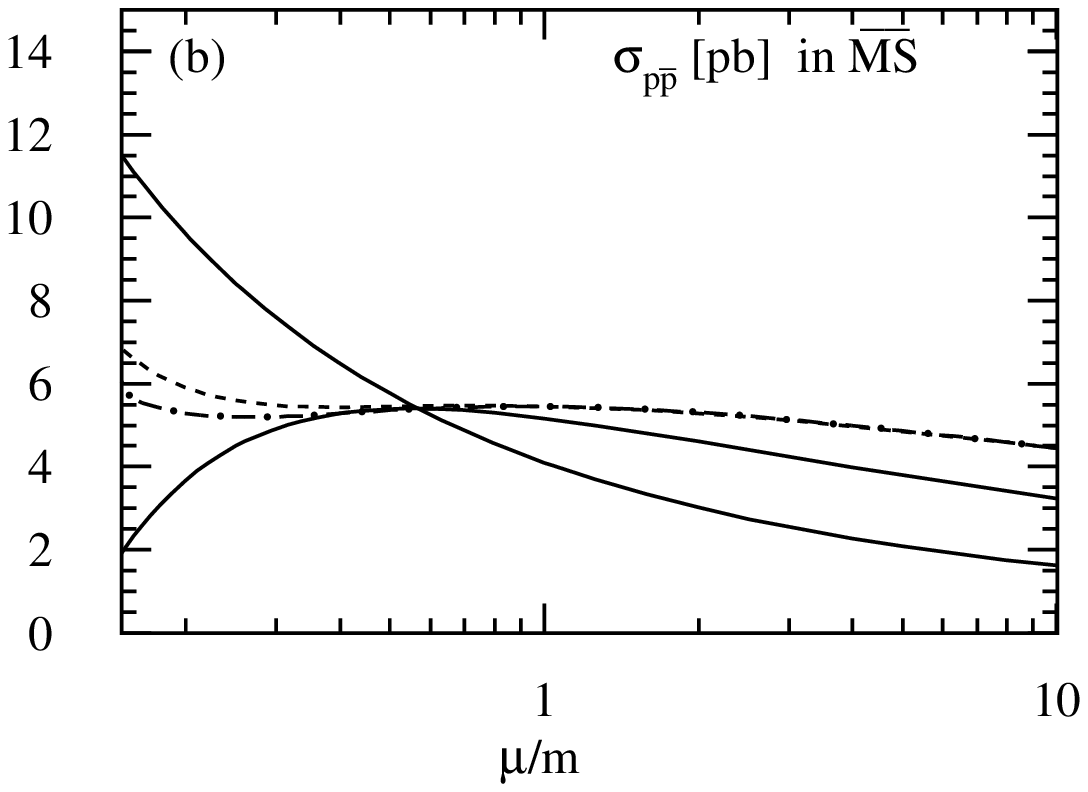,%
bbllx=50pt,bblly=110pt,bburx=285pt,bbury=450pt,angle=270,width=8.25cm}
\caption[dum]{\label{hadro-sc1} {\small{
(a) The $\mu$-dependence of the top quark cross section 
at the Tevatron with $\sqrt{S}=1.8$ TeV and $m=175$ GeV
for the sum of the $q \bar{q}$ and $gg$ channels in the 
$\overline{\rm MS}$ scheme.
We show the Born (upper solid line at small $\mu/m$) and the exact NLO 
(lower solid line at small $\mu/m$) cross sections, 
the 1PI approximate NNLL-NNLO cross section (dashed line)
and the NNLO estimate with only $f^{(2,0)}_{q{\Bar{q}}}$ and $f^{(2,0)}_{gg}$  
NNLL approximate 
(dashed-dotted line).
(b) The same as (a) in PIM kinematics.
}}}
\end{center}
\end{figure}

Next, in view of the upgrade in energy for the Tevatron
from $\sqrt{S}=1.8$  to 2.0 TeV, 
we investigate the $\sqrt{S}$ dependence of the top quark production
cross section.
In Fig.~\ref{hadro-scms1} we present the inclusive cross section for the sum 
of the $q \bar{q}$ and $gg$ channels in the $\overline{\rm MS}$ scheme 
at NLO and NNLO as a function 
of $\sqrt{S}$. 
The NNLO curves result from adding the NNLL NNLO corrections
to the exact NLO cross section.  We have normalized all calculations
to the value of the exact NLO cross section at $\m=m$.
Comparing 1PI and PIM kinematics we find that at lower energies, where 
the $q \bar q$ channel is dominant, the NNLO results 
are 10-30\% larger than at NLO in both kinematics.
As $\sqrt{S}$ increases, the $gg$ channel, with its larger corrections, grows
in importance.
The cross sections are also more sensitive to the large $\eta$ behavior
of the scaling functions.
This leads to large kinematics differences at energies above 5 TeV
although the scale uncertainties remain small.
In PIM kinematics this is due 
to $f^{(2,0)}_{gg}$ which is large and negative for $\eta > 0.1$
at NNLL, thus reducing the NNLO results relative to the 
exact NLO cross section.

\begin{figure}[htp]
\begin{center}
\epsfig{file=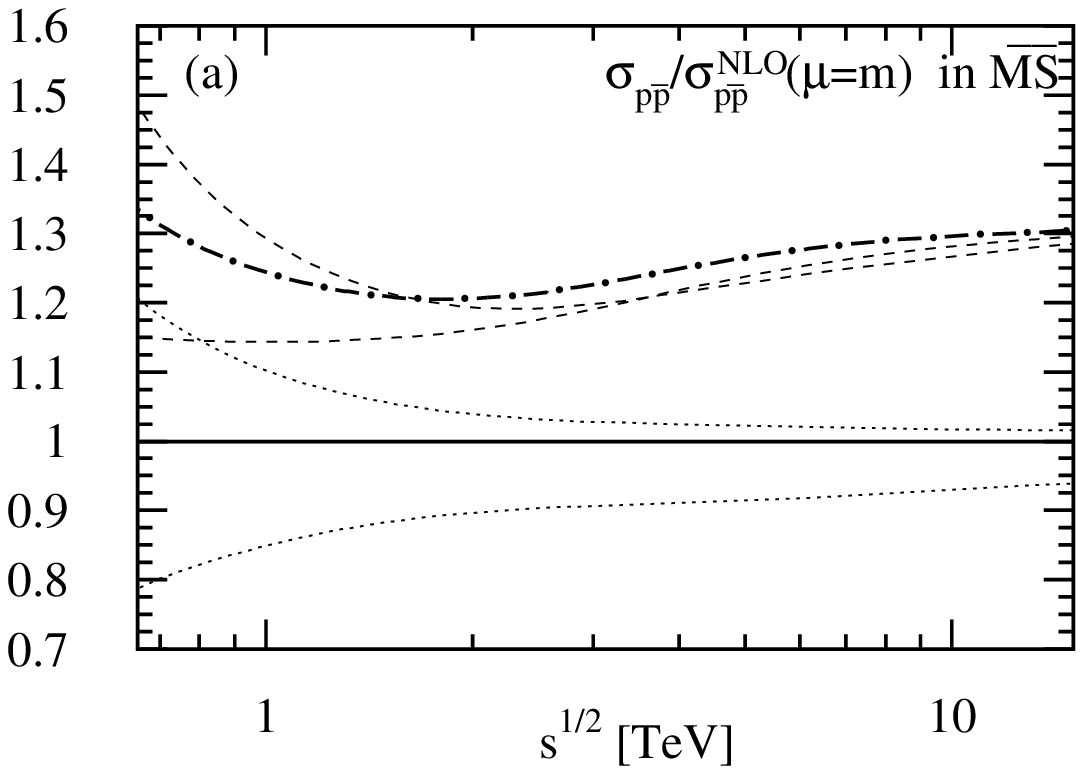,%
bbllx=50pt,bblly=110pt,bburx=285pt,bbury=450pt,angle=270,width=8.25cm}
\epsfig{file=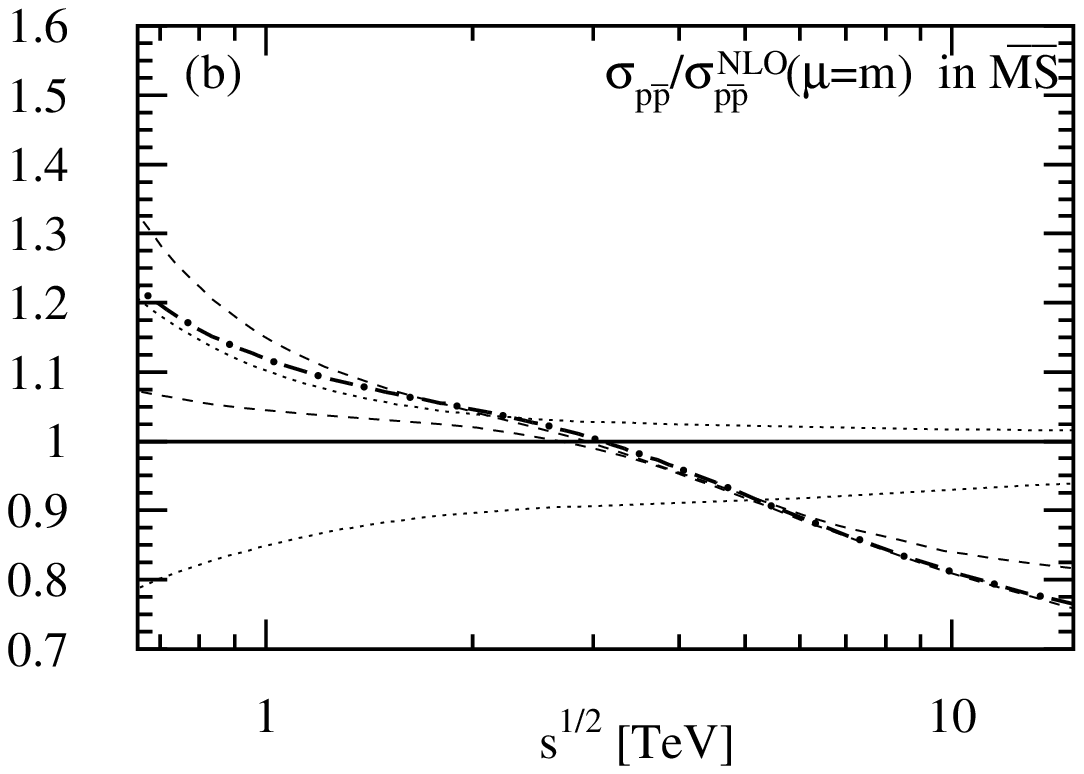,%
bbllx=50pt,bblly=110pt,bburx=285pt,bbury=450pt,angle=270,width=8.25cm}
\caption[dum]{\label{hadro-scms1} {\small{
(a) The top quark cross section with $m=175$ GeV
at $p {\bar p}$ colliders for the sum of 
the $q \bar{q}$ and $gg$ channels in the $\overline{\rm MS}$ scheme 
as a function of $\sqrt{S}$ normalized to the exact NLO cross section 
at $\m=m$.
We show the ratios of the exact NLO cross sections 
for $\mu=m$ (solid line), $m/2$ (upper dotted line), and $2m$ 
(lower dotted line), and the ratios of the 1PI approximate NNLL-NNLO 
cross sections 
for $\mu=m$ (dashed-dotted line), $m/2$ (upper dashed line), and $2m$ 
(lower dashed line).
(b) The same as (a) in PIM kinematics.
}}}
\end{center}
\end{figure}

We now present the values of the NLO and NNLO 
total $\overline{\rm MS}$ cross sections  
for top quark production at the Tevatron with 
$\sqrt{S}=1.8$ TeV and 2.0 TeV for $m=175$ GeV and  
$\mu=m$, $m/2$, and $2m$ in Tables~\ref{tab:1} and \ref{tab:2}.
They detail the effects of both multiplying
the approximate scaling functions $f^{(2,k)}_{ij},\, k=0,1,2$,
with the damping factor $1/\sqrt{1+\eta}$
and using the approximate or exact (Eqs.~(\ref{ex-f11qq})--(\ref{ex-f22gg}))
scaling functions $f^{(2,1)}_{ij}$ and $f^{(2,2)}_{ij}$
on the NNLO results. Note that all NNLO results 
in the figures correspond to the option AP+DF, 
except in Fig.~\ref{hadro-sc1} where we also show 
NNLO results corresponding to the EX+DF choice (dash-dotted
lines).
Comparing the NNLO and NLO cross sections in the
tables shows that the NNLO corrections are much smaller 
in PIM kinematics than in 1PI kinematics,
as already shown in Figs.~\ref{hadro-scms1}(a) and (b).
\begin{table}[htbp]
\begin{center}
\begin{tabular}{|c|c|c|c|} \hline
\multicolumn{4}{|c|}{$\sqrt{S}=1.8$ TeV} \\ \hline
   &$\mu=m/2$  &$\mu=m$ &$\mu=2m$ \\ \hline
NLO   & 5.39 & 5.16 & 4.61 \\ \hline
\multicolumn{4}{|c|}{NNLO (1PI)} \\ \hline
AP    & 6.27 & 6.32 & 6.12 \\ 
AP+DF & 6.19 & 6.22 & 5.97 \\ 
EX    & 6.91 & 6.32 & 5.84 \\ 
EX+DF & 6.73 & 6.22 & 5.78 \\ \hline
\multicolumn{4}{|c|}{NNLO (PIM)} \\ \hline
AP    & 5.37 & 5.35 & 5.26 \\ 
AP+DF & 5.45 & 5.45 & 5.29 \\ 
EX    & 5.18 & 5.35 & 5.26 \\ 
EX+DF & 5.36 & 5.45 & 5.31 \\ \hline
\end{tabular}
\caption[]{{\small{The hadronic $t \overline t$ production cross sections
in pb for $p \overline p$ collisions in the $\overline{\rm MS}$ scheme
with $\sqrt{S} = 1.8$ TeV and $m=175$ GeV, for $\mu=m, m/2$, and $2m$.
The labelling of the NNLO results corresponds to 
with or without the damping
factor (DF) $1/\sqrt{1+\eta}$, and using exact (EX) or NNLL approximate (AP)
scaling functions $f^{(2,1)}_{ij}$ and $f^{(2,2)}_{ij}$.}}}
\label{tab:1}
\end{center}
\end{table}                     
\begin{table}[htbp]
\begin{center}
\begin{tabular}{|c|c|c|c|} \hline
\multicolumn{4}{|c|}{$\sqrt{S}=2.0$ TeV} \\ \hline
   &$\mu=m/2$  &$\mu=m$ &$\mu=2m$ \\ \hline
 NLO  & 7.37 & 7.10 & 6.36 \\ \hline
\multicolumn{4}{|c|}{NNLO (1PI)} \\ \hline
AP    & 8.58 & 8.71 & 8.46 \\ 
AP+DF & 8.47 & 8.56 & 8.24 \\ 
EX    & 9.47 & 8.71 & 8.06 \\ 
EX+DF & 9.21 & 8.56 & 7.97 \\ \hline
\multicolumn{4}{|c|}{NNLO (PIM)} \\ \hline
AP    & 7.27 & 7.27 & 7.17 \\ 
AP+DF & 7.40 & 7.43 & 7.24 \\ 
EX    & 6.88 & 7.27 & 7.20 \\ 
EX+DF & 7.20 & 7.43 & 7.29 \\ \hline
\end{tabular}
\caption[]{{\small{The hadronic $t \overline t$ production cross sections
in pb for $p \overline p$ collisions in the $\overline{\rm MS}$ scheme
with $\sqrt{S} = 2.0$ TeV and $m=175$ GeV, for $\mu=m, m/2$, and $2m$.
The labelling of the NNLO results corresponds to with or without the damping
factor (DF) $1/\sqrt{1+\eta}$, and using exact (EX) or NNLL approximate (AP)
scaling functions $f^{(2,1)}_{ij}$ and $f^{(2,2)}_{ij}$.}}}
\label{tab:2}
\end{center}
\end{table}                     
We see that for the top cross section the
effect of the damping factor is rather small. 
We also note that the use of exact $f^{(2,1)}_{ij}$ and $f^{(2,2)}_{ij}$
affects the $\mu=m/2$ case the most. The difference between the exact
and approximate calculation is larger for 
1PI than PIM kinematics, as can also be observed in Fig.~\ref{hadro-sc1}.

We note that AP results in 1PI kinematics were also presented 
in Ref. \cite{Kidonakis:2000ui}. There are some small numerical 
differences with the results in this paper stemming mainly from using  
slightly different analytical expressions, all equivalent at threshold. 
To be specific, the expressions $t_1^2+u_1^2$ and $s^2-2t_1u_1$ are equivalent 
at threshold and either choice can be made in our NNLO expansions. 
Different choices produce slightly 
different numerical results away from threshold and the variation in  
these results represents a small but inherent uncertainty in the 
cross sections.  

Let us finally comment on 
the applicability of our results for top quark production 
at the LHC where $\sqrt{S}=14$ TeV. 
At this $pp$ collider the $gg$ channel is dominant
(about 90\% of the total) because
only sea quarks contribute to the antiquark distributions
in the $q \bar q$ channel.
Although top production at these energies is far from the hadronic threshold 
region, it might be close enough to partonic threshold 
for threshold resummation to be relevant since the
gluon flux may favor small values of 
$s=x_1 x_2 S \ll S$ (see the 
discussion in Refs.~\cite{Catani:2000jh,Catani:2000zg}).
Fig.~\ref{fluxlhc} indeed confirms this,
but also shows that the hadronic cross section
is sensitive to
the high-energy behaviour of the scaling functions.
Thus, estimates of the inclusive top cross section 
at the LHC based on the threshold approximation alone 
are unreliable.
\begin{figure}[htp]
\begin{center}
\epsfig{file=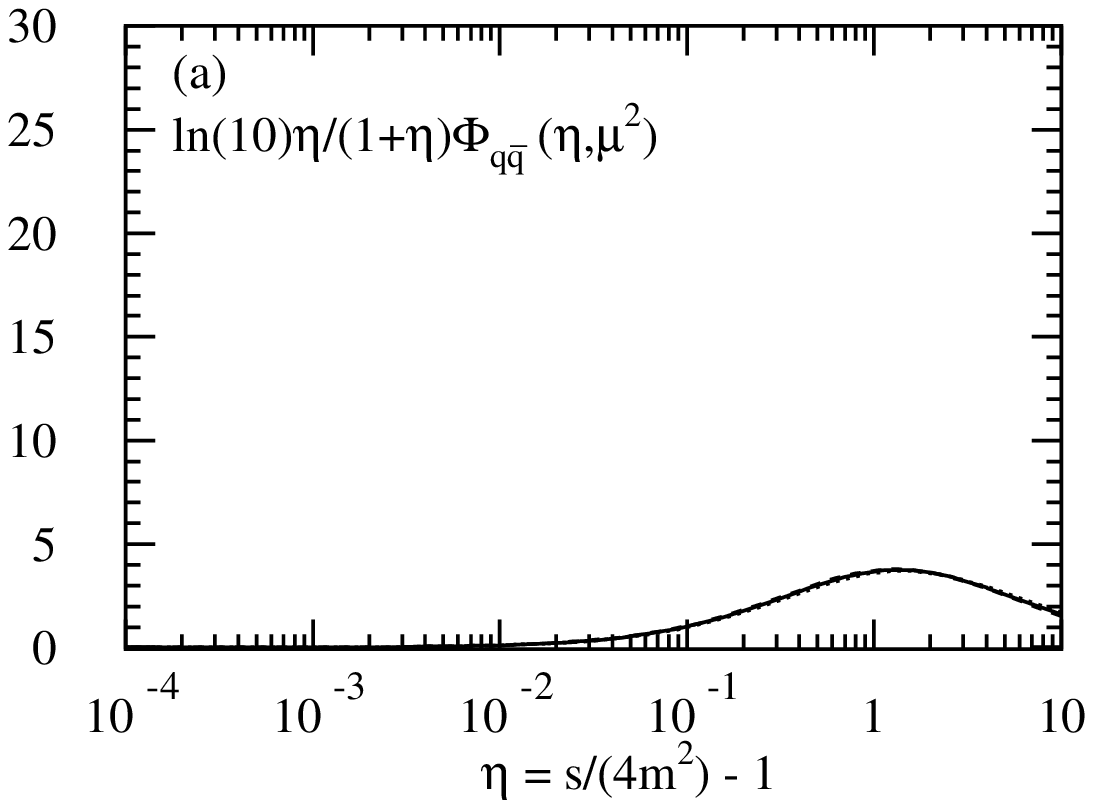,%
bbllx=50pt,bblly=110pt,bburx=285pt,bbury=450pt,angle=270,width=8.25cm}
\epsfig{file=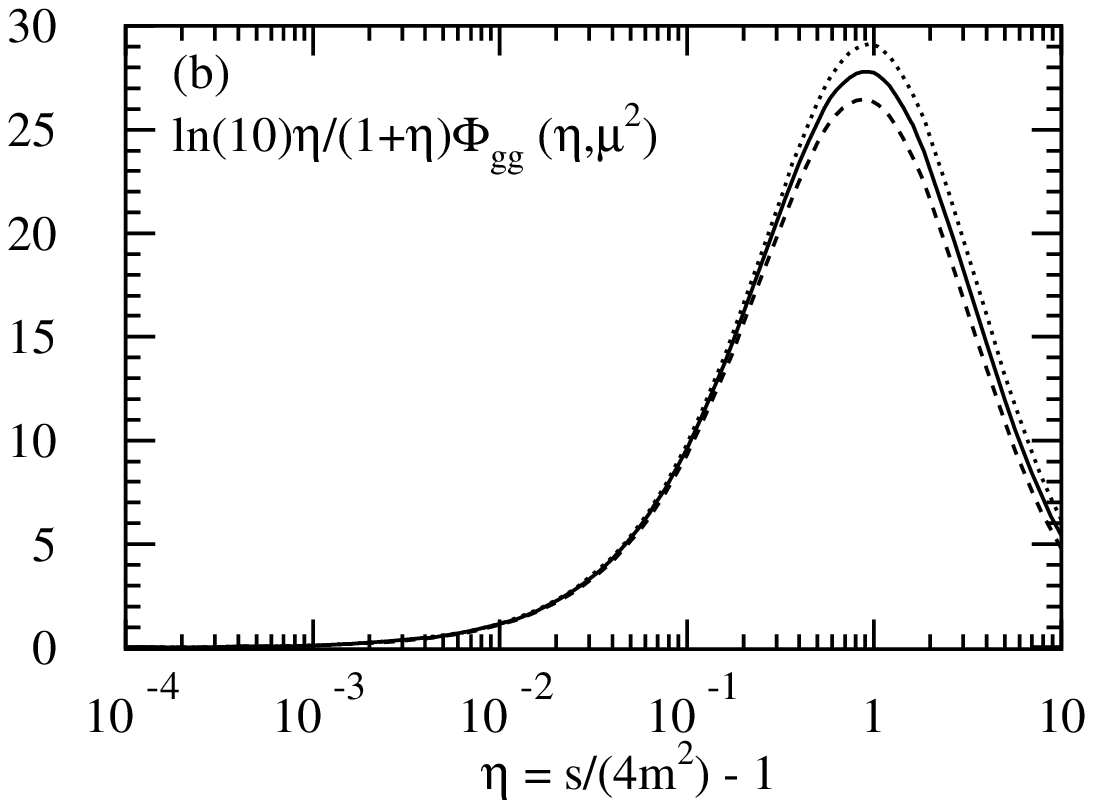,%
bbllx=50pt,bblly=110pt,bburx=285pt,bbury=450pt,angle=270,width=8.25cm}
\caption[dum]{\label{fluxlhc} {\small{
(a) The $q\Bar{q}$ parton flux factor, $\ln(10)\eta/(1+\eta)\Phi_{q{\Bar{q}}}$,
for the CTEQ5M parametrization at
the LHC ($\sqrt{S}=14$ TeV and $m=175$ GeV).
We show results for 
$\mu=m$ (solid curve), 
$\mu=m/2$ (dotted curve), and 
$\mu=2m$ (dashed curve).
(b) Same as (a) for the $gg$ parton flux factor, $\ln(10)\eta/(1+\eta)\Phi_{gg}$.
}}}
\end{center}
\end{figure}

\subsection{Results for $b\bar{b}$ production at HERA-B}

In this section we present the inclusive cross section
for bottom quark production at fixed-target $pp$ experiments,
in particular, the HERA-B experiment. The energy of the 
proton beam at HERA-B is 920 GeV so that $\sqrt{S}=41.6$ GeV.
Here the $gg$ channel is dominant (about 70\% of the total cross section),
with the $q \bar q$ channel contributing the remainder.
The $g q$ and $g \bar q$ channels are again negligible,
of the order of a few percent. Figure~\ref{fluxxtevqq} shows that 
the $\eta\;\ltap \; 1$ region is dominant 
in the convolution with the parton densities, Eq.~(\ref{totalhadroncrs}).

In Fig.~\ref{hadro-bbbar} we present our NLO and NNLL-NNLO results 
for the $b$ quark production cross section in fixed-target $pp$ interactions
as a function of beam energy in the range 200-1200 GeV$/c$
with $m=4.75$ GeV. 
Comparing 1PI and PIM kinematics we find that,
particularly at high energies, the NNLO predictions are different. 
Again, this can be attributed to increased sensitivity to the high-energy 
asymptotics of the scaling functions.
\begin{figure}[htp]
\begin{center}
\epsfig{file=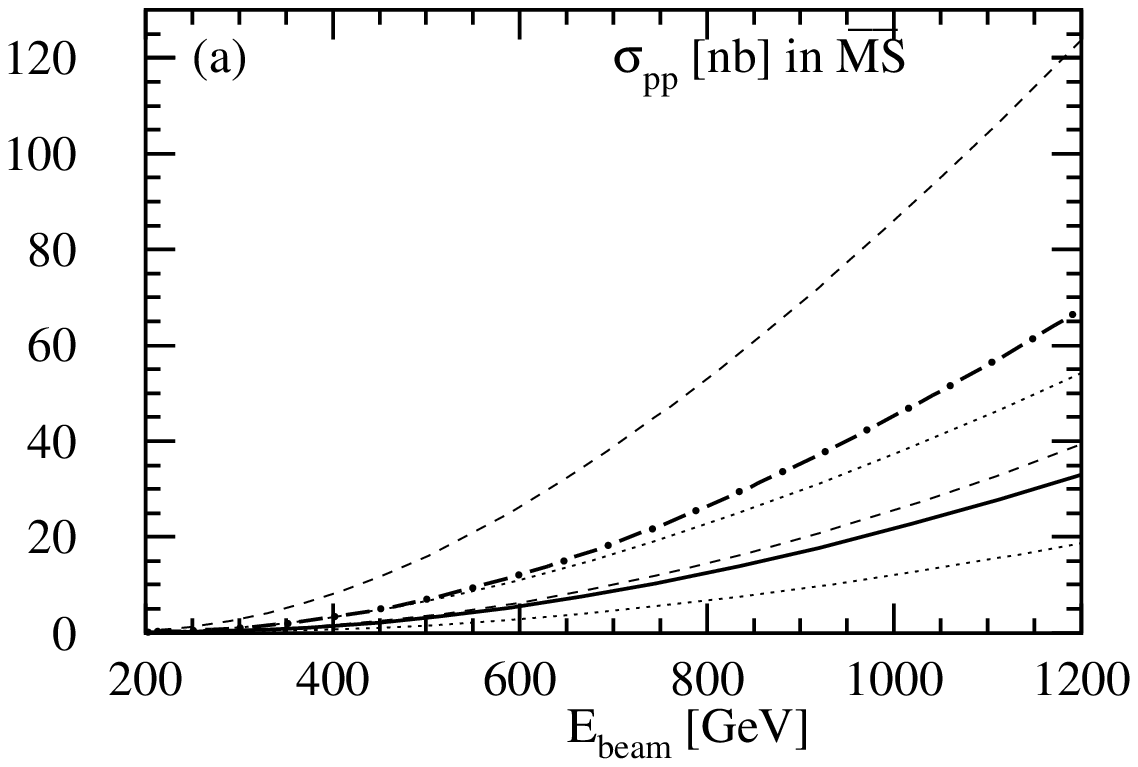,%
bbllx=50pt,bblly=110pt,bburx=285pt,bbury=450pt,angle=270,width=8.25cm}
\epsfig{file=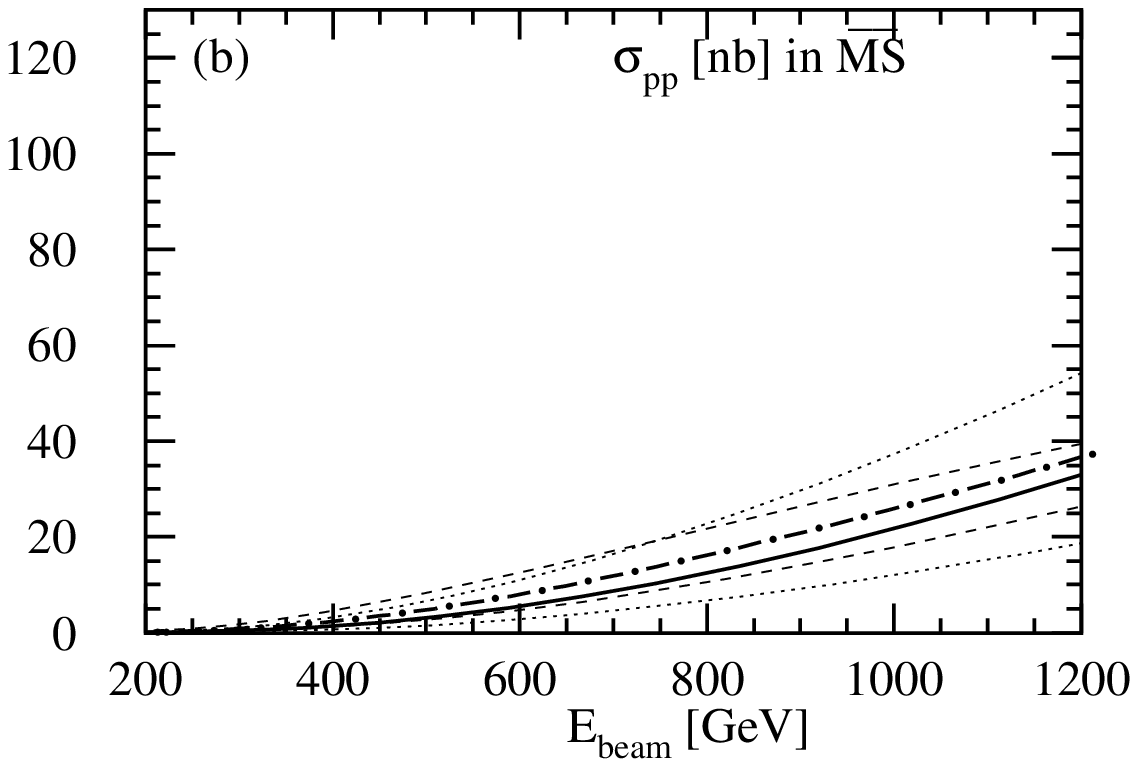,%
bbllx=50pt,bblly=110pt,bburx=285pt,bbury=450pt,angle=270,width=8.25cm}
\caption[dum]{\label{hadro-bbbar} {\small{
(a)  The total $\overline{\rm MS}$ bottom quark cross section 
at fixed-target $pp$ experiments with $m=4.75$ GeV.
We show the exact NLO cross section 
for $\mu=m$ (solid line), $m/2$ (upper dotted line), and $2m$ 
(lower dotted line), and the NNLL-NNLO cross section
for $\mu=m$ (dashed-dotted line), $m/2$ (upper dashed line), and $2m$ 
(lower dashed line).
(b) The same as (a) in PIM kinematics. 
}}}
\end{center}
\end{figure}
\begin{table}[htbp]
\begin{center}
\begin{tabular}{|c|c|c|c|} \hline
\multicolumn{4}{|c|}{$\sqrt{S}=41.6$ GeV} \\ \hline
   &$\mu=m/2$  &$\mu=m$ &$\mu=2m$ \\ \hline
NLO   & 31.1 & 17.8 &  9.7 \\ \hline
\multicolumn{4}{|c|}{NNLO (1PI)} \\ \hline
AP    & 77.1 & 39.1 & 21.9 \\ 
AP+DF & 72.2 & 37.2 & 20.8 \\ 
EX    & 91.1 & 39.1 & 20.8 \\ 
EX+DF & 80.7 & 37.2 & 20.3 \\ \hline
\multicolumn{4}{|c|}{NNLO (PIM)} \\ \hline
AP    & 18.1 & 19.0 & 14.0 \\ 
AP+DF & 27.3 & 21.8 & 14.7 \\ 
EX    &-19.8 & 19.0 & 15.6 \\ 
EX+DF & -2.5 & 21.8 & 16.2 \\ \hline
\end{tabular}
\caption[]{{\small{The hadronic $b \overline b$ production cross sections
in nb at HERA-B in the $\overline{\rm MS}$ scheme
with $\sqrt{S} = 41.6$ GeV and $m=4.75$ GeV for $\mu=m,m/2$, and $2m$.
The labelling of the NNLO results corresponds to 
with or without the damping
factor (DF) $1/\sqrt{1+\eta}$, and using exact (EX) or NNLL approximate (AP)
scaling functions $f^{(2,1)}_{ij}$ and $f^{(2,2)}_{ij}$.}}}
\label{tab:3}
\end{center}
\end{table}                     
In Table~\ref{tab:3} we list the NLO and
NNLO inclusive $b$ quark production cross sections.
We show the effects of both using the damping factor $1/\sqrt{1+\eta}$
and the approximate or exact (Eqs.~(\ref{ex-f11qq})--(\ref{ex-f22gg}))
scaling functions $f^{(2,1)}_{ij}$ and $f^{(2,2)}_{ij}$
on the NNLO results.

We observe that the damping factor has a significant effect
at small scales. The use of exact scaling functions
also has a dramatic effect since it causes the 
NNLO bottom cross section in PIM kinematics to become negative at $\mu=m/2$. 
The 1PI inclusive cross section at NNLO is
significantly larger than the PIM cross section. 
Only the PIM NNLO results show a significant
reduction in scale dependence. Note also
that for most options the scale uncertainty is larger than 
the kinematics ambiguity. In general,
we see that the theoretical control over this cross section is still
rather poor.

\section{Conclusions}

In this paper we have resummed threshold enhancements
to double-differential heavy quark hadroproduction
cross sections, both in single-particle inclusive
and pair-invariant mass kinematics and in
the $q\bar q$ and $gg$ production channels. We used
these resummed expressions to construct analytic
approximations for the heavy quark cross sections up to NNLO,
including the first three powers (NNLL) of the large
logarithms. This involved including color-coherence
effects and soft radiation attached to
all the one-loop virtual corrections, the contribution
of which we determined via matching conditions.
We derived exact results for the scale-dependent terms 
at NNLO using renormalization group methods.

We examined the magnitude of the NNLO correction to 
the inclusive cross section at the parton and 
hadron levels and studied its variations due
to kinematics choice, factorization scheme choice,
and changes in the renormalization/factorization scale
for top quark production at the Tevatron and bottom
quark production at HERA-B.
We found that the scale dependence of the cross section
after including the NNLO correction is typically substantially reduced,
as expected on general grounds \cite{Oderda:1999im,Sterman:2000pu}.

We now provide what are, in our judgement, reasonable values of the
approximate NNLO top cross section at 1.8 and 2.0 TeV and the
approximate NNLO bottom cross section at 41.6 GeV.  These values,
based on a combination of the results and their uncertainties, are
obtained in the following procedure.  For each energy, we take the
AP+DF results for each kinematics choice, given in the tables in
section 6.  The scale uncertainty for 1PI or PIM is given by the
maximum and minimum values of the cross sections in each row. 
The central value of the cross section is the average of the $\mu=m$ 
values in 1PI and PIM
kinematics.  Two errors are assigned to the resulting cross section.
The first is due to the kinematics-induced ambiguity and is the
difference between the central value and the $\mu=m$ values
obtained in 1PI and PIM kinematics alone.  The second is the weighted
average of the scale uncertainties in the two kinematics, giving more weight
to smaller scale uncertainties.  
Thus we obtain the following NNLO top
production cross sections at the Tevatron,
\begin{equation}
  \label{eq:34}
  \sigma_{t\bar{t}}(1.8\,\mathrm{TeV})
 = 5.8
\pm 0.4 
\pm 0.1 \;\;\mathrm{pb}\,,
\end{equation}
and
\begin{equation}
\label{eq:35}
  \sigma_{t\bar{t}}(2.0\,\mathrm{TeV})
 = 8.0  
\pm 0.6
\pm 0.1 \;\;\mathrm{pb}\,.
\end{equation}
The HERA-B cross section to NNLO is
\begin{equation}
\label{eq:36}
  \sigma_{b\bar{b}}(41.6\,\mathrm{GeV})
 = 30 
\pm  8 
\pm 10 \;\;\mathrm{nb}\,.
\end{equation}
Recall that the first set of errors indicates the kinematics
ambiguity while the second is an estimate of the scale uncertainty.
Note that the scale uncertainty is considerably smaller than the
kinematics uncertainty for top production.
There are other sources of uncertainties in the total cross section
such as parton distribution function uncertainties 
\cite{Alekhin:1999za,Botje:2000dj,Stump:2001gu,Pumplin:2001ct,Giele:2001mr}, 
ambiguities in the analytical
expressions near threshold, as discussed in the previous section, 
and contributions from subleading logarithms \cite{Kidonakis:2000ui}. 
Contributions from even higher orders may also be
significant \cite{Kidonakis:2000ui}. Due to the relative large 
uncertainties in the cross section we give one less significant figure
in our combined results than in the tables of the previous section.

We note that the central values for the cross sections
do not change at all for top production and change only slightly
for bottom production if we use any of the other options
in the tables. The kinematics uncertainty also changes only slightly
for the other options. The scale uncertainty is increased substantially, 
however, if we use the EX or EX+DF options.  
The combined results presented here reflect our
subjective judgment - more unbiased results are presented in section 6.

In conclusion, we have provided, using soft gluon resummation
techniques, improved finite-order perturbative estimates for heavy
quark production along with their uncertainties.  We hope that our
results will thereby allow more meaningful comparisons with
measurements.

\subsection*{Acknowledgments}
We would like to thank S. Catani, V. Del Duca, J. Owens, J. Smith, G. Sterman,
and A. Waldron for fruitful discussions.
The work of N.K. is supported in part by the U.S. Department of Energy. 
The work of E.L. is supported by the Foundation for 
Fundamental Research of Matter 
(FOM) and the National Organization for Scientific Research (NWO).
The work of S.M. is supported by BMBF under grant BMBF-05HT9VKA and by 
DFG under contract FOR~264/2-1. The work of R.V. is supported in part by the 
Division of Nuclear Physics of the Office of High Energy and Nuclear Physics
of the U.S. Department of Energy under Contract No. DE-AC-03-76SF00098.

\def\appendix{\setcounter{section}{0} \setcounter{equation}{0}
  \def\thesection{\Alph{section}}
  \def\theequation{\Alph{section}.\arabic{equation}}}
\appendix

\section{NNLL matching terms for PIM kinematics}

Here we collect the terms required for NLO matching
in PIM kinematics. We first define the Born functions that occur in these
terms.
Recall that in our notation the superscript
$(n)$ indicates  the power,
$(\alpha_s(\mu_R)/\pi)^n$, in the expansion of the corresponding
quantity.

The $q{\Bar{q}}$ channel 1PI Born function is
\begin{eqnarray}
\label{born-amp-squared}
F^{\rm{Born}}_{q\Bar{q}}(s,t_1,u_1) &=&
\pi\, \alpha_s^2(\m_R)\, \frac{C_F}{N_c}\, 
\left( \frac{t_1^2 + u_1^2}{s^2} + \frac{2 m^2}{s} \right) \, .
\end{eqnarray}
The PIM Born function is 
\begin{eqnarray}
\label{eq:27}
F^{\rm{Born}}_{q\Bar{q}}(s,M^2,{\rm{cos}}\q)&=&
\frac{\beta}{2}F^{\rm{Born}}_{q\Bar{q}}(s,t_1,u_1)|_{\pim}\,,
\end{eqnarray}
where the subscript PIM indicates that 
the expressions for $t_1$ and $u_1$ 
in Eq.~(\ref{tupidef}) should be used with $\beta_M$ replaced by $\beta$.

The $gg$ channel 1PI Born function is
\begin{eqnarray}
\label{eq:31}
F^{\rm{Born}}_{gg}(s,t_1,u_1)
&=& \frac{ \alpha_s^2(\m_R)\, \pi}{2(N_c^2-1)^2}\, 
B_{gg}\, 
\left[ C_O \left( 1 - \frac{2 t_1 u_1}{s^2}\right) - C_K \right] \, ,
\\[1ex]
B_{gg} &=&  \frac{u_1}{t_1} + \frac{t_1}{u_1} + 
\frac{4 s m^2}{t_1 u_1}\left(1 - \frac{s m^2}{t_1 u_1}\right)\, .
\end{eqnarray}
The color factors $C_O, C_K$ and, 
for later use, $C_{\rm{QED}}$, are defined as 
\begin{eqnarray}
C_O \,=\, N_c (N_c^2-1) \, , \,\,\,\,\,\,\,
C_K \,=\, \frac{N_c^2-1}{N_c} \, , \,\,\,\,\,\,\,
C_{\rm{QED}} \,=\, \frac{N_c^4-1}{N_c^2} \, .
\end{eqnarray}
The PIM Born function is
\begin{eqnarray}
\label{eq:16}
F^{\rm{Born}}_{gg}(s,M^2,{\rm{cos}}\q)&=&\frac{\beta}{2}F^{\rm
{Born}}_{gg}(s,t_1,u_1)|_{\pim}\,.
\end{eqnarray}
We again indicate that the expressions
in Eq.~(\ref{tupidef}) should be used for $t_1$ and $u_1$ 
with $\beta_M$ replaced by $\beta$.
We also define for PIM kinematics
\begin{eqnarray}
  \label{eq:28}
B'_{gg} = \frac{\beta}{2}B_{gg}  \,.
\end{eqnarray}

Let us now turn to the one-loop ${\rm S+MF}$ cross sections,
as defined in section \ref{sec:nnll-nnlo-expansions}.
The $\msb$ scheme 1PI $q\Bar{q}$ result is
\begin{eqnarray}
{\lefteqn{
\label{eq:18}
s^2 \frac{d^2\sigma^{(1),{\rm S+MF}}_{q\Bar{q}}(s,t_1,u_1)}{dt_1\, du_1} 
\,=\,  F^{\rm{Born}}_{q\Bar{q}}(s,t_1,u_1)\, 
\,\d(s_4)\, }}
\\[1ex]
& &\hspace*{8mm}\times\,
  \Biggl[ 
\left(C_F-\frac{C_A}{2}\right) \frac{1-2m^2/s}{\beta}
\left\{ 
2\,{\rm Li}_2(x)+2\,{\rm Li}_2(-x) + 2\ln(x)\ln(1-x^2)
- \ln^2(x) - \zeta_2 \right\} 
\NO\\[1ex]
& &\hspace*{12mm}+ \, C_F\Biggl\{ 
1-{3\over 2} \zeta_2 +4\ln(x)\ln\left( \frac{t_1}{u_1}\right) 
+4 {\rm Li}_2\left(1-\frac{u_1}{x t_1}\right) -4 {\rm Li}_2\left(1-\frac{t_1}{x u_1}\right)
\NO\\[1ex]
& &\hspace*{12mm}+ {\rm Li}_2\left(1-\frac{s m^2}{t_1 u_1}\right)
+ {1\over 2} \ln^2\left(\frac{s m^2}{t_1 u_1}\right) 
- {\rm Re}L_\beta 
\Biggr\}\NO\\[1ex]
& &\hspace*{12mm}+ \,{1\over 2} C_A\Biggl\{ 
-3\ln(x)\ln\left( \frac{t_1}{u_1}\right)  -{1\over 2}\ln^2x
-3 {\rm Li}_2\left(1-\frac{u_1}{x t_1}\right) +3 {\rm Li}_2\left(1-\frac{t_1}{x u_1}\right)
\NO\\[1ex]
& &\hspace*{12mm}- {\rm Li}_2\left(1-\frac{s m^2}{t_1 u_1}\right)
-{1\over 2} \ln^2\left(\frac{s m^2}{t_1 u_1}\right)+ {1\over 2}\ln^2\left( \frac{t_1}{u_1}\right)
\Biggr\}
\Biggr] \,  \NO
\end{eqnarray}
with $x \equiv (1-\beta)/(1+\beta)$.
The $\msb$ scheme PIM $q\Bar{q}$ result is
\begin{eqnarray}
{\lefteqn{
\label{eq:20}
s \frac{d^2\sigma^{(1),{\rm S+MF}}_{q\Bar{q}}(s,M^2,{\rm{cos}}\q)}{dM^2\, d{\rm{cos}}\q}
\,=\, \frac{1}{s} F^{\rm{Born}}_{q\Bar{q}}(s,M^2,{\rm{cos}}\q)\, \d(1-z)
\, }}  
\\[1ex]
& &\hspace*{8mm}\times\,
 \Biggl[ 
\left(C_F-\frac{C_A}{2}\right) \frac{1-2m^2/s}{\beta}\Biggl\{ 
- {\rm Li}_2\left(\frac{2\beta}{1+\beta}  \right) +
{\rm Li}_2\left(\frac{-2\beta}{1-\beta}  \right) -
\ln(x)\ln\left(\frac{s}{m^2}\right)\Biggr\} \NO\\[1ex]
& &\hspace*{12mm}
+ C_F\Biggl\{ 
-\frac{1}{\beta}\ln(x) -\frac{3}{2}\zeta_2
-4\ln\left(\frac{t_1}{u_1}\right)\ln\left(\frac{s}{m^2}\right)
-\ln\left(\frac{s}{m^2}\right) 
+4{\rm Li}_2\left(1-\frac{2t_1}{s(1-\beta)}\right)
\NO\\[1ex]
& &\hspace*{12mm}
+4{\rm Li}_2\left(1-\frac{2t_1}{s(1+\beta)}\right)
-4{\rm Li}_2\left(1-\frac{2u_1}{s(1-\beta)}\right)
-4{\rm Li}_2\left(1-\frac{2u_1}{s(1+\beta)}\right)
\NO\\[1ex]
& &\hspace*{12mm}
+4\ln\left(\frac{-2t_1}{s(1+\beta)}\right) \ln\left(\frac{-2t_1}{s(1-\beta)}\right)
-4\ln\left(\frac{-2u_1}{s(1+\beta)}\right)\ln\left(\frac{-2u_1}{s(1-\beta)}\right)
+\frac{1}{2}\ln^2\left(\frac{s}{m^2}\right)
\Biggr\}
\NO\\[1ex] 
& &\hspace*{12mm} + C_A\,\Biggl\{
\frac{1}{4}\ln^2(x) +2\ln\left(\frac{-t_1}{m^2}\right)\ln\left(\frac{s}{m^2}\right)
-\ln\left(\frac{-u_1}{m^2}\right)\ln\left(\frac{s}{m^2}\right)
-2\,{\rm Li}_2\left(1-\frac{2t_1}{s(1-\beta)}\right)
\NO\\[1ex]
& &\hspace*{12mm} 
-2\,{\rm Li}_2\left(1-\frac{2t_1}{s(1+\beta)}\right)
+{\rm Li}_2\left(1-\frac{2u_1}{s(1-\beta)}\right)
+{\rm Li}_2\left(1-\frac{2u_1}{s(1+\beta)}\right)
\NO\\[1ex]
& &\hspace*{12mm} 
+\ln\left(\frac{-2u_1}{s(1+\beta)}\right)\ln\left(\frac{-2u_1}{s(1-\beta)}\right)
-2\ln\left(\frac{-2t_1}{s(1+\beta)}\right)\ln\left(\frac{-2t_1}{s(1-\beta)}\right)
-\frac{1}{2}\ln^2\left(\frac{s}{m^2}\right)\Biggr\}
\Biggr] 
 \, . \NO
\end{eqnarray}
The $\msb$ scheme 1PI $gg$ result is
\begin{eqnarray}
\label{eq:21}
{\lefteqn{
s^2 \frac{d^2\sigma^{(1),{\rm S+MF}}_{gg}(s,t_1,u_1)}{dt_1\, du_1} 
\,= \, \frac{\alpha_s^2(\mu_R)\pi}{2(N_c^2-1)^2}B_{gg}\,\d(s_4)\,}}
\\[1ex]
& &\hspace*{8mm}\times\, 
\Biggl[  
{1\over 2} \left( C_{\rm QED} - 
 N_c C_K \left( 1 - \frac{2 t_1 u_1}{s^2}\right) \right) \frac{1-2m^2/s}{\beta}
\left\{ 
2\,{\rm Li}_2(x)+2\,{\rm Li}_2(-x) 
\right.
\NO\\[1ex]
& &\hspace*{18mm} 
\left.
+ 2\ln(x)\ln(1-x^2) - \ln^2(x) - \zeta_2 \right\} 
+ {1\over 2} C_{\rm QED} \left\{ 1 - {\rm Re}L_\beta \right\}
\NO\\[1ex]
& &\hspace*{12mm} 
+ {1\over 2} N_c C_O \Biggl\{ 
\left( 1 - \frac{2 t_1 u_1}{s^2}\right) 
\left( 1 -{\rm Re}L_\beta -3\zeta_2 
- {1\over 2}\ln^2(x) 
+{1\over 2}\ln^2\left( \frac{t_1}{u_1}\right) 
+ {\rm Li}_2\left(1-\frac{s m^2}{t_1 u_1}\right) 
\right.
\NO\\[1ex]
& &\hspace*{18mm}
\left.
+{1\over 2} \ln^2\left(\frac{s m^2}{t_1 u_1}\right)
 \right)
+ \frac{(t_1^2 - u_1^2)}{s^2}
\left(
\ln(x)\ln\left(\frac{t_1}{u_1}\right) 
+{\rm Li}_2\left(1-\frac{u_1}{x t_1}\right) - {\rm Li}_2\left(1-\frac{t_1}{x u_1}\right)
\right)
\Biggr\} 
 \NO\\[1ex]
& &\hspace*{12mm}
 + {1\over 2} N_c C_K \Biggl\{ 
- 2 + 2 {\rm Re}L_\beta  + 3\zeta_2 
+ \ln^2(x) 
- \ln^2\left( \frac{t_1}{u_1}\right) 
- \left( 1 - \frac{2 t_1 u_1}{s^2}\right) 
\left( 1 - {\rm Re}L_\beta \right)
\Biggr\}  
\Biggr]
\, .\NO                       
\end{eqnarray}
The $\msb$ scheme PIM $gg$ result is
\begin{eqnarray}
\label{eq:22}
{\lefteqn{
s \frac{d^2\sigma^{(1),{\rm S+MF}}_{gg}(s,M^2,{\rm{cos}}\q)}{dM^2\, d{\rm{cos}}\q}
\,=\, \frac{1}{s}  \frac{\alpha_s^2(\mu_R)\pi}{2(N_c^2-1)^2}B'_{gg}     \, \d(1-z)
\, }}  
\\[1ex]
& &\hspace*{8mm}\times\,
 \Biggl[ 
{1\over 2} \left( C_{\rm QED} - 
  N_c C_K \left( 1 - \frac{2 t_1 u_1}{s^2}\right) \right) \frac{1-2m^2/s}{\beta}
\Biggl\{ 
- {\rm Li}_2\left(\frac{2\beta}{1+\beta}  \right) +
{\rm Li}_2\left(\frac{-2\beta}{1-\beta}  \right) 
 \NO\\[1ex]
& &\hspace*{12mm} 
-
\ln(x)\ln\left(\frac{s}{m^2}\right)\Biggr\} 
+\frac{1}{2} C_{\rm QED}\,\Biggl(
-\frac{1}{\beta}\ln x-\ln\left(\frac{s}{m^2}\right)
\Biggr)
\NO\\[1ex]
& &\hspace*{12mm} 
+ \frac{1}{2} N_c C_O\,\Biggl\{
\left(1-2\frac{t_1 u_1}{s^2}\right) \Biggl(
-\frac{1}{\beta}\ln(x) -3\zeta_2 + \frac{1}{2}\ln^2(x)
+ \ln\left(\frac{t_1 u_1}{m^4}\right)\ln\left(\frac{s}{m^2}\right)
\NO\\[1ex]
& &\hspace*{12mm} 
-\ln\left(\frac{s}{m^2}\right) \Biggr)
- \frac{(t_1^2 - u_1^2)}{s^2}
\ln\left(\frac{t_1}{u_1}\right) \ln\left(\frac{s}{m^2}\right)
-2\frac{u_1^2}{s^2}{\rm Li}_2\left(1-\frac{2t_1}{s(1-\beta)}\right)
\NO\\[1ex]
& &\hspace*{12mm} 
-2\frac{u_1^2}{s^2}{\rm Li}_2\left(1-\frac{2t_1}{s(1+\beta)}\right)
-2\frac{t_1^2}{s^2}{\rm Li}_2\left(1-\frac{2u_1}{s(1-\beta)}\right)
-2\frac{t_1^2}{s^2}{\rm Li}_2\left(1-\frac{2u_1}{s(1+\beta)}\right)
\NO\\[1ex]
& &\hspace*{12mm}  
-2\frac{u_1^2}{s^2}\ln\left(\frac{-2t_1}{s(1+\beta)}\right)\ln\left(\frac{-2t_1}{s(1-\beta)}\right)
-2\frac{t_1^2}{s^2}\ln\left(\frac{-2u_1}{s(1+\beta)}\right)\ln\left(\frac{-2u_1}{s(1-\beta)}\right)
\Biggr\}  \NO\\[1ex] 
& &\hspace*{12mm} 
+\frac{1}{2} N_c C_K\,\Biggl\{
2\frac{1}{\beta}\ln(x)
+3\zeta_2 -\ln^2(x) 
- 2 \ln\left(\frac{t_1 u_1}{m^4}\right)\ln\left(\frac{s}{m^2}\right)
+ 2\ln\left(\frac{s}{m^2}\right) + \ln^2\left(\frac{s}{m^2}\right)
\NO\\[1ex] 
& &\hspace*{12mm} 
+2\,{\rm Li}_2\left(1-\frac{2t_1}{s(1-\beta)}\right)
+2\,{\rm Li}_2\left(1-\frac{2t_1}{s(1+\beta)}\right)
+2\,{\rm Li}_2\left(1-\frac{2u_1}{s(1-\beta)}\right)
\NO\\[1ex]
& &\hspace*{12mm} 
+2\,{\rm Li}_2\left(1-\frac{2u_1}{s(1+\beta)}\right)
+2\ln\left(\frac{-2t_1}{s(1+\beta)}\right)\ln\left(\frac{-2t_1}{s(1-\beta)}\right)
\NO\\[1ex]
& &\hspace*{12mm} 
+2\ln\left(\frac{-2u_1}{s(1+\beta)}\right) \ln\left(\frac{-2u_1}{s(1-\beta)}\right)
+ \left(1-2\frac{t_1 u_1}{s^2}\right)
\left(\frac{1}{\beta}\ln(x) + \ln\left(\frac{s}{m^2}\right) \right)\Biggr\} 
\Biggr] 
 \, . \NO
\end{eqnarray}

\section{NNLL differential partonic heavy quark cross sections to NNLO}
\setcounter{equation}{0}                          

In this appendix  we collect all our results for 
the NNLL-NNLO differential partonic heavy quark cross sections.
We first make some general observations.
The NLO 1PI results agree with those in
Refs.~\cite{Beenakker:1989bq,Beenakker:1991ma,Meng:1990rp}.
The NLO PIM results are consistent with Ref.~\cite{Mangano:1992jk}.
The $q{\Bar{q}}$ channel results contain terms 
that are antisymmetric under $t_1,u_1$ interchange,
a consequence of charge conjugation asymmetry of the initial state.
The $gg$ channel results are symmetric under 
$t_1 \leftrightarrow u_1$. 
In the $q{\Bar{q}}$ channel the Born function factors
out as a whole for LL and NLL terms.
In the $gg$ channel the Born function
function factors out as a whole only for LL terms.
We also use the notation that the superscript
$(n)$ denotes the coefficient
of $(\alpha_s(\mu_R)/\pi)^n$ in the expansion of the corresponding
quantity.

\subsection{The $q{\Bar{q}}$ channel in 1PI kinematics}

In the $q{\bar q}$ channel, Eq. (\ref{eq:7}),
the Born cross section is
\begin{eqnarray}
\label{qqbar-born}
s^2 \frac{d^2\sigma^{(0)}_{q\Bar{q}}(s,t_1,u_1)}{dt_1\, du_1} 
= \o^{(0)}_{q\Bar{q}}(s,t_1,u_1)
&=& \d(s+t_1+u_1)\,  F^{\rm{Born}}_{q\Bar{q}}(s,t_1,u_1)\, 
\end{eqnarray}
with $F^{\rm{Born}}_{q\Bar{q}}(s,t_1,u_1)$ given in
Eq.~(\ref{born-amp-squared}).
The $\msb$ one-loop NNLL corrections are
\begin{eqnarray}
{\lefteqn{
\label{qqbar-one-loop}
s^2 \frac{d^2\sigma^{(1)}_{q\Bar{q}}(s,t_1,u_1)}{dt_1\, du_1} 
= \o^{(1)}_{q\Bar{q}}(s_4,s,t_1,u_1)
\,=\,  F^{\rm{Born}}_{q\Bar{q}}(s,t_1,u_1)\, 
\, }}
\\[1ex]
& &\hspace*{8mm}\times\,\Biggl[ 
\,4  C_F\, \left[{\ln(s_4/m^2)\over s_4}\right]_+   
\NO\\[1ex]
& &\hspace*{12mm} 
{}+ \left[1\over s_4\right]_+ \Biggl\{ 2C_F\, 
\left( 4 \ln\left( \frac{u_1}{t_1} \right) 
+ \ln\left( \frac{s m^2}{t_1 u_1} \right)   
- {\rm{Re}}L_\b - 1 - \ln\left(\frac{\mu^2}{m^2} \right) \right)  
\NO\\[1ex]
& &\hspace*{22mm}
+\, C_A \left( -3 \ln\left( \frac{u_1}{t_1} \right) 
- \ln\left( \frac{s m^2}{t_1 u_1} \right) + {\rm{Re}}L_\b \right) \Biggr\} 
\NO\\[1ex]
& &\hspace*{3mm}{} +\, \d(s_4)\, \Biggl\{ C_F\,
\ln\left(\frac{\mu^2}{m^2}\right) \Biggl(
\ln\left( \frac{t_1u_1}{m^4}\right) - \frac{3}{2} 
\Biggr)
+ 2 \ln\left(\frac{\m_R^2}{m^2}\right) b_2 
\Biggr\} 
\Biggr] +\,\,
T^{(1)}_{q\Bar{q}\, \opi} (s,t_1,u_1)
\, . \NO
\end{eqnarray}
The $\msb$ two-loop NLL corrections are
\begin{eqnarray}
{\lefteqn{
\label{qqbar-two-loop}
s^2 \frac{d^2\sigma^{(2)}_{q\Bar{q}}(s,t_1,u_1)}{dt_1\, du_1} 
= \o^{(2)}_{q\Bar{q}}(s_4,s,t_1,u_1)
\,=\,  F^{\rm{Born}}_{q\Bar{q}}(s,t_1,u_1)\, \, }}
\\[1ex]
& &\hspace*{8mm}\times\,
\Biggl[ 
\,8 C_F^2\, \left[{\ln^3(s_4/m^2)\over s_4}\right]_+ \, 
\NO\\[1ex] 
&&\hspace*{12mm}
{}+ \left[{\ln^2(s_4/m^2)\over s_4}\right]_+ \Biggl\{ 
12 C_F^2 \left( 4 \ln\left(\frac{u_1}{t_1}\right) 
+ \ln\left(\frac{s m^2}{t_1 u_1}\right) 
- {\rm{Re}}L_\b - 1  
\right. 
\NO\\[1ex] 
&&\hspace*{21mm}
\left.
{}- \ln\left(\frac{\mu^2}{m^2}\right) \right) 
- 4 C_F b_2  
+ 6 C_A C_F \left( - 3 \ln\left(\frac{u_1}{t_1}\right)  
-  \ln\left(\frac{s m^2}{t_1 u_1}\right) 
+  {\rm{Re}}L_{\beta}  \right) 
\Biggr\} 
\NO\\[1ex] 
&&\hspace*{12mm}
+ \left[{\ln(s_4/m^2)\over s_4}\right]_+ 
\Biggl\{ 
4 C_F^2
\ln\left(\frac{\mu^2}{m^2}\right)  
 \Biggl( \ln\left(\frac{t_1u_1}{m^4}\right)  
\NO\\[1ex] 
&&\hspace*{21mm}
\left.
{}-8 \ln\left(\frac{u_1}{t_1}\right) - 2 \ln\left(\frac{s m^2}{t_1 u_1}\right) 
    + 2 {\rm{Re}}L_{\beta} + \frac{1}{2}  
+  \ln\left(\frac{\mu^2}{m^2}\right) \right) 
+ 12 
\ln\left(\frac{\m_R^2}{m^2}\right)  
C_F b_2 
\NO\\[1ex] 
&&\hspace*{21mm}
{}+ 4 C_A C_F
\ln\left(\frac{\mu^2}{m^2}\right)  
 \left( 3 \ln\left(\frac{u_1}{t_1}\right)  
+  \ln\left(\frac{s m^2}{t_1 u_1}\right) -  {\rm{Re}}L_{\beta}  \right)
\Biggr\} 
\NO\\[1ex] 
&&\hspace*{12mm}
{}+ \left[{1\over s_4}\right]_+ \ln\left(\frac{\mu^2}{m^2}\right) 
\Biggl\{ 
\ln\left(\frac{\mu^2}{m^2}\right) 
\Biggl( C_F b_2 + 3 C_F^2 
- 2 C_F^2 \ln\left(\frac{t_1u_1}{m^4}\right) \Biggr)
{}- 6 \ln\left(\frac{\m_R^2}{m^2}\right) C_F b_2 
\Biggr\}
\Biggr] \, .\NO
\end{eqnarray}
To achieve NNLL accuracy one must add
\begin{eqnarray}
\label{qqbar-two-loop-nnll}
& &
2\, C_F\, 
 \hat{T}^{(1)}_{q\Bar{q}\, \opi} (s,t_1,u_1)
\Biggl\{
2 \left[{\ln(s_4/m^2)\over s_4}\right]_+
- \left[{1\over s_4}\right]_+ \ln\left(\frac{\mu^2}{m^2}\right)
\Biggr\}
\,+\, 
 F^{\rm{Born}}_{q\Bar{q}}(s,t_1,u_1)
\NO\\[1ex]
& &\hspace*{6mm}\times\, \Biggl[ \left[{\ln(s_4/m^2)\over s_4}\right]_+ 
\Biggl\{
2 C_F K  + 4 C_F b_2 \ln\left(\frac{t_1 u_1}{m^4}\right) 
        - 16 C_F^2 \z_2
        + 4 C_F^2 \ln^2\left(\frac{t_1u_1}{m^4}\right)  
\NO\\[1ex]
& &\hspace*{21mm}
- 8 C_F \left( {\rm{Re}}\Gamma^{(1)}_{22} - C_F
+ C_F \ln\left(2\sqrt{\n_1\n_2} \frac{s}{m^2}\right)
\right) 
        \left( \ln\left(\frac{t_1u_1}{m^4}\right) 
        + \frac{1}{2 C_F} b_2 \right)
\NO\\[1ex]
& &\hspace*{21mm}
      {}  + 4 \Gamma^{(1)}_{12} \Gamma^{(1)}_{21} 
      {}  + 4 \left({\rm{Re}} \Gamma^{(1)}_{22} - C_F
+ C_F \ln\left(2\sqrt{\n_1\n_2} \frac{s}{m^2}\right)
\right)^2  
\Biggr\}\, 
\NO\\[1ex]
& &\hspace*{12mm}+ \left[{1 \over s_4}\right]_+ \ln\left(\frac{\mu^2}{m^2}\right)
\Biggl\{  - C_F  K + 8 C_F^2 \zeta_2 + 3 C_F^2 \ln\left(\frac{t_1u_1}{m^4}\right)
          - 2 C_F^2 \ln^2\left(\frac{t_1u_1}{m^4}\right)
\NO\\
& &\hspace*{21mm}
        - 3 C_F \left( {\rm{Re}} \Gamma^{(1)}_{22} - C_F 
          + C_F \ln\left(2\sqrt{\n_1\n_2} \frac{s}{m^2}\right)\right)
\NO\\
& &\hspace*{21mm}
+ 2 C_F \ln\left(\frac{t_1u_1}{m^4}\right) \left( {\rm{Re}} \Gamma^{(1)}_{22} - C_F 
          + C_F \ln\left(2\sqrt{\n_1\n_2} \frac{s}{m^2}\right)\right)
\Biggr\} 
\NO\\
& &
\hspace*{12mm}+ \left[{1 \over s_4}\right]_+ \ln\left(\frac{\mu_R^2}{m^2}\right)
\Biggl\{ 6 b_2 \left( {\rm{Re}} \Gamma^{(1)}_{22} - C_F 
          + C_F \ln\left(2\sqrt{\n_1\n_2} \frac{s}{m^2}\right)\right)
 -  6 C_F b_2 \ln\left(\frac{t_1u_1}{m^4}\right)
\Biggr\}
\NO\\[1ex]
& &\hspace*{12mm}+ \delta(s_4) 
\Biggl\{ \ln^2\left(\frac{\mu^2}{m^2}\right)
\Biggl( - 2  C_F^2 \zeta_2 + \frac{3}{4} C_F b_2  + \frac{9}{8} C_F^2 
- \frac{1}{2} C_F b_2 \ln\left(\frac{t_1u_1}{m^4}\right)
- \frac{3}{2} C_F^2 \ln\left(\frac{t_1u_1}{m^4}\right) 
\NO\\
& &\hspace*{21mm}
+ \frac{1}{2} C_F^2 \ln^2\left(\frac{t_1u_1}{m^4}\right)  \Biggr)
+ \ln\left(\frac{\mu_R^2}{m^2}\right) \ln\left(\frac{\mu^2}{m^2}\right) \Biggl( 
- \frac{9}{2} C_F b_2 + 3 C_F b_2 \ln\left(\frac{t_1u_1}{m^4}\right) \Biggr)
\NO\\
& &\hspace*{21mm}
+ 3 b_2^2 \ln^2\left(\frac{\mu_R^2}{m^2}\right) 
\Biggr\} \Biggr] \, 
\end{eqnarray}
where we suppressed 
the $q\Bar{q}$ channel label on the soft anomalous dimension matrix
elements.

We note that at NNLL accuracy we have derived all NNLO soft plus virtual
terms involving the scale, except for $\delta(s_4)$ terms involving
single logarithms of the scale. We can go beyond NNLL accuracy 
and also derive these terms in the partonic
cross section by requiring that the scale dependence in the hadronic cross 
section cancel out. These terms are
\beqa
&&  F^{\rm{Born}}_{q\Bar{q}}(s,t_1,u_1)
\delta(s_4)\left\{\ln\left(\frac{\mu_R^2}{m^2}\right)
\left[\frac{b_3}{128}+\frac{3}{16} b_2 
{\hat {T'}}^{(1)}_{q{\bar q}\,\rm 1PI}(s,t_1,u_1)\right]\right.
\nonumber \\ \hspace{10mm} && 
{}+\ln\left(\frac{\mu_F^2}{m^2}\right)
\left[-8C_F^2 \zeta_3+4C_F\zeta_2
\left({\rm Re}{\Gamma}_{22}^{(1)}-C_F+C_F\ln(2\sqrt{\nu_1\nu_2})
-C_F\ln\left(\frac{t_1u_1}{sm^2}\right)\right)\right.
\nonumber \\ && \hspace{20mm} \left. \left.
{}+C_F\left(\ln\left(\frac{t_1u_1}{m^4}\right)-\frac{3}{2}\right) 
{\hat {T'}}^{(1)}_{q{\bar q}\,\rm 1PI}(s,t_1,u_1)
+\frac{1}{2}C_F K \ln\left(\frac{t_1u_1}{m^4}\right)
\right]\right\}\,,
\eeqa
with ${\hat {T'}}^{(1)}_{q{\bar q}\,\rm 1PI}(s,t_1,u_1)
={\hat T}^{(1)}_{q{\bar q}\,\rm 1PI}(s,t_1,u_1)/
F^{\rm{Born}}_{q\Bar{q}}(s,t_1,u_1)$.
We have checked that these results are consistent with the 
exact expressions in Eq. (\ref{ex-f21qq}) 
and with the expansion of the resummed cross 
section beyond NNLL accuracy as discussed in Ref. \cite{Kidonakis:2000ui}.

The DIS one-loop NNLL corrections are
\begin{eqnarray}
\label{DIS-qqbar-one-loop}
{\lefteqn{
s^2 \frac{d^2\sigma^{(1)}_{q\Bar{q}}(s,t_1,u_1)}{dt_1\, du_1} \Biggr|_{\rm{DIS}}
\,=\, 
s^2 \frac{d^2\sigma^{(1)}_{q\Bar{q}}(s,t_1,u_1)}{dt_1\, du_1} \Biggr|_{\msb}
\, + 
 F^{\rm{Born}}_{q\Bar{q}}(s,t_1,u_1)\, 
\,}} 
\\[1ex]
\NO
& &\hspace*{5mm}
\times\, 
\Biggl[
\,{}- 2 C_F\,  \left[{\ln(s_4/m^2)\over s_4}\right]_+  
+C_F\, \left[1\over s_4\right]_+ 
\Biggl\{
\ln\left( \frac{t_1u_1}{m^4} \right) + \frac{3}{2} 
\Biggr\} \\[1ex]
& &\hspace*{5mm}
+ C_F\,\delta(s_4)\,\Biggl\{
- \frac{1}{2} \ln^2\left(\frac{t_1u_1}{m^4}\right)
- \frac{3}{4} \ln\left(\frac{t_1u_1}{m^4}\right)
+ \ln\left(\frac{-t_1}{m^2}\right)\ln\left(\frac{-u_1}{m^2}\right)
+ \frac{9}{2} + 2\z_2
\Biggr\}
\Biggr] \, . \nonumber
\end{eqnarray}
The DIS two-loop NLL corrections are
\begin{eqnarray}
\label{DIS-qqbar-two-loop}
{\lefteqn{
s^2 \frac{d^2\sigma^{(2)}_{q\Bar{q}}(s,t_1,u_1)}{dt_1\, du_1} \Biggr|_{\rm{DIS}}
\,=\,  
s^2 \frac{d^2\sigma^{(2)}_{q\Bar{q}}(s,t_1,u_1)}{dt_1\, du_1} \Biggr|_{\msb}
\, + 
 F^{\rm{Born}}_{q\Bar{q}}(s,t_1,u_1)\, 
\, }}
\\[1ex]
\NO
& &\hspace*{8mm}
\times\, 
\Biggl[
\,{}- 6 C_F^2 \left[{\ln^3(s_4/m^2)\over s_4}\right]_+ 
\\[1ex]
& &\hspace*{12mm} 
+ \left[{\ln^2(s_4/m^2)\over s_4}\right]_+  
\Biggl\{
6 C_F^2\,
\left( 
\frac{1}{2}\ln\left( \frac{t_1u_1}{m^4} \right) 
- 4  \ln\left( \frac{u_1}{t_1} \right) 
-  \ln\left( \frac{s m^2}{t_1 u_1} \right)   
+  {\rm{Re}}L_\b + \frac{21}{12}  
\right.
\NO\\[1ex]
& &\hspace*{22mm}
\left.
{}
+ \ln\left(\frac{\mu^2}{m^2} \right) \right)  
+ C_F b_2  
{}+ 3 C_A C_F\, \left( 3 \ln\left( \frac{u_1}{t_1} \right) 
+ \ln\left( \frac{s m^2}{t_1 u_1} \right) - {\rm{Re}}L_\b \right)
\Biggr\} 
\NO\\[1ex]
\NO
& &\hspace*{12mm} 
+ \left[{\ln(s_4/m^2)\over s_4}\right]_+  
\Biggl\{
\ln\left(\frac{\mu^2}{m^2}\right) 
\Biggl(
- 6 C_F^2\, \ln\left( \frac{t_1u_1}{m^4} \right)  
- 3 C_F^2
\Biggr)
{}-6 \ln\left(\frac{\m_R^2}{m^2}\right) C_F\,b_2
\Biggr\} 
\Biggr] \, . \NO
\end{eqnarray}
To achieve  NNLL accuracy for the DIS scheme at two loops we have to add 
to the NNLL $\msb$ terms
\begin{eqnarray}
\label{DIS-qqbar-two-loop-nnll}
{\lefteqn{
s^2 \frac{d^2\sigma^{(2)}_{q\Bar{q}}(s,t_1,u_1)}{dt_1\, du_1} \Biggr|_{\rm{DIS}}^{\rm NNLL \; terms}
\,=\,  
s^2 \frac{d^2\sigma^{(2)}_{q\Bar{q}}(s,t_1,u_1)}{dt_1\, du_1} \Biggr|_{\msb}^{\rm NNLL \; terms} 
\,}}\\[1ex]
& &
\NO
{}-2\, C_F\, 
{\hat T}^{(1)}_{q\Bar{q}\, \opi} (s,t_1,u_1) \Biggr|_{\msb} 
 \left[{\ln(s_4/m^2)\over s_4}\right]_+
\,+\,
 F^{\rm{Born}}_{q\Bar{q}}(s,t_1,u_1)\, 
\\[1ex]
\NO
& &
\times\,\Biggl[ \left[{\ln(s_4/m^2)\over s_4}\right]_+\,
\Biggl\{{}- C_F K 
        + 4 C_F \left( {\rm{Re}}\Gamma^{(1)}_{22} - C_F
+ C_F \ln\left(2\sqrt{\n_1\n_2} \frac{s}{m^2}\right)
\right) 
        \left( \ln\left(\frac{t_1u_1}{m^4}\right) 
        + \frac{3}{2} \right) 
\NO\\[1ex]
& &\hspace*{6mm}
      {}  - C_F b_2 \ln\left(\frac{t_1u_1}{m^4}\right) 
       + 2 C_F b_2 \ln\left(\frac{s}{m^2}\right) 
        - \frac{3}{2} C_F b_2 + 16 C_F^2 \z_2 
        - 4 C_F^2 \ln^2\left(\frac{t_1u_1}{m^4}\right) 
\NO\\[1ex]
& &\hspace*{6mm}
      {}  - \frac{9}{2} C_F^2 \ln\left(\frac{t_1u_1}{m^4}\right)
        + 2 C_F^2 \ln\left(\frac{-t_1}{m^2}\right)\ln\left(\frac{-u_1}{m^2}\right)
        + \frac{45}{4} C_F^2    
\Biggr\} 
\NO\\
& &
+ \left[{1\over s_4}\right]_+ \,
\Biggl\{  
\Biggl(- 8 C_F^2 \z_2 - \frac{45}{4} C_F^2 
+ 2 C_F^2 \ln^2\left(\frac{t_1u_1}{m^4}\right)
- 2 C_F^2 \ln\left(\frac{-t_1}{m^2}\right)\ln\left(\frac{-u_1}{m^2}\right)
\NO\\
& &\hspace*{6mm}
{} + \frac{3}{2} C_F^2 \ln\left(\frac{t_1u_1}{m^4}\right)
\Biggr) \ln\left(\frac{\mu^2}{m^2}\right) 
+ \Biggl( \frac{9}{2} C_F b_2  + 3 C_F b_2  \ln\left(\frac{t_1u_1}{m^4}\right) 
\Biggr)
\ln\left(\frac{\mu_R^2}{m^2}\right) 
\Biggr\} \Biggr]
\, .\NO
\end{eqnarray}

\subsection{The $gg$ channel in 1PI kinematics}

In the $gg$ channel, Eq.~(\ref{eq:8}),
the Born cross section is 
\begin{eqnarray}
s^2 \frac{d^2\sigma^{(0)}_{gg}(s,t_1,u_1)}{dt_1\, du_1} 
= \o^{(0)}_{gg}(s,t_1,u_1)
&=& \d(s+t_1+u_1)\, F^{\rm{Born}}_{gg}(s,t_1,u_1)\, 
\end{eqnarray}
with $F^{\rm{Born}}_{gg}(s,t_1,u_1)$ given in Eq.~(\ref{eq:31}).
The $\msb$ one-loop NNLL corrections are
\begin{eqnarray}
\label{gg-one-loop}
{\lefteqn{
s^2 \frac{d^2\sigma^{(1)}_{gg}(s,t_1,u_1)}{dt_1\, du_1} 
=\o^{(1)}_{gg}(s_4,s,t_1,u_1) }}
\\[1ex]
&\,= &
 \Biggl[ 
\,4 N_c\, \left[{\ln(s_4/m^2)\over s_4}\right]_+\, 
F^{\rm{Born}}_{gg}(s,t_1,u_1)\, 
\NO \\[1ex]
&&\hspace*{6mm}
{}+ \left[1\over s_4\right]_+ \Biggl\{  
\frac{ \alpha_s^2(\m_R)\, \pi}{2(N_c^2-1)^2}\, B_{gg}\, 
\Biggl(
N_c C_O \left\{ \left(
\ln\left(\frac{s m^2}{t_1 u_1}\right) 
- 1\, \right) \left( 1 - \frac{2 t_1 u_1}{s^2}\right) 
\right.
\NO \\[1ex]
&&\hspace*{18mm}
\left.
{}+  \ln\left(\frac{u_1}{t_1}\right)   \frac{(t_1^2-u_1^2)}{s^2} 
\right\} 
+ N_c C_K
\left\{ ({\rm{Re}}L_{\beta} + 1 ) 
\left( 1 - \frac{2 t_1 u_1}{s^2}\right)
+ 2\, \right\} 
\NO \\[1ex]
&&\hspace*{18mm}
{}- C_{\rm{QED}} \left( {\rm{Re}}L_{\beta} + 1 \right)
\Biggr)
- 2 N_c\, F^{\rm{Born}}_{gg}(s,t_1,u_1)\, 
    \ln\left(\frac{\mu^2}{m^2}\right) \Biggr\} 
\NO\\[1ex]
& &\hspace*{6mm} {} +\, 
\d(s_4)\, F^{\rm{Born}}_{gg}(s,t_1,u_1)\,
 \Biggl\{  N_c\,
\ln\left(\frac{\mu^2}{m^2}\right) 
 \ln\left( \frac{t_1u_1}{m^4}\right) 
+ 2 \ln\left(\frac{\m_R^2}{\mu^2}\right) b_2
\Biggr\}\Biggr]   {} +\, 
T^{(1)}_{gg\, \opi} (s,t_1,u_1)
\, . \NO
\end{eqnarray}
The $\msb$ two-loop NLL corrections are
\begin{eqnarray}
\label{gg-two-loop}
{\lefteqn{
s^2 \frac{d^2\sigma^{(2)}_{gg}(s,t_1,u_1)}{dt_1\, du_1}= 
\o^{(2)}_{gg}(s_4,s,t_1,u_1)}}
\\[1ex]
&\,= &
 \Biggl[ 
\,8 N_c^2\, \left[{\ln^3(s_4/m^2)\over s_4}\right]_+\, 
F^{\rm{Born}}_{gg}(s,t_1,u_1)\, 
+ \left[{\ln^2(s_4/m^2)\over s_4}\right]_+ \Biggl\{  
\frac{ \alpha_s^2(\m_R)\, \pi}{2(N_c^2-1)^2}\, B_{gg} \, 6 N_c
\NO \\[1ex]
&&\hspace*{17mm}
\times\,\Biggl( N_c C_O \left\{ \left(
\ln\left(\frac{s m^2}{t_1 u_1}\right) 
- 1\, \right) \left( 1 - \frac{2 t_1 u_1}{s^2}\right)
+ \, \ln\left(\frac{u_1}{t_1}\right)   \frac{(t_1^2-u_1^2)}{s^2} \right\}
\NO \\[1ex]
&&\hspace*{21mm}
{} + N_c\, C_K
\Biggl\{ 
({\rm{Re}}L_{\beta} + 1 ) \left( 1 - \frac{2 t_1 u_1}{s^2}\right)
+ 2 \, \Biggr\}- C_{\rm{QED}} \left( {\rm{Re}}L_{\beta} + 1 \right)
\Biggr) 
\NO \\[1ex]
&&\hspace*{21mm} 
{}- F^{\rm{Born}}_{gg}(s,t_1,u_1)\, 
\left(12 N_c^2 \ln\left(\frac{\mu^2}{m^2}\right) 
+4 b_2 N_c \right) \Biggr\} 
\NO \\[1ex]
&&\hspace*{12mm}
{}+ \left[{\ln(s_4/m^2)\over s_4}\right]_+ 
\Biggl\{  
\frac{ \alpha_s^2(\m_R)\, \pi}{2(N_c^2-1)^2}\, B_{gg}\, 
\ln\left(\frac{\mu^2}{m^2}\right) 4 N_c
\Biggl(  
N_c\, C_O  
\ln\left(\frac{u_1}{t_1}\right) \frac{(u_1^2-t_1^2)}{s^2}
\NO \\[1ex]
&&\hspace*{21mm}
{}- N_c\, C_K \left\{ 
({\rm{Re}}L_{\beta} + 1 ) \left( 1 - \frac{2 t_1 u_1}{s^2}\right) +
\ln\left(\frac{s m^2}{t_1 u_1}\right) + 1  \right\} 
\NO \\[1ex]
&&\hspace*{21mm}
{}+ C_{\rm{QED}} \left( {\rm{Re}}L_{\beta} + 1 \right)
\Biggr) 
+ 4 N_c\, F^{\rm{Born}}_{gg}(s,t_1,u_1)\, 
\ln\left(\frac{\mu^2}{m^2}\right)
\Biggl( 
N_c\, \ln\left(\frac{t_1u_1}{m^4}\right)  
\NO \\[1ex]
&&\hspace*{21mm}
{}- N_c\, \ln\left(\frac{s m^2}{t_1 u_1}\right) - 2 b_2  + N_c 
+ N_c\, \ln\left(\frac{\mu^2}{m^2}\right) \Biggr) 
\NO \\[1ex]
&&\hspace*{21mm}
{}+ 12 N_c\, F^{\rm{Born}}_{gg}(s,t_1,u_1)\, 
\ln\left(\frac{\m_R^2}{m^2}\right) b_2  \Biggr\} 
\NO \\[1ex]
& &\hspace*{12mm}+\, 
\left[{1 \over s_4}\right]_+\, F^{\rm{Born}}_{gg}(s,t_1,u_1)\,
\ln\left(\frac{\mu^2}{m^2}\right) 
\Biggl\{ 
\ln\left(\frac{\mu^2}{m^2}\right) 
\Biggl(
- 2 N_c^2\, \ln\left(\frac{t_1u_1}{m^4}\right) 
\NO \\[1ex]
&&\hspace*{21mm}
{}+ 5 b_2 N_c \Biggr)
- 6 \ln\left(\frac{\m_R^2}{m^2}\right) b_2 N_c
 \Biggr\}
 \Biggr] \, . \NO
\end{eqnarray}
To achieve NNLL accuracy one must add
\begin{eqnarray}
\label{gg-two-loop-nnll}
& &
2\, C_A\, 
\hat{T}^{(1)}_{gg\, \opi} (s,t_1,u_1)
\Biggl\{
2 \left[{\ln(s_4/m^2)\over s_4}\right]_+
- \left[{1\over s_4}\right]_+ \ln\left(\frac{\mu^2}{m^2}\right)
\Biggr\}
\\[1ex]
& &
\,+\,   \left[{\ln(s_4/m^2)\over s_4}\right]_+\,
\Biggl( \,
F^{\rm{Born}}_{gg}(s,t_1,u_1)\, \Biggl\{
      {}  - 16 N_c^2 \z_2  + 2 N_c K 
        + 4 N_c^2 \ln^2\left(\frac{t_1u_1}{m^4}\right)  
\NO \\[1ex]
&&\hspace*{10mm}
     {}   + 4 N_c b_2 \ln\left(\frac{t_1u_1}{m^4}\right) \Biggr\} 
\NO \\[1ex]
&&\hspace*{6mm}
{}+ \frac{\a_s^2(\m_R) \pi}{2 (N_c^2-1)}\, B_{gg}\, 
\left( 1 - \frac{2 t_1 u_1}{s^2}\right)
\Biggl\{     
        \left( N_c + \frac{1}{4} N_c^3 \right) \left( \Gamma^{(1)}_{31} \right)^2 
\NO \\[1ex]
&&\hspace*{10mm}
     {}   - 8 N_c^2 \left( {\rm{Re}}\Gamma^{(1)}_{22} - C_A
+ C_A \ln\left(2\sqrt{\n_1\n_2} \frac{s}{m^2}\right)
\right) 
\left( \ln\left(\frac{t_1u_1}{m^4}\right) 
+ \frac{1}{2N_c} b_2 \right) 
\NO \\[1ex]
&&\hspace*{10mm}
      {}  + 4 N_c \left({\rm{Re}} \Gamma^{(1)}_{22} - C_A
+ C_A \ln\left(2\sqrt{\n_1\n_2} \frac{s}{m^2}\right)
\right)^2 
\Biggr\} 
\NO \\[1ex]
&&\hspace*{6mm}
{}+ \frac{\a_s^2(\m_R) \pi}{2 (N_c^2-1)}\, B_{gg}\, 
\frac{(t_1^2-u_1^2)}{s^2}
\Biggl\{  
        4 \left({\rm{Re}} \Gamma^{(1)}_{11} - C_A
+ C_A \ln\left(2\sqrt{\n_1\n_2} \frac{s}{m^2}\right)
\right) \Gamma^{(1)}_{31} 
\NO \\[1ex]
&&\hspace*{10mm}
  {}      + 2  \left( N_c^2 - 2 \right) 
         \Gamma^{(1)}_{31} \left( {\rm{Re}} \Gamma^{(1)}_{22} - C_A
+ C_A \ln\left(2\sqrt{\n_1\n_2} \frac{s}{m^2}\right)
\right) 
\NO \\[1ex]
&&\hspace*{10mm}
    {}    - 2 N_c^3 \Gamma^{(1)}_{31} 
\left( \ln\left(\frac{t_1u_1}{m^4}\right)  + \frac{1}{2N_c} b_2 \right) 
\Biggr\} 
\NO \\[1ex]
&&\hspace*{6mm}
{} + \frac{\a_s^2(\m_R) \pi}{2 (N_c^2-1)}\, B_{gg}\, 
\Biggl\{   
-  \frac{N_c}{2} \left( \Gamma^{(1)}_{31} \right)^2 
+ 4 \frac{1}{N_c} \left(  {\rm{Re}} \Gamma^{(1)}_{11} - C_A
+ C_A \ln\left(2\sqrt{\n_1\n_2} \frac{s}{m^2}\right)
\right)^2 
\NO \\[1ex]
&&\hspace*{10mm}
   {}     - 8 \left({\rm{Re}}\Gamma^{(1)}_{11} - C_A
+ C_A \ln\left(2\sqrt{\n_1\n_2} \frac{s}{m^2}\right)
\right) 
\left( \ln\left(\frac{t_1u_1}{m^4}\right) 
+   \frac{1}{2 N_c} b_2
\right) 
\NO \\[1ex]
&&\hspace*{10mm}
     {}   + 16 \left( {\rm{Re}}\Gamma^{(1)}_{22} - C_A
+ C_A \ln\left(2\sqrt{\n_1\n_2} \frac{s}{m^2}\right)
\right) 
\left( \ln\left(\frac{t_1u_1}{m^4}\right) 
+ \frac{1}{2N_c} b_2 \right) 
\NO \\[1ex]
&&\hspace*{10mm}
     {}   - 8 \frac{1}{N_c}  \left( {\rm{Re}} \Gamma^{(1)}_{22} - C_A
+ C_A \ln\left(2\sqrt{\n_1\n_2} \frac{s}{m^2}\right)
\right)^2 
\Biggr\} \Biggl)
\NO\\[1ex]
& &+ \left[{1\over s_4}\right]_+ 
\Biggl( F^{\rm{Born}}_{gg}(s,t_1,u_1) 
\Biggl\{ 
\ln\left(\frac{\mu^2}{m^2}\right)
\Biggl(  - N_c  K + 8 N_c^2 \zeta_2 
+ 4 N_c b_2 \ln\left(\frac{t_1u_1}{m^4}\right) 
\NO\\[1ex]
& &\hspace*{21mm}
- 2 N_c^2 \ln^2\left(\frac{t_1u_1}{m^4}\right) \Biggr)
- 6 N_c b_2 \ln\left(\frac{t_1u_1}{m^4}\right) 
\ln\left(\frac{\mu_R^2}{m^2}\right) 
\Biggr\} 
\NO \\[1ex]
&&\hspace*{6mm}
+ \frac{\a_s^2(\m_R) \pi}{2 (N_c^2-1)}\, B_{gg}\, 
\left( 1 - \frac{2 t_1 u_1}{s^2}\right)
\Biggl\{ 6 \ln\left(\frac{\mu_R^2}{m^2}\right) 
+ 4 
\Biggl( \frac{1}{2} \frac{N_c}{b_2} \ln\left(\frac{t_1u_1}{m^4}\right)  - 1 \Biggr)
\ln\left(\frac{\mu^2}{m^2}\right) \Biggr\} N_c b_2 
\NO \\[1ex]
&&\hspace*{21mm}
\times
\left( {\rm{Re}} \Gamma^{(1)}_{22} - C_A
+ C_A \ln\left(2\sqrt{\n_1\n_2} \frac{s}{m^2}\right)
\right)
\NO \\[1ex]
&&\hspace*{6mm}
+ \frac{\a_s^2(\m_R) \pi}{2 (N_c^2-1)}\, B_{gg}\, 
\frac{(t_1^2-u_1^2)}{s^2} 
\Biggl\{
\frac{3}{2} \ln\left(\frac{\mu_R^2}{m^2}\right) + 
\Biggl( \frac{1}{2} \frac{N_c}{b_2} \ln\left(\frac{t_1u_1}{m^4}\right)  - 1 \Biggr)
\ln\left(\frac{\mu^2}{m^2}\right) \Biggr\} N_c^2 b_2 \Gamma^{(1)}_{31} 
\NO \\[1ex]
&&\hspace*{6mm}
+ \frac{\a_s^2(\m_R) \pi}{2 (N_c^2-1)}\, B_{gg}\, 
\Biggl\{ 6 \ln\left(\frac{\mu_R^2}{m^2}\right)
+ 4 
\Biggl( \frac{1}{2} \frac{N_c}{b_2} \ln\left(\frac{t_1u_1}{m^4}\right)  - 1 \Biggr)
\ln\left(\frac{\mu^2}{m^2}\right) \Biggr\} \frac{b_2}{N_c}
\NO \\[1ex]
&&\hspace*{21mm}
\times \Biggl\{
\left( {\rm{Re}} \Gamma^{(1)}_{11} - C_A
+ C_A \ln\left(2\sqrt{\n_1\n_2} \frac{s}{m^2}\right)
\right)
\NO\\[1ex]
& &\hspace*{21mm}
- 2
\left( {\rm{Re}} \Gamma^{(1)}_{22} - C_A
+ C_A \ln\left(2\sqrt{\n_1\n_2} \frac{s}{m^2}\right)
\right) \Biggr\}
\Biggr)
\NO\\[1ex]
& &+ \delta(s_4) F^{\rm{Born}}_{gg}(s,t_1,u_1)\,
\Biggl\{  \Biggl( - 2 \z_2 N_c^2
- \frac{5}{2} N_c b_2 \ln\left(\frac{t_1u_1}{m^4}\right)  
+ \frac{1}{2} N_c^2 \ln^2\left(\frac{t_1u_1}{m^4}\right)  
\Biggr)
\ln^2\left(\frac{\mu^2}{m^2}\right) 
\NO \\[1ex]
&&\hspace*{21mm} 
+ 3 N_c b_2 \ln\left(\frac{t_1u_1}{m^4}\right) 
\ln\left(\frac{\mu^2}{m^2}\right) \ln\left(\frac{\mu_R^2}{m^2}\right) 
+ 3 b_2^2 \ln^2\left(\frac{\mu_R^2}{\mu^2}\right) 
\Biggr\}\, ,
\end{eqnarray}
where we suppressed the $gg$ channel label on the soft anomalous 
dimension matrix elements.

As we discussed in the previous subsection, 
we can also derive the NNLO $\delta(s_4)$ terms involving 
single logarithms of the scale. These terms are 
\beqa
&&  \hspace{-10mm} F^{\rm Born}_{gg}(s,t_1,u_1) 
\delta(s_4)\left\{\ln\left(\frac{\mu_F^2}{m^2}\right)
\left[-8\zeta_3 C_A^2+\left(C_A\ln\left(\frac{t_1u_1}{m^4}\right)
-\frac{b_2}{8}\right){\hat {T'}}^{(1)}_{gg\,\rm 1PI}(s,t_1,u_1)
+C_A\frac{K}{2}\ln\left(\frac{t_1u_1}{m^4}\right)\right]\right.
\nonumber \\ && \hspace{25mm} \left.
{}+\ln\left(\frac{\mu_R^2}{m^2}\right)
\left[\frac{b_3}{128}+\frac{3}{16}b_2 
{\hat {T'}}^{(1)}_{gg\,\rm 1PI}(s,t_1,u_1)\right]\right\}
\nonumber \\ &&
{}+\frac{\alpha_s^2(\mu_R)\pi}{(N_c^2-1)^2} B_{gg}
\ln\left(\frac{\mu_F^2}{m^2}\right) 2C_A \zeta_2 \, \delta(s_4) 
\left\{N_c(N_c^2-1)\frac{(t_1^2+u_1^2)}{s^2}
\left[\left(-C_F+\frac{C_A}{2}\right)
{\rm Re} L_{\beta}\right. \right.
\nonumber \\ && \hspace{45mm} \left.
{}+\frac{C_A}{2}\ln\left(\frac{m^2s}{t_1u_1}\right)
-C_F\right]+\frac{(N_c^2-1)}{N_c}(C_F-C_A) {\rm Re} L_{\beta}
\nonumber \\ && \hspace{45mm} \left.
{}+C_F \frac{(N_c^2-1)}{N_c}
+\frac{N_c^2}{2}(N_c^2-1)
\ln\left(\frac{u_1}{t_1}\right)\frac{(t_1^2-u_1^2)}{s^2} \right\}\,,
\eeqa
with ${\hat {T'}}^{(1)}_{gg\,\rm 1PI}(s,t_1,u_1)
={\hat T}^{(1)}_{gg\,\rm 1PI}(s,t_1,u_1)/
F^{\rm{Born}}_{gg}(s,t_1,u_1)$.
We have checked that these results are consistent with the 
exact expressions in Eq. (\ref{ex-f21gg}) 
and with the expansion of the resummed cross 
section beyond NNLL accuracy.

\subsection{The $q{\Bar{q}}$ channel in PIM kinematics}

In the $q{\Bar{q}}$ channel, Eq.~(\ref{q-qbar-annihilation-0_PIM}), 
the Born cross section is
\begin{eqnarray}
\label{qqbar-born_PIM}
s\frac{d^2\sigma^{(0)}_{q\Bar{q}}(s,M^2,{\rm{cos}}\q)}{dM^2\, d{\rm{cos}}\q}
= \frac{\beta}{2} \o^{(0)}_{q\Bar{q}}(s,t_1,u_1)|_{\pim}
&=& \frac{1}{s}\d(1-z)\,  F^{\rm{Born}}_{q\Bar{q}}(s,M^2,{\rm{cos}}\q)\, 
\end{eqnarray}
with $F^{\rm{Born}}_{q\Bar{q}}(s,M^2,{\rm{cos}}\q)$ defined
in Eq.~(\ref{eq:27}).

The $\msb$ one-loop NNLL corrections are
\begin{eqnarray}
{\lefteqn{
\label{qqbar-one-loop_PIM}
s \frac{d^2\sigma^{(1)}_{q\Bar{q}}(s,M^2,{\rm{cos}}\q)}{dM^2\, d{\rm{cos}}\q}
=\o^{(1)}_{q\Bar{q}}(1-z,s,M^2,{\rm{cos}}\q) =
 \frac{\beta}{2} \o^{(1)}_{q\Bar{q}}(s_4,s,t_1,u_1)|_{\pim} \, }}
\\[1ex]\nonumber
& &\hspace*{8mm}
\,=\, \frac{1}{s} F^{\rm{Born}}_{q\Bar{q}}(s,M^2,{\rm{cos}}\q))\, 
\\[1ex]
& &\hspace*{8mm}\times\,\Biggl[ 
\,4  C_F\, \left[{\ln(1-z)\over 1-z}\right]_+  
\NO\\[1ex]
& &\hspace*{12mm} 
{}+ \left[1\over 1-z\right]_+ \Biggl\{ 2C_F\, 
\left( 4 \ln\left( \frac{u_1}{t_1} \right) 
+ \ln\left( \frac{s}{m^2} \right)   
- {\rm{Re}}L_\b - 1 - \ln\left(\frac{\mu^2}{m^2} \right) \right)  
\NO\\[1ex]
& &\hspace*{22mm}
+\, C_A \left( -3 \ln\left( \frac{u_1}{t_1} \right) 
- \ln\left( \frac{s m^2}{t_1 u_1} \right) + {\rm{Re}}L_\b \right) \Biggr\} 
\NO\\[1ex]
& &\hspace*{13mm} {}+\, \d(1-z)\, 
\Biggl\{ 
 - \frac{3}{2} \ln\left(\frac{\mu^2}{m^2}\right) C_F + 
2 \ln\left(\frac{\m_R^2}{m^2}\right) b_2 
\Biggr\} \Biggr] \, 
\NO\\[1ex]
& &\hspace*{13mm}+\,\,
T^{(1)}_{q\Bar{q}\, \pim} (M^2,\cos\theta). \NO
\end{eqnarray}
The $\msb$ two-loop NLL corrections are
\begin{eqnarray}
{\lefteqn{
\label{qqbar-two-loop_PIM}
s\frac{d^2\sigma^{(2)}_{q\Bar{q}}(s,M^2,{\rm{cos}}\q)}{dM^2\, d{\rm{cos}}\q}
=  \o^{(2)}_{q\Bar{q}}(1-z,s,M^2,{\rm{cos}}\q)
=\frac{\beta}{2}\o^{(2)}_{q\Bar{q}}(s_4,s,t_1,u_1)|_{\pim}\, }}
\\[1ex]\nonumber
& &\hspace*{8mm}
\,=\, \frac{1}{s} F^{\rm{Born}}_{q\Bar{q}}(s,M^2,{\rm{cos}}\q)\, 
\\[1ex]
& &\hspace*{8mm}\times\,
\Biggl[ 
\,8 C_F^2\, \left[{\ln^3(1-z)\over 1-z}\right]_+ \, 
\NO\\[1ex] 
&&\hspace*{12mm}
{} + \left[{\ln^2(1-z)\over 1-z}\right]_+ \Biggl\{ 
12 C_F^2 \left( 4 \ln\left(\frac{u_1}{t_1}\right) 
+ \ln\left(\frac{s}{m^2}\right) 
- {\rm{Re}}L_\b - 1  
\right. 
\NO\\[1ex] 
&&\hspace*{21mm}
\left.
{}-  \ln\left(\frac{\mu^2}{m^2}\right) \right) 
- 4 C_F b_2  
+ 6 C_A C_F \left( - 3 \ln\left(\frac{u_1}{t_1}\right)  
- \ln\left(\frac{s m^2}{t_1 u_1}\right) 
+ {\rm{Re}}L_{\beta}  \right) 
\Biggr\} 
\NO\\[1ex] 
&&\hspace*{12mm}
{}+ \left[{\ln(1-z)\over 1-z}\right]_+ \!
\Biggl\{ 
4 C_F^2
\ln\left(\frac{\mu^2}{m^2}\right) \!  
 \Biggl( -8  \ln\left(\frac{u_1}{t_1}\right) - 2 \ln\left(\frac{s}{m^2}\right) 
+ 2 {\rm{Re}}L_{\beta} + \frac{1}{2} 
+ \ln\left(\frac{\mu^2}{m^2}\right) \Biggr) 
\NO\\[1ex] 
&&\hspace*{21mm}
{}+ 12 \ln\left(\frac{\m_R^2}{m^2}\right) C_F b_2 + 
4 C_A C_F \ln\left(\frac{\mu^2}{m^2}\right)  \left( 3 \ln\left(\frac{u_1}{t_1}\right)  
+ \ln\left(\frac{s m^2}{t_1 u_1}\right) - {\rm{Re}}L_{\beta}  \right)
\Biggr\} 
\NO\\[1ex] 
&&\hspace*{12mm}
{}+ \left[{1\over 1-z}\right]_+ \ln\left(\frac{\mu^2}{m^2}\right)
\Biggl\{ 
\ln\left(\frac{\mu^2}{m^2}\right)
\Biggl( C_F b_2 + 3 C_F^2 
\Biggr)
- 6 \ln\left(\frac{\m_R^2}{m^2}\right) C_F b_2
\Biggr\}
\Biggr] \, .\NO
\end{eqnarray}
To achieve NNLL accuracy to two-loops one must add
\begin{eqnarray}
\label{qqbar-two-loop_PIM-nnll}
& &
2 C_F\, \hat{T}^{(1)}_{q\Bar{q}\, \pim} (M^2,{\rm{cos}}\q)
\Biggl\{
2 \left[{\ln(1-z)\over 1-z}\right]_+ 
- \left[{1\over 1-z}\right]_+ \ln\left(\frac{\mu^2}{m^2}\right)
\Biggr\}
\,+\,
 \frac{1}{s}F^{\rm{Born}}_{q\Bar{q}}(s,M^2,{\rm{cos}}\q)\, 
\NO\\[1ex]
& &\hspace*{5mm}\times\,\Biggl[ \left[{\ln(1-z)\over 1-z}\right]_+
\Biggl\{ 
 2 C_F K - 16 C_F^2 \z_2 
        - 4 \left( {\rm{Re}}\Gamma^{(1)}_{22} - C_F 
+ C_F \ln\left(2\sqrt{\n_1\n_2} \frac{s}{m^2}\right)\right) b_2 
\NO\\
& &\hspace*{18mm} + 4 \Gamma^{(1)}_{12}\Gamma^{(1)}_{21}  
      {}  + 4 \left( {\rm{Re}} \Gamma^{(1)}_{22} - C_F 
+ C_F \ln\left(2\sqrt{\n_1\n_2} \frac{s}{m^2}\right)\right)^2 
\Biggr\}
\NO\\[1ex]
& &\hspace*{9mm}+ \left[{1\over 1-z}\right]_+ \ln\left(\frac{\mu^2}{m^2}\right)
\Biggl\{  - C_F  K + 8 C_F^2 \zeta_2 
        - 3 C_F \left( {\rm{Re}} \Gamma^{(1)}_{22} - C_F 
          + C_F \ln\left(2\sqrt{\n_1\n_2} \frac{s}{m^2}\right)\right)
\Biggr\} 
\NO\\
& &
\hspace*{9mm}+ \left[{1\over 1-z}\right]_+ \ln\left(\frac{\mu_R^2}{m^2}\right)
\Biggl\{ 6 b_2 \left( {\rm{Re}} \Gamma^{(1)}_{22} - C_F 
          + C_F \ln\left(2\sqrt{\n_1\n_2} \frac{s}{m^2}\right)\right)
\Biggr\}
\NO\\[1ex]
& &\hspace*{9mm}+ \delta(1-z) 
\Biggl\{ \ln^2\left(\frac{\mu^2}{m^2}\right)
\Biggl( - 2  C_F^2 \zeta_2 + \frac{3}{4} C_F b_2  + \frac{9}{8} C_F^2  \Biggr)
- \frac{9}{2} C_F b_2 \ln\left(\frac{\mu_R^2}{m^2}\right) \ln\left(\frac{\mu^2}{m^2}\right) 
\NO\\
& &\hspace*{18mm}
+ 3 b_2^2 \ln^2\left(\frac{\mu_R^2}{m^2}\right) 
\Biggr\} \Biggr] \, .
\end{eqnarray}

As for 1PI kinematics, we note that at NNLL accuracy we have derived 
all NNLO soft plus virtual
terms involving the scale, except for $\delta(1-z)$ terms involving
single logarithms of the scale. We can go beyond NNLL accuracy
and derive these terms in the partonic
cross section by requiring that the scale dependence in the hadronic cross 
section cancel out. These terms are
\beqa
&& \hspace{-5mm} \frac{1}{s} F^{\rm{Born}}_{q\Bar{q}}(s,M^2,{\rm{cos}}\q)
\delta(1-z)\left\{\ln\left(\frac{\mu_R^2}{m^2}\right)
\left[\frac{b_3}{128}+\frac{3}{16} b_2 
{\hat {T'}}^{(1)}_{q{\bar q}\,\rm PIM}(M^2,{\rm{cos}}\q)\right]\right.
\nonumber \\ && \hspace{-10mm} \left.
{}+\ln\left(\frac{\mu_F^2}{m^2}\right)
\left[-8C_F^2 \zeta_3+4C_F\zeta_2
\left({\rm Re}\Gamma_{22}^{(1)}-C_F
+C_F\ln\left(2\sqrt{\nu_1\nu_2}\frac{s}{m^2}\right)\right)
-\frac{3}{2}C_F {\hat {T'}}^{(1)}_{q{\bar q}\,\rm PIM}(M^2,{\rm{cos}}\q)
\right]\right\}\,,
\nonumber \\
\eeqa
with ${\hat {T'}}^{(1)}_{q{\bar q}\,\rm PIM}(M^2,{\rm{cos}}\q)
=s\, {\hat T}^{(1)}_{q{\bar q}\,\rm PIM}(M^2,{\rm{cos}}\q)/
F^{\rm{Born}}_{q\Bar{q}}(s,M^2,{\rm{cos}}\q)$.
Again, we have checked that these results are consistent with 
Eq. (\ref{ex-f21qq}) and with the expansion of the resummed cross 
section beyond NNLL accuracy.

The DIS one-loop NNLL corrections are
\begin{eqnarray}
\label{DIS-qqbar-one-loop_PIM}
s \frac{d^2\sigma^{(1)}_{q\Bar{q}}(s,M^2,{\rm{cos}}\q)}{dM^2\, d{\rm{cos}}\q} \Biggr|_{\rm{DIS}}
&=& 
s \frac{d^2\sigma^{(1)}_{q\Bar{q}}(s,M^2,{\rm{cos}}\q)}{dM^2\, d{\rm{cos}}\q} \Biggr|_{\msb}
\, + 
\frac{1}{s} F^{\rm{Born}}_{q\Bar{q}}(s,M^2,{\rm{cos}}\q)\, 
\,
\\[1ex]
& &\hspace*{-20mm}
\times\, 
\Biggl[
\,{}- 2 C_F\,  \left[{\ln(1-z)\over 1-z} \right]_+  
+\frac{3}{2} C_F\, \left[1\over 1-z \right]_+ 
+C_F\, \d(1-z)\, 
\left(
\frac{9}{2} + 2\z_2
\right)
\Biggr] \, .\nonumber
\end{eqnarray}
The DIS two-loop NLL corrections are
\begin{eqnarray}
\label{DIS-qqbar-two-loop_PIM}
{\lefteqn{
s \frac{d^2\sigma^{(2)}_{q\Bar{q}}(s,M^2,{\rm{cos}}\q)}{dM^2\, d{\rm{cos}}\q} \Biggr|_{\rm{DIS}}
\,=\,  
s \frac{d^2\sigma^{(2)}_{q\Bar{q}}(s,M^2,{\rm{cos}}\q)}{dM^2\, d{\rm{cos}}\q} \Biggr|_{\msb}
\, + 
 \frac{1}{s}F^{\rm{Born}}_{q\Bar{q}}(s,M^2,{\rm{cos}}\q)\, 
\, }}
\\[1ex]
\NO
& &\hspace*{8mm}
\times\, 
\Biggl[
\,{}- 6 C_F^2 \left[{\ln^3(1-z)\over 1-z}\right]_+ 
\\[1ex]
& &\hspace*{12mm} 
{}+ \left[{\ln^2(1-z)\over 1-z}\right]_+  
\Biggl\{
6 C_F^2
\left( \frac{21}{12} + {\rm{Re}}L_\b
- 4 \ln\left( \frac{u_1}{t_1} \right)
- \ln\left( \frac{s}{m^2} \right)   
+ \ln\left(\frac{\mu^2}{m^2} \right) \right) 
\NO\\[1ex]
& &\hspace*{22mm}
+\, C_F b_2 
+\, 3 C_A C_F\, \left( 3 \ln\left( \frac{u_1}{t_1} \right) 
+ \ln\left( \frac{s m^2}{t_1 u_1} \right) - {\rm{Re}}L_\b \right)
\Biggr\} 
\NO\\[1ex]
\NO
& &\hspace*{12mm} 
{}+ \left[{\ln(1-z)\over 1-z}\right]_+  
\Biggl\{
- 3 \ln\left(\frac{\mu^2}{m^2}\right) C_F^2
- 6 \ln\left(\frac{\m_R^2}{m^2}\right)  C_F\,b_2 
\Biggr\} 
\Biggr] \, .
\end{eqnarray}
To achieve  NNLL accuracy for the DIS scheme at two loops we have to add
to the NNLL $\msb$ terms 
\begin{eqnarray}
\label{DIS-qqbar-two-loop_PIM-nnll}
{\lefteqn{
s \frac{d^2\sigma^{(2)}_{q\Bar{q}}(s,M^2,{\rm{cos}}\q)}{dM^2\, d{\rm{cos}}\q}
 \Biggr|_{\rm{DIS}}^{\rm NNLL \; terms}
\,=\,  
s \frac{d^2\sigma^{(2)}_{q\Bar{q}}(s,M^2,{\rm{cos}}\q)}{dM^2\, d{\rm{cos}}\q} 
\Biggr|_{\msb}^{\rm NNLL \; terms}
\,}}
\\[1ex]
\NO
& &
{}-2\, C_F\, 
{\hat T}^{(1)}_{q\Bar{q}\, \pim} (M^2,{\rm{cos}}\q) \Biggr|_{\msb} 
 \left[{\ln(1-z)\over 1-z}\right]_+
\,+\,
\frac{1}{s} F^{\rm{Born}}_{q\Bar{q}}(s,M^2,{\rm{cos}}\q)\,  
\\[1ex]
\NO
& &\hspace*{6mm}\times\, \Biggl[ 
\left[{\ln(1-z)\over 1-z}\right]_+ \,
\Biggl\{  
{}  - C_F K + 6 C_F \left( {\rm{Re}}\Gamma^{(1)}_{22} - C_F
+ C_F \ln\left(2\sqrt{\n_1\n_2} \frac{s}{m^2}\right)
\right) 
\NO\\
& &\hspace*{21mm}
{}- \frac{3}{2} C_F b_2  + 2 C_F b_2 \ln\left(\frac{s}{m^2}\right) +
         16 C_F^2 \z_2 + \frac{45}{4} C_F^2 
\Biggr\}
\NO\\
& &\hspace*{12mm}
+ \left[{1\over 1-z}\right]_+ \,
\Biggl\{  
\Biggl(- 8 C_F^2 \z_2 - \frac{45}{4} C_F^2 \Biggr) \ln\left(\frac{\mu^2}{m^2}\right) 
+ \frac{9}{2} C_F b_2 \ln\left(\frac{\mu_R^2}{m^2}\right) 
\Biggr\} \Biggr]
\, .\NO
\end{eqnarray}

\subsection{The $gg$ channel in PIM kinematics}

In the $gg$ channel, Eq. (\ref{gluon-gluon-fusion-0_PIM}), 
the Born cross section is 
\begin{eqnarray}
s \frac{d^2\sigma^{(0)}_{gg}(s,M^2,{\rm{cos}}\q)}{dM^2\, d{\rm{cos}}\q}
= \frac{\beta}{2}\o^{(0)}_{gg}(s,t_1,u_1)|_{\rm PIM}
&=& \frac{1}{s}\d(1-z)\,   F^{\rm{Born}}_{gg}(s,M^2,{\rm{cos}}\q)\, 
\end{eqnarray}
with $F^{\rm{Born}}_{gg}(s,M^2,{\rm{cos}}\q)$ defined
in Eq.~(\ref{eq:16}).
The $\msb$ one-loop NNLL corrections are
\begin{eqnarray}
\label{gg-one-loop_PIM}
{\lefteqn{
s \frac{d^2\sigma^{(1)}_{gg}(s,M^2,{\rm{cos}}\q)}{dM^2\, d{\rm{cos}}\q}
= \o^{(1)}_{gg}(1-z,s,M^2,{\rm{cos}}\q)
= \frac{\beta}{2}\o^{(1)}_{gg}(s_4,s,t_1,u_1)|_{\pim}
\, }}
\\[1ex]
&=&
\frac{1}{s} \Biggl[ 
\,4 N_c\, \left[{\ln(1-z)\over 1-z}\right]_+\, 
F^{\rm{Born}}_{gg}(s,M^2,{\rm{cos}}\q)\, 
\NO \\[1ex]
&&\hspace*{12mm}
{}+ \left[{1\over 1-z}\right]_+ \Biggl\{  
\frac{ \alpha_s^2(\m_R)\, \pi}{2(N_c^2-1)^2}\, B'_{gg} \, 
\Biggl(
N_c C_O \left\{ \left( 2 \ln\left(\frac{s}{m^2}\right) 
\right.
\right.
\NO \\[1ex]
&&\hspace*{21mm}
\left. \left.
{}- \ln\left(\frac{s m^2}{t_1 u_1}\right) 
- 1\, \right) \left( 1 - \frac{2 t_1 u_1}{s^2}\right) 
+  \ln\left(\frac{u_1}{t_1}\right)   \frac{(t_1^2-u_1^2)}{s^2} 
\right\} 
\NO \\[1ex]
&&\hspace*{21mm}
{}+ N_c C_K
\left\{ ({\rm{Re}}L_{\beta} + 1 ) 
\left( 1 - \frac{2 t_1 u_1}{s^2}\right)
- 2 \ln\left(\frac{t_1u_1}{m^4}\right) 
+ 2\, \right\} 
\NO \\[1ex]
&&\hspace*{21mm}
{}- C_{\rm{QED}} \left( {\rm{Re}}L_{\beta} + 1 \right)
\Biggr)
- 2 N_c\, F^{\rm{Born}}_{gg}(s,M^2,{\rm{cos}}\q)\, 
    \ln\left(\frac{\mu^2}{m^2}\right) \Biggr\} 
\NO \\[1ex]
&&\hspace*{12mm}
+\, \d(1-z) \left\{2\,F^{\rm{Born}}_{gg}(s,M^2,{\rm{cos}}\q)\, 
    \ln\left(\frac{\m_R^2}{\mu^2}\right) b_2 \right\}\Biggr] 
+ T^{(1)}_{gg\, \pim}(M^2,{\rm{cos}}\q)
\, . \NO
\end{eqnarray}
The $\msb$ two-loop NLL corrections are
\begin{eqnarray}
\label{gg-two-loop_PIM}
{\lefteqn{
s \frac{d^2\sigma^{(2)}_{gg}(s,M^2,{\rm{cos}}\q)}{dM^2\, d{\rm{cos}}\q}
= \o^{(2)}_{gg}(1-z,s,M^2,{\rm{cos}}\q)
= \frac{\beta}{2}\o^{(2)}_{gg}(s_4,s,t_1,u_1)|_{\pim}
\, }}
\\[1ex]
&=&
\frac{1}{s} \Biggl[ 
\,8 N_c^2\, \left[{\ln^3(1-z)\over 1-z}\right]_+\, 
F^{\rm{Born}}_{gg}(s,M^2,{\rm{cos}}\q)\, 
\NO \\[1ex]
&&\hspace*{12mm}
{}+ \left[{\ln^2(1-z)\over 1-z}\right]_+ \Biggl\{  
\frac{ \alpha_s^2(\m_R)\, \pi}{2(N_c^2-1)^2}\, B'_{gg} 6 N_c\,
\Biggl( N_c C_O \left\{ \left(
2 \, \ln\left(\frac{s}{m^2} \right)
\right.
\right.
\NO \\[1ex]
&&\hspace*{21mm}
\left. \left.
- \, \ln\left(\frac{s m^2}{t_1 u_1}\right) 
- 1\, \right) \left( 1 - \frac{2 t_1 u_1}{s^2}\right)
+ \, \ln\left(\frac{u_1}{t_1}\right)   \frac{(t_1^2-u_1^2)}{s^2} 
\right\}
\NO \\[1ex]
&&\hspace*{21mm} 
{} + N_c\, C_K
\Biggl\{ 
({\rm{Re}}L_{\beta} + 1)  \left( 1 - \frac{2 t_1 u_1}{s^2}\right) 
{}- 2\, \ln\left(\frac{t_1u_1}{m^4}\right) 
+ 2 \, \Biggr\}- C_{\rm{QED}} \left( {\rm{Re}}L_{\beta} + 1 \right)
\Biggr) 
\NO \\[1ex]
&&\hspace*{21mm}
{} - F^{\rm{Born}}_{gg}(s,M^2,{\rm{cos}}\q)\, 
\left(12 N_c^2 \ln\left(\frac{\mu^2}{m^2}\right)+ 4 b_2 N_c \right) \Biggr\} 
\NO \\[1ex]
&&\hspace*{12mm}
{}+ \left[{\ln(1-z)\over 1-z}\right]_+ 
 \Biggl\{  
\frac{ \alpha_s^2(\m_R)\, \pi}{2(N_c^2-1)^2}\, B'_{gg}\, 
\ln\left(\frac{\mu^2}{m^2}\right) 4 N_c
\Biggl(  
N_c\, C_O  
\ln\left(\frac{u_1}{t_1}\right) \frac{(u_1^2-t_1^2)}{s^2}
\NO \\[1ex]
&&\hspace*{21mm}
{} - N_c\, C_K \left\{ 
({\rm{Re}}L_{\beta} + 1 ) \left( 1 - \frac{2 t_1 u_1}{s^2}\right) +
\ln\left(\frac{s m^2}{t_1 u_1}\right) + 1  \right\} 
\NO \\[1ex]
&&\hspace*{21mm}
{}+ C_{\rm{QED}} \left( {\rm{Re}}L_{\beta} + 1 \right)
\Biggr) 
+ 4 N_c\, F^{\rm{Born}}_{gg}(s,M^2,{\rm{cos}}\q)\, 
\ln\left(\frac{\mu^2}{m^2}\right)
\Biggl( 
- 2 N_c\, \ln\left(\frac{s}{m^2}\right)  
\NO \\[1ex]
&&\hspace*{21mm}
{} + N_c\, \ln\left(\frac{s m^2}{t_1 u_1}\right) - 2 b_2  + N_c 
+ N_c\, \ln\left(\frac{\mu^2}{m^2}\right) \Biggr)
\NO \\[1ex]
&&\hspace*{21mm}
{} + 12 N_c\, F^{\rm{Born}}_{gg}(s,M^2,{\rm{cos}}\q)\, 
\ln\left(\frac{\m_R^2}{m^2}\right) b_2
\Biggr\} 
\NO \\[1ex]
& &\hspace*{12mm}+\, 
\left[{1\over 1-z}\right]_+\, N_c\, F^{\rm{Born}}_{gg}(s,M^2,{\rm{cos}}\q)\,
\ln\left(\frac{\mu^2}{m^2}\right) 
\Biggl\{
5
\ln\left(\frac{\mu^2}{m^2}\right) 
- 6 
\ln\left(\frac{\m_R^2}{m^2}\right) 
\Biggr\}  b_2\, 
\Biggr] \, . \NO
\end{eqnarray}
To achieve NNLL accuracy to two-loops one must add
\begin{eqnarray}
\label{gg-two-loop_PIM-nnll}
& &
2\, C_A\, 
\hat{T}^{(1)}_{gg\, \pim} (M^2,{\rm{cos}}\q)
\Biggl\{
2 \left[{\ln(1-z)\over 1-z}\right]_+ 
- \left[{1\over 1-z}\right]_+ \ln\left(\frac{\mu^2}{m^2}\right)
\Biggr\}
\NO\\[1ex]
& &+\,
\frac{1}{s} \left[{\ln(1-z)\over 1-z}\right]_+\, 
\Biggl( \,
F^{\rm{Born}}_{gg}(s,M^2,{\rm{cos}}\q)\, ( - 16 \z_2 N_c^2 + 2 N_c\, K )
\NO \\[1ex]
&&\hspace*{6mm}
{}+ \frac{\a_s^2(\m_R) \pi}{2 (N_c^2-1)}\, B'_{gg}\, 
\left( 1 - \frac{2 t_1 u_1}{s^2}\right)
\Biggl\{  \left( N_c + \frac{1}{4} N_c^3 \right) \left( \Gamma^{(1)}_{31} \right)^2 
\NO \\[1ex]
&&\hspace*{21mm}
  {}      - 4 N_c b_2 \left({\rm{Re}}\Gamma^{(1)}_{22} - C_A
+ C_A \ln\left(2\sqrt{\n_1\n_2} \frac{s}{m^2}\right)
\right) 
\NO \\[1ex]
&&\hspace*{21mm}
    {}    + 4 N_c \left({\rm{Re}} \Gamma^{(1)}_{22} - C_A
+ C_A \ln\left(2\sqrt{\n_1\n_2} \frac{s}{m^2}\right)
\right)^2 
\Biggr\} 
\NO \\[1ex]
&&\hspace*{6mm}
{}+\frac{\a_s^2(\m_R) \pi}{2 (N_c^2-1)}\, B'_{gg}\, 
\frac{(t_1^2-u_1^2)}{s^2}
\Biggl\{  
          4 \left({\rm{Re}} \Gamma^{(1)}_{11} - C_A
+ C_A \ln\left(2\sqrt{\n_1\n_2} \frac{s}{m^2}\right)
\right) \Gamma^{(1)}_{31} 
\NO \\[1ex]
&&\hspace*{21mm}
  {}      + 2  (N_c^2 - 2)\Gamma^{(1)}_{31} \left( {\rm{Re}} \Gamma^{(1)}_{22} - C_A
+ C_A \ln\left(2\sqrt{\n_1\n_2} \frac{s}{m^2}\right)
\right) 
        - N_c^2 b_2 \Gamma^{(1)}_{31}  
\Biggr\} 
\NO \\[1ex]
&&\hspace*{6mm}
{} + \frac{\a_s^2(\m_R) \pi}{2 (N_c^2-1)}\, B'_{gg}\, 
\Biggl\{   
         - 4 \frac{1}{N_c}  b_2 \left( {\rm{Re}}\Gamma^{(1)}_{11} - C_A
+ C_A \ln\left(2\sqrt{\n_1\n_2} \frac{s}{m^2}\right)
\right)
\NO \\[1ex]
&&\hspace*{21mm}
   {}     + 4 \frac{1}{N_c}  \left({\rm{Re}} \Gamma^{(1)}_{11} - C_A
+ C_A \ln\left(2\sqrt{\n_1\n_2} \frac{s}{m^2}\right)
\right)^2 
\NO \\[1ex]
&&\hspace*{21mm}
     {}   - \frac{N_c}{2} \left( \Gamma^{(1)}_{31} \right)^2 
        + 8 \frac{1}{N_c}  b_2 \left( {\rm{Re}}\Gamma^{(1)}_{22} - C_A
+ C_A \ln\left(2\sqrt{\n_1\n_2} \frac{s}{m^2}\right)
\right) 
\NO \\[1ex]
&&\hspace*{21mm}
     {}   - 8 \frac{1}{N_c} \left( {\rm{Re}} \Gamma^{(1)}_{22} - C_A
+ C_A \ln\left(2\sqrt{\n_1\n_2} \frac{s}{m^2}\right)
\right)^2  
\Biggr\} \Biggr)
\NO\\[1ex]
& &+ \frac{1}{s} \left[{1\over 1-z}\right]_+ 
\Biggl( F^{\rm{Born}}_{gg}(s,M^2,{\rm{cos}}\q) 
\ln\left(\frac{\mu^2}{m^2}\right) (- N_c  K + 8 N_c^2 \zeta_2 )
\NO \\[1ex]
&&\hspace*{6mm}
+ \frac{\a_s^2(\m_R) \pi}{2 (N_c^2-1)}\, B'_{gg}\, 
\left( 1 - \frac{2 t_1 u_1}{s^2}\right)
\Biggl\{ 6 \ln\left(\frac{\mu_R^2}{m^2}\right) 
- 4 \ln\left(\frac{\mu^2}{m^2}\right) \Biggr\} N_c b_2 
\NO \\[1ex]
&&\hspace*{21mm}
\times
\left( {\rm{Re}} \Gamma^{(1)}_{22} - C_A
+ C_A \ln\left(2\sqrt{\n_1\n_2} \frac{s}{m^2}\right)
\right)
\NO \\[1ex]
&&\hspace*{6mm}
+ \frac{\a_s^2(\m_R) \pi}{2 (N_c^2-1)}\, B'_{gg}\, 
\frac{(t_1^2-u_1^2)}{s^2}
\Biggl\{ \frac{3}{2} \ln\left(\frac{\mu_R^2}{m^2}\right) 
- \ln\left(\frac{\mu^2}{m^2}\right) \Biggr\} N_c^2 b_2 \Gamma^{(1)}_{31}  
\NO \\[1ex]
&&\hspace*{6mm}
+ \frac{\a_s^2(\m_R) \pi}{2 (N_c^2-1)}\, B'_{gg}\, 
\Biggl\{ 6 \ln\left(\frac{\mu_R^2}{m^2}\right) 
- 4 \ln\left(\frac{\mu^2}{m^2}\right) \Biggr\} \frac{b_2}{N_c}
\NO \\[1ex]
&&\hspace*{21mm}
\times \Biggl\{
\left( {\rm{Re}} \Gamma^{(1)}_{11} - C_A
+ C_A \ln\left(2\sqrt{\n_1\n_2} \frac{s}{m^2}\right)
\right)
\NO\\[1ex]
& &\hspace*{21mm}
- 2
\left( {\rm{Re}} \Gamma^{(1)}_{22} - C_A
+ C_A \ln\left(2\sqrt{\n_1\n_2} \frac{s}{m^2}\right)
\right) \Biggr\}
\Biggr)
\NO\\[1ex]
& &+ \frac{1}{s} \delta(1-z) F^{\rm{Born}}_{gg}(s,M^2,{\rm{cos}}\q)\,
\Biggl\{  - 2 \z_2 N_c^2 \ln^2\left(\frac{\mu^2}{m^2}\right) 
+ 3 b_2^2 \ln^2\left(\frac{\mu_R^2}{\mu^2}\right) 
\Biggr\}\, .
\end{eqnarray}

As we discussed in the previous subsection, 
we can also derive the NNLO $\delta(1-z)$ terms 
involving single logarithms of the 
scale. These terms are 
\beqa
&& \frac{1}{s} F^{\rm Born}_{gg}(s,M^2,{\rm{cos}}\q) 
\delta(1-z)\left\{\ln\left(\frac{\mu_F^2}{m^2}\right)
\left[-8\zeta_3 C_A^2-4 C_A^2 \zeta_2 \ln\left(\frac{m^2}{s}\right)
-\frac{b_2}{8} {\hat {T'}}^{(1)}_{gg\,\rm PIM}(M^2,{\rm{cos}}\q)
\right]\right.
\nonumber \\ && \hspace{55mm} \left.
{}+\ln\left(\frac{\mu_R^2}{m^2}\right)
\left[\frac{b_3}{128}+\frac{3}{16}b_2 
{\hat {T'}}^{(1)}_{gg\,\rm PIM}(M^2,{\rm{cos}}\q)\right]\right\}
\nonumber \\ &&
{}+\frac{\alpha_s^2 \pi}{(N_c^2-1)^2} B'_{gg}
\ln\left(\frac{\mu_F^2}{m^2}\right) 2C_A \zeta_2 \, \delta(1-z) 
\left\{N_c(N_c^2-1)\frac{(t_1^2+u_1^2)}{s^2}
\left[\left(-C_F+\frac{C_A}{2}\right)
{\rm Re} L_{\beta}\right. \right.
\nonumber \\ && \hspace{20mm} \left.
{}+\frac{C_A}{2}\ln\left(\frac{t_1u_1}{m^2s}\right)
-C_F\right]+\frac{(N_c^2-1)}{N_c}(C_F-C_A) {\rm Re} L_{\beta}
\nonumber \\ && \hspace{20mm} \left.
{}-(N_c^2-1)\ln\left(\frac{t_1u_1}{m^2s}\right)
+C_F \frac{(N_c^2-1)}{N_c}
+\frac{N_c^2}{2}(N_c^2-1)
\ln\left(\frac{u_1}{t_1}\right)\frac{(t_1^2-u_1^2)}{s^2} \right\} \,,
\eeqa
with ${\hat {T'}}^{(1)}_{gg\,\rm PIM}(M^2,{\rm{cos}}\q)
=s\, {\hat T}^{(1)}_{gg\,\rm PIM}(M^2,{\rm{cos}}\q)/
F^{\rm{Born}}_{gg}(s,M^2,{\rm{cos}}\q)$.
Again, we have checked that these results are consistent with 
Eq. (\ref{ex-f21gg}) and with the expansion of the resummed cross 
section beyond NNLL accuracy.

\subsection{Results for $s\rightarrow 4m^2$}
\label{sec:results-sright-4m2}

Here we present the scaling functions
of section \ref{sec:scalingfunctions}
up to two loops for the inclusive cross section
for $s\rightarrow 4m^2$. They may be obtained
from the results in sections B.1 to B.4 via
Eqs. (\ref{eq:11}) and (\ref{totalpartoncrs_PIM}),
keeping only terms that behave as $\ln(\beta)$.
We checked that we find the same results for both kinematics.
Our results are as follows. 

For the $q\bar{q}$ channel in the ${\msb}$ scheme 
\begin{eqnarray}
\label{qqbar-beta-one-loop}
f^{(1,0)}_{q\Bar{q}} &=& 
\frac{1}{4\p^2}\, f^{(0,0)}_{q\Bar{q}}
\Biggr\{  8 C_F\,  \ln^2\left(\b\right) 
    + 
    \left( 24 C_F \ln(2) - 16 C_F - 2 C_A 
    \right)\, \ln\left(\b \right)
\Biggr\}\, ,
\end{eqnarray}
\begin{eqnarray}
\label{qqbar-beta-two-loop}
f^{(2,0)}_{q\Bar{q}} &=& 
\frac{1}{16\p^4}\, f^{(0,0)}_{q\Bar{q}}
\Biggr\{  32 C_F^2\,  \ln^4\left(\b\right) 
    + \frac{16}{3} C_F 
    \left(  36 C_F \ln(2) - 24 C_F - 3 C_A  - 2 b_2
    \right)\, \ln^3\left(\b \right)
\Biggr\}\, ,
\nonumber \\ &&
\end{eqnarray}
while in the DIS scheme we have
\begin{eqnarray}
\label{DIS-qqbar-beta-one-loop}
f^{(1,0)}_{q\Bar{q}} &=& 
\frac{1}{4\p^2}\, f^{(0,0)}_{q\Bar{q}}
\Biggr\{  4 C_F\,  \ln^2\left(\b\right) 
    + 
    \left( 16 C_F \ln(2) - 5 C_F - 2 C_A 
    \right)\, \ln\left(\b \right)
\Biggr\}\, ,
\end{eqnarray}
\begin{eqnarray}
\label{DIS-qqbar-beta-two-loop}
f^{(2,0)}_{q\Bar{q}} &=& 
\frac{1}{16\p^4}\, f^{(0,0)}_{q\Bar{q}}
\Biggr\{   8 C_F^2\,  \ln^4\left(\b\right) 
    + 4 C_F 
    \left( 16 C_F \ln(2)  - 5 C_F - 2 C_A - 2 b_2
    \right)\, \ln^3\left(\b \right)
\Biggr\}\,.
\nonumber \\ &&
\end{eqnarray}
For the $gg$ channel we find
\begin{eqnarray}
\label{gg-beta-one-loop}
f^{(1,0)}_{gg} &=& 
\frac{1}{4\p^2}\, f^{(0,0)}_{gg}
\Biggr\{  8 C_A\,  \ln^2\left(\b\right) 
    + \left( 
       24 C_A \ln(2) - 18 C_A + \frac{4 C_A}{N_c^2-2} 
    \right)\, \ln\left(\b \right)
\Biggr\}\, ,
\end{eqnarray}
\begin{eqnarray}
\label{gg-beta-two-loop}
f^{(2,0)}_{gg} &=& 
\frac{1}{16\p^4}\, f^{(0,0)}_{gg}
\Biggr\{   32 C_A^2\,  \ln^4\left(\b\right) 
    + \frac{16}{3} C_A 
    \left( 36 C_A \ln(2) - 27 C_A - 2 b_2
    + \frac{6 C_A}{N_c^2-2} 
    \right)\, \ln^3\left(\b \right)
\Biggr\}\, .
\nonumber \\ &&
\end{eqnarray}

\end{document}